\documentstyle[11pt]{article}
\makeatletter \@addtoreset{equation}{section} \makeatother

\newcommand{\noi}{\vspace{12pt}\noindent}
\newcommand{\beq}{\begin{equation}}
\newcommand{\eeq}{\end{equation}}
\newcommand{\bea}{\begin{eqnarray}}
\newcommand{\eea}{\end{eqnarray}}

\newcommand{\e}[1]{{(\ref{#1})}}
\newcommand{\eq}[1]{{eq.\ (\ref{#1})}}
\newcommand{\es}[2]{{(\ref{#1}) and (\ref{#2})}}
\newcommand{\eqs}[2]{{eqs.\ (\ref{#1}) and (\ref{#2})}}
\newcommand{\Ref}[1]{{Ref.~\cite{#1}}}

\newcommand{\mb}[1]{{\mbox{${#1}$}}}

\newcommand{\ie}{{${ i.e.\ }$}}

\newcommand{\cf}{{cf.\ }}
\newcommand{\wrt}{{with respect to }}
\newcommand{\wtho}{{with the help of }}
\newcommand{\lhs}{{left-hand side }}
\newcommand{\rhs}{{right-hand side }}
\newcommand{\hla}{{homotopy Lie algebra }}
\newcommand{\hlas}{{homotopy Lie algebras }}
\newcommand{\odd}{{\rm odd}}
\newcommand{\even}{{\rm even}}
\newcommand{\ad}{{\rm ad}}

\newcommand{\Sym}{{\rm Sym}}
\newcommand{\nth}{\mb{n^{\prime}}{\em th} }

\newcommand{\Hf}{{\frac{1}{2}}}

\newcommand{\soprod}{{\odot}}
\newcommand{\oprod}{{\otimes}}
\newcommand{\bb}{{b}}
\newcommand{\cc}{{c}}
\newcommand{\BB}{{B}}
\newcommand{\PP}{{P}}
\newcommand{\cE}{{\cal E}}

\newcommand{\ddd}{{\rm d}}
\newcommand{\DD}{{D}^{}}
\newcommand{\HH}{{H}^{}}
\newcommand{\QQ}{{Q}^{}}
\newcommand{\SS}{{S}^{}}
\newcommand{\XX}{{X}^{}}
\newcommand{\YY}{{Y}^{}}

\newcommand{\twobyone}[2]{\left(\begin{array}{c}{#1} \cr
                                {#2} \end{array} \right)}

\newcommand{\deder}[1]{{ 
 {\stackrel{\raise.1ex\hbox{$\leftarrow$}}{\delta^r}   } 
\over {   \delta {#1}}  }}
\newcommand{\dedel}[1]{{ 
 {\stackrel{\lower.3ex \hbox{$\rightarrow$}}{\delta^{\ell}}   }
 \over {   \delta {#1}}  }}
\newcommand{\papar}[1]{{ 
 {\stackrel{\raise.1ex\hbox{$\leftarrow$}}{\partial^r}   } 
\over {   \partial {#1}}  }}
\newcommand{\papal}[1]{{ 
 {\stackrel{\lower.3ex \hbox{$\rightarrow$}}{\partial^{\ell}}   }
 \over {   \partial {#1}}  }}
\newcommand{\ddr}[1]{{ 
 {\stackrel{\raise.1ex\hbox{$\leftarrow$}}{\delta^r}   } 
\over {   \delta {#1}}  }}
\newcommand{\ddl}[1]{{ 
 {\stackrel{\lower.3ex \hbox{$\rightarrow$}}{\delta^{\ell}}   }
 \over {   \delta {#1}}  }}

\newcommand{\proofbox}{\begin{flushright}
${\,\lower0.9pt\vbox{\hrule \hbox{\vrule
height 0.2 cm \hskip 0.2 cm \vrule height 0.2 cm}\hrule}\,}$
\end{flushright}}

\newtheorem{theorem}{Theorem}[section]

\newtheorem{corollary}[theorem]{Corollary}
\newtheorem{definition}[theorem]{Definition}

\newtheorem{proposition}[theorem]{Proposition}

\begin{document}
\thispagestyle{empty}
\vspace{3cm}
%\begin{center}
\title{\Large{\bf Non-Commutative Batalin-Vilkovisky Algebras, \\
Homotopy Lie Algebras and the Courant Bracket}}
%\end{center}
\vspace{2cm}
%\begin{center}
\author{{\sc K.~Bering}$^1$\\Institute for Theoretical Physics \& Astrophysics
\\Masaryk University\\Kotl\'a\v{r}sk\'a 2\\CZ-611 37 Brno\\Czech Republic}
%\end{center}
\maketitle

\begin{abstract}
We consider two different constructions of higher brackets. First, based on
a Grassmann-odd, nilpotent $\Delta$ operator, we define a non-commutative
generalization of the higher Koszul brackets, which are used in a generalized
Batalin-Vilkovisky algebra, and we show that they form a homotopy Lie algebra.
Secondly, we investigate higher, so-called derived brackets built from
symmetrized, nested Lie brackets with a fixed nilpotent Lie algebra element
$Q$. We find the most general Jacobi-like identity that such a hierarchy
satisfies. The numerical coefficients in front of each term in these
generalized Jacobi identities are related to the Bernoulli numbers. We suggest
that the definition of a homotopy Lie algebra should be enlarged to
accommodate this important case. Finally, we consider the Courant bracket as
an example of a derived bracket. We extend it to the ``big bracket'' of
exterior forms and multi-vectors, and give closed formulas for the higher
Courant brackets.
\end{abstract}

\vspace{10mm}

\vfill
\begin{quote}
%PACS number(s): \\
\hrule width 5.cm \vskip 2.mm \noindent 
$^{1}${\small E-mail:~{\tt bering@physics.muni.cz}} \\ 
\end{quote}

\vfill
\newpage

\tableofcontents

\setcounter{equation}{0}
\section{Introduction}

\noi
It is well-known \cite{voronov03} that in general the symmetrized, multiple
nested Lie brackets 
\beq
[[\ldots,[[Q,a_{1}],a_{2}],\ldots],a_{n}]~,
\label{symnest}
\eeq
where \mb{Q} is a fixed nilpotent Lie algebra element \mb{[Q,Q]=0}, do not
obey the original \hla definition of Lada and Stasheff \cite{ladastasheff93}.
Several papers have been devoted to tackle this in special situations. {}For
instance, Voronov considers the projection of above nested, so-called derived
brackets \e{symnest} into an Abelian subalgebra 
\cite{voronov03,voronov04k,voronov04,akman05}. In this paper we stay in the
non-Abelian setting and observe that although a multiple nested bracket
\e{symnest} does not obey the generalized Jacobi identities of Lada and
Stasheff \cite{ladastasheff93}, it is -- after all -- very close. It turns out
 that one may organize the nested Lie brackets \e{symnest} in such a way that
all the terms in the generalized Jacobi identities of Lada and Stasheff
appear, but as a caveat, with different numerical prefactors related to the
Bernoulli numbers.

\noi
The paper is organized as follows. In Section~\ref{secdefshliealg} we widen
the definition of a \hla by basing it on a generalized bracket product to
allow for more general prefactors. In Section~\ref{seckoszul} we consider the 
Koszul bracket hierarchy \cite{koszul85,akman97,ad96,bda96}, and solve a 
long-standing problem of providing an ordering prescription for a 
construction of higher Koszul brackets for a non-commutative algebra
\mb{\cal A}, in such a way that the higher brackets form a homotopy
Lie algebra. It turns out that in the non-commutative Koszul construction, 
the emerging \hla is of the original type considered by Lada and Stasheff 
\cite{ladastasheff93}, \cf Theorem~\ref{theoremkoszul}. On the other hand,
the new types of (generalized) \hlas with non-trivial prefactors will be
essential for the
derived bracket hierarchies \e{symnest} studied in Section~\ref{secderoperad}.
Section~\ref{seccourant} is devoted to the Courant bracket \cite{courant90},
which is a two-bracket defined on a direct sum of the tangent and the 
cotangent bundle \mb{TM\oplus T^{*}M} over a manifold \mb{M}, 
\beq
[X\oplus\xi,Y\oplus\eta]_\HH
~=~[X,Y]\oplus\left({\cal L}_\XX\eta-{\cal L}_\YY\xi
+\Hf d\left(i_\YY\xi-i_\XX\eta\right)+i_\XX i_\YY H\right)~,
\label{twistedcourantbracket}
\eeq
where \mb{H} is a closed ``twisting'' three-form. This bracket
has many interesting applications, for instance Hitchin's generalized complex
geometry \cite{gualtieri04,hitchin03}. In hindsight, the importance of the
Courant bracket can be traced to the fact that it belongs to a derived \hla 
\cite{roytenberg98,roytenberg99,roytenberg02} related to the exterior de Rham
complex. Section~\ref{secsuppl} contains further theoretical aspects of
homotopy Lie algebras. The pre-Lie property of a bracket product is
investigated in Subsection~\ref{secprelie}, and the co-algebraic structures
are studied in Subsection~\ref{seccoproduct}-\ref{secprops}. {}Finally,
Section~\ref{secconclusion} has our conclusions.

\section{Homotopy Lie Algebras}
\label{secdefshliealg}

\noi
Let \mb{\Sym^{\bullet}_{\epsilon}{\cal A}:=T^{\bullet}{\cal A} / I} denote a
graded\footnote{Adjectives from supermathematics such as ``graded'',
``super'', etc., are from now on implicitly implied. We will also follow 
commonly accepted superconventions, such as, Grassmann parities are only 
defined modulo \mb{2}, and ``nilpotent'' means ``nilpotent of order \mb{2}''.

} symmetric tensor algebra
over a graded vector space \mb{{\cal A}}, where 
\mb{I\subseteq~T^{\bullet}{\cal A}\equiv\bigoplus_{n\geq 0}T^{n}{\cal A}}
is the two-sided ideal generated by the set 
\beq
\left\{\left.
b\oprod a-(-1)^{(\epsilon_{a}+\epsilon)(\epsilon_{b}+\epsilon)}
a\oprod b \right| a,b\in{\cal A}  \right\}~
\subseteq~T^{2}{\cal A}~\equiv~{\cal A}\oprod{\cal A}~.
\label{ideal}
\eeq
Here \mb{\epsilon\in\{0,1\}} modulo \mb{2} is a fixed ``suspension parity''. 
We let the symbols ``\mb{\oprod}'' and ``\mb{\soprod}'' denote the
un-symmetrized and the symmetrized tensor product in the tensor algebras
\mb{T^{\bullet}{\cal A}} and \mb{\Sym^{\bullet}_{\epsilon}{\cal A}}, 
respectively. In practice we shall focus on the symmetric tensor product
``\mb{\soprod}'', and the only important thing is, that two arbitrary vectors 
\mb{a,b \in{\cal A}}, with Grassmann parities \mb{\epsilon_{a}} and 
\mb{\epsilon_{b}}, commute in \mb{\Sym^{2}_{\epsilon}{\cal A}} up to the
following sign convention:
\beq
b\soprod a~=~(-1)^{(\epsilon_{a}+\epsilon)(\epsilon_{b}+\epsilon)}
a\soprod b~\in~\Sym^{2}_{\epsilon}{\cal A}~. \label{permutationsign}
\eeq
A \mb{\bullet}-bracket \mb{\Phi:~\Sym^{\bullet}_{\epsilon}{\cal A}\to{\cal A}}
is a collection of multi-linear \mb{n}-brackets 
\mb{\Phi^{n}:~\Sym^{n}_{\epsilon}{\cal A} \to {\cal A}}, where 
\mb{n\in\{0,1,2, \ldots\}} runs over the non-negative integers. In addition
a \mb{\bullet}-bracket \mb{\Phi} carries an intrinsic Grassmann 
parity \mb{\epsilon_{\Phi}\in\{0,1\}}. Detailed explanations of sign 
conventions are relegated to Subsection~\ref{secsigconv}. We now introduce a 
\mb{\bullet}-bracket product denoted with a ``\mb{\circ}''.

\begin{definition}
Let there be given a set of complex numbers \mb{c^{n}_{k}} where
\mb{n\!\geq\! k\!\geq\! 0}. The {\bf ``\mb{\circ}'' product} 
\mb{\Phi\circ\Phi^{\prime}:~\Sym^{\bullet}_{\epsilon}{\cal A}\to{\cal A}}
of two \mb{\bullet}-brackets
\mb{\Phi,\Phi^{\prime}:~\Sym^{\bullet}_{\epsilon}{\cal A}\to{\cal A}}
is then defined as
\beq
  (\Phi\circ\Phi^{\prime})^{n}(a_{1},\ldots,a_{n})
~:=~\sum_{k=0}^{n}\frac{c^{n}_{k}}{k!(n\!-\!k)!}
\sum_{\pi\in S_{n}}(-1)^{\epsilon_{\pi,a}}
\Phi^{n-k+1}\left(\Phi^{\prime k}(a_{\pi(1)}, \ldots, a_{\pi(k)}),
a_{\pi(k+1)}, \ldots, a_{\pi(n)}\right) \label{operadproduct}
\eeq
for \mb{n\in\{0,1,2, \ldots\}}.
\end{definition}

\begin{definition}
The ``\mb{\circ}'' product is {\bf non-degenerate} if the complex 
coefficients \mb{c^{n}_{k}}, \mb{n\geq k\geq 0}, satisfy
\beq
\forall n\in\{0,1,2,\ldots\}~\exists k\in\{0,1,\ldots,n\}:~c^{n}_{k}~\neq~0~.
\label{nondegenerateproduct}
\eeq
\end{definition}

\noi
A priori we shall not assume any other properties of this product, like for
instance associativity or a pre-Lie property. See Subsection~\ref{secprelie}
for further discussions of potential product properties. The aim of this 
paper is to determine values of the \mb{c^{n}_{k}} coefficients that lead to
useful products, guided by important examples. We first generalize an
important definition of Lada and Stasheff \cite{ladastasheff93}.  

\begin{definition}
A vector space \mb{{\cal A}} with a Grassmann-odd  \mb{\bullet}-bracket 
\mb{\Phi:~\Sym^{\bullet}_{\epsilon}{\cal A}\to{\cal A}} is a {\bf (generalized)
\hla} if the \mb{\bullet}-bracket \mb{\Phi} is nilpotent \wrt a 
non-degenerate ``\mb{\circ}'' product, 
\beq
\Phi\circ\Phi~=~0~,~~~~~~~~~~~~~~~~~~~\epsilon_{\Phi}~=~1~.\label{nilprel}
\eeq
\end{definition}

\noi
The infinite hierarchy of nilpotency relations behind \e{nilprel} are
also known as ``generalized Jacobi identities'' \cite{ladamarkl95} or ``main
identities'' \cite{zwiebach93,bda96}. The first few relations will be 
displayed in detail in Subsection~\ref{seclikealg}. In the original \hla
definition of Lada and Stasheff \cite{ladastasheff93} the product 
coefficients are fixed to be
\beq
c^{n}_{k}~=~1~,
\label{stasheffchoice}
\eeq
\cf Subsection~\ref{seclikealg}.
We shall {\em not} assume \e{stasheffchoice} because important examples are
incompatible with this restriction, \cf Section~\ref{secderoperad}. Instead
we adapt the non-degeneracy condition \e{nondegenerateproduct}. 
The \mb{A_{\infty}} definition \cite{stasheff63} can similarly be generalized.
We note that we shall in general lose an auxiliary description of a \hla in
terms of a nilpotent co-derivation, \cf Subsection~\ref{secprops}.

\noi
Our goal is to determine universal values of the \mb{c^{n}_{k}} 
coefficients that generate important classes of homotopy Lie 
algebras. By the word ``universal'' we mean that a particular set of 
\mb{c^{n}_{k}} coefficients works within an entire class of 
\mb{\bullet}-brackets. {}For instance, the Bernoulli numbers will
play an important r\^ole for the so-called derived brackets, \cf
Section~\ref{secderoperad}.

\noi
The above algebraic \hla construction has a geometric
generalization to vector bundles \mb{E=\coprod_{p\in M}E_{p}} over a manifold
\mb{M}, where each fiber space \mb{E_{p}} is a homotopy Lie algebra. However,
for most of this paper, it is enough to work at the level of a single fiber.

\subsection{Sign Conventions}
\label{secsigconv}

\noi
The sign factor \mb{(-1)^{\epsilon_{\pi,a}}} in the product definition
\e{operadproduct} arises from introducing a sign \e{permutationsign} each
time two neighboring elements of the symmetric tensor 
\mb{a_{1}\soprod\ldots\soprod a_{n}} are exchanged to form a permuted 
tensor \mb{a_{\pi(1)}\soprod\ldots\soprod a_{\pi(n)}}, \ie
working in \mb{\Sym^{n}_{\epsilon}{\cal A}}, we have
\beq
a_{\pi(1)}\soprod\ldots\soprod a_{\pi(n)}
~=~(-1)^{\epsilon_{\pi,a}}a_{1}\soprod\ldots\soprod a_{n}~\in~
\Sym^{n}_{\epsilon}{\cal A}~.
\eeq
In detail, the sign conventions are 
\bea
\epsilon_{\Phi\circ \Phi^{\prime}}
&=&\epsilon_{\Phi}+\epsilon_{\Phi^{\prime}}~,\label{evencircle}\\
\epsilon(\Phi^{n}(a_{1},\ldots,a_{n}))
&=&\sum_{i=1}^{n}\epsilon_{a_{i}}+(n-1)\epsilon+\epsilon_{\Phi}~,\\
\Phi^{n}(a_{1},\ldots,a_{i},a_{i+1},\ldots,a_{n})
&=&(-1)^{(\epsilon_{i}+\epsilon)(\epsilon_{i+1}+\epsilon)}
\Phi^{n}(a_{1},\ldots,a_{i+1}, a_{i},\ldots,a_{n})~, \\
\Phi^{n}(a_{1},\ldots,a_{i}\lambda,a_{i+1},\ldots,a_{n})
&=&(-1)^{\epsilon \epsilon_{\lambda}} 
\Phi^{n}(a_{1},\ldots,a_{i},\lambda a_{i+1},\ldots,a_{n})~, \\
\Phi^{n}(\lambda a_{1},a_{2},\ldots,a_{n})
&=&(-1)^{\epsilon_{\lambda}\epsilon_{\Phi}}\lambda 
\Phi^{n}( a_{1},a_{2},\ldots,a_{n})~,\\
\Phi^{n}(a_{1},\ldots,a_{n}\lambda)
&=& \Phi^{n}(a_{1},\ldots,a_{n})\lambda~, \\
a\lambda&=&(-1)^{\epsilon_{\lambda} \epsilon_{a}}\lambda a~,
\eea 
where \mb{\lambda} is a supernumber of Grassmann parity 
\mb{\epsilon_{\lambda}}. It is useful to memorize these sign conventions by
saying that a symbol ``\mb{\Phi}'' carries Grassmann parity
\mb{\epsilon_{\Phi}}, while a ``comma'' and  tensor-symbols ``\mb{\oprod}''
and ``\mb{\soprod}'' carry Grassmann parity \mb{\epsilon}.
Note however that the zero-bracket \mb{\Phi^{0}} has Grassmann parity 
\mb{\epsilon_{\Phi}\!+\!\epsilon}. The bracket product ``\mb{\circ}'' is 
Grassmann-even, \mb{\epsilon(\circ)=0}, \cf \eq{evencircle}.
While the sign implementation may vary with authors and applications, we
stress that the Grassmann-odd nature of a \mb{\bullet}-bracket \mb{\Phi}
is an unavoidable, characteristic feature of a homotopy Lie algebra,
\cf \eq{nilprel}. 

\noi
We remark that one could in principle bring different
kinds of Grassmann parities \mb{\epsilon^{(i)}} into play, where an upper index
\mb{i\in I} labels the different species. In that case the 
\eq{permutationsign} should be replace by
\beq
b\soprod a~=~a\soprod b\prod_{i\in I}
(-1)^{(\epsilon^{(i)}_{a}+\epsilon^{(i)})(\epsilon^{(i)}_{b}+\epsilon^{(i)})}
~. \label{permutationsignmorespecies}
\eeq
As an example the exterior form degree could be assigned to a different type
of parity. This could provide more flexible conventions for certain systems.
Nevertheless, we shall only consider one type of parity in this paper for the 
sake of simplicity.

\subsection{Connection to Lie Algebras}
\label{seclikealg}

\noi
The importance of the \hla construction is underscored by the fact that the 
two-bracket \mb{\Phi^{2}(a,b)} of a Grassmann-odd \mb{\bullet}-bracket 
\mb{\Phi} gives rise to a Lie-like bracket \mb{[\cdot,\cdot]} of opposite 
parity \mb{\epsilon^{\prime}:=1-\epsilon},
\beq
[a,b]~:=~(-1)^{(\epsilon_{a}+\epsilon^{\prime})}\Phi^{2}(a,b)
~,~~~~~~~~~~~~~~a,b\in{\cal A}~,~~~~~~~~~~~~\epsilon_{\Phi}=1~.
\label{lielikebracket}
\eeq
(This particular choice of sign is natural for a derived bracket,
\cf Section~\ref{secderoperad}. Note that in the context of the Koszul 
bracket hierarchy and Batalin-Vilkovisky algebras the opposite sign convention
is usually adapted, \ie 
\mb{[a,b]:=(-1)^{(\epsilon_{a}+\epsilon)}\Phi^{2}(a,b)}, \cf 
Section~\ref{seckoszul}.) The bracket \e{lielikebracket} satisfies
 bi-linearity and skewsymmetry
\bea
\epsilon([a,b])&=&\epsilon_{a}+\epsilon_{b}+\epsilon^{\prime}~,\\
{} [ \lambda a,b \mu ]&=&\lambda [ a,b] \mu~, \\
{} [a\lambda, b]&=&(-1)^{\epsilon^{\prime}\epsilon_{\lambda}}[a,\lambda b]~,\\
{} [b,a]&=&-(-1)^{(\epsilon_{a}+\epsilon^{\prime})
(\epsilon_{b}+\epsilon^{\prime})}[a,b]~,
\eea
where \mb{\lambda,\mu} are supernumbers. The failure (if any) of the Jacobi 
identity 
\beq
\sum_{a,b,c~{\rm cycl.}}
(-1)^{(\epsilon_{a}+\epsilon^{\prime})(\epsilon_{c}+\epsilon^{\prime})}
[[a,b],c]~=~(-1)^{\epsilon_{b}+\epsilon^{\prime}
+(\epsilon_{a}+\epsilon^{\prime})(\epsilon_{c}+\epsilon^{\prime})}
{\rm Jac}(a,b,c)\label{jacobiator2}
\eeq
is measured by the Jacobiator 
\mb{{\rm Jac}:~\Sym^{3}_{\epsilon}{\cal A}\to{\cal A}}, defined as
\beq
{\rm Jac}(a_{1},a_{2},a_{3})~:=~
\Hf \sum_{\pi\in S_{3}}(-1)^{\epsilon_{\pi,a}}
\Phi^{2}\left(\Phi^{2}(a_{\pi(1)},a_{\pi(2)}),a_{\pi(3)}\right)~.
\label{jacobiator1}
\eeq
The first few nilpotency relations \e{nilprel} are
\bea
c^{0}_{0}~\Phi^{1}(\Phi^{0})&=&~0~,\label{nilprel0} \\
c^{1}_{0}~\Phi^{2}(\Phi^{0},a)
+c^{1}_{1}~\Phi^{1}\left(\Phi^{1}(a)\right)&=&~0~, \label{nilprel1} \\
c^{2}_{0}~\Phi^{3}(\Phi^{0},a,b)
+c^{2}_{1}\left[\Phi^{2}\left(\Phi^{1}(a),b\right)
+(-1)^{\epsilon_{a}+\epsilon}
\Phi^{2}\left(a,\Phi^{1}(b)\right) \right]
+c^{2}_{2}~\Phi^{1}\left(\Phi^{2}(a,b)\right)&=&~0~, \label{nilprel2} \\
c^{3}_{0}~\Phi^{4}(\Phi^{0},a,b,c) 
+c^{3}_{2}~{\rm Jac}(a,b,c)
+c^{3}_{3}~\Phi^{1}\left(\Phi^{3}(a,b,c)\right)&& \cr
+c^{3}_{1}\left[\Phi^{3}\left(\Phi^{1}(a),b,c\right)
+(-1)^{\epsilon_{a}+\epsilon}
\Phi^{3}\left(a,\Phi^{1}(b),c\right) 
+(-1)^{\epsilon_{a}+\epsilon_{b}}
\Phi^{3}\left(a,b,\Phi^{1}(c)\right)\right]
&=&~0~,\label{nilprel3} 
\eea
and so forth. If all the \mb{c^{n}_{k}} coefficients are equal to \mb{1}
this becomes the \hla of Lada and Stasheff \cite{ladastasheff93}.
(We shall ignore the fact that Lada and Stasheff \cite{ladastasheff93} do not
include a zero-bracket \mb{\Phi^{0}} in the definition and they use
another sign convention.)
If \mb{c^{1}_{0}=0} the one-bracket \mb{\Phi^{1}} becomes nilpotent,
\cf \eq{nilprel1}, so in this case (ignoring the fact that we have not
defined an integer grading) the one-bracket \mb{\Phi^{1}} essentially gives
rise to a complex \mb{({\cal A},\Phi^{1})}. Note that the Jacobi identity
\e{nilprel3} is modified by the presence of higher brackets.
In this paper we work under the hypothesis that the characteristic features
of a \hla is formed by the Grassmann-odd and nilpotent nature of the 
\mb{\bullet}-bracket \mb{\Phi}, \ie the {\em mere existence} of the 
\mb{c^{n}_{k}} coefficients, rather than what particular values those 
\mb{c^{n}_{k}} coefficients might have.

\subsection{Rescaling}
\label{secrenormalization}

\noi
A couple of general remarks about the prefactors in the ``\mb{\circ}''
product \e{operadproduct} is in order. {}First of all, the denominator
\mb{k!(n\!-\!k)!} has been included to conform with standard practices.
(Traditionally \hlas are explained via un-shuffles \cite{ladastasheff93},
a notion we shall not use in this paper. The combinatorial factor
\mb{k!(n\!-\!k)!} disappears when recast in the language of un-shuffles.)
It is convenient to introduce an equivalent scaled set of
coefficients \mb{b^{n}_{k}} that are always assumed to be equal to the
\mb{c^{n}_{k}} coefficients multiplied with the binomial coefficients,
\beq
  b^{n}_{k}~\equiv~\twobyone{n}{k}c^{n}_{k}~,~~~~~~~~~~~~~~0\leq k\leq n~.
\label{bcdef}
\eeq
We shall often switch back and forth between the ``\mb{b}'' and the ``\mb{c}''
picture using \eq{bcdef}. 

\noi
Secondly, we remark that the nilpotency relations \e{nilprelaaa}, and hence the
coefficients \mb{c^{n}_{k}}, may always be trivially scaled 
\beq
c^{n}_{k}~\longrightarrow~ \lambda_{n} c^{n}_{k}~,\label{trivialscaling}
\eeq 
where \mb{\lambda_{n}}, \mb{n\in\{0,1,2, \ldots\}},  are non-zero complex
numbers. Also, if one allows for a re-normalization of the bracket definition
\mb{\Phi^{n}\to\Phi^{n}/ \lambda_{n}}, and one scales the coefficients
\mb{c^{n}_{k}\to \lambda_{k}\lambda_{n-k+1} c^{n}_{k}} accordingly,
the nilpotency relations are not changed. 
In practice, one works with a fixed convention for the normalization
of the brackets, so the latter type of scaling is usually not an issue,
while the former type \e{trivialscaling} is a trivial ambiguity
inherent in the definition \e{nilprel}. 
When we in the following make uniqueness claims about the \mb{c^{n}_{k}}
coefficients in various situations, it should always be understood modulo
the trivial scaling \e{trivialscaling}.

\subsection{Polarization}
\label{secpolarization}

\noi
The product definition \e{operadproduct} may equivalently be written in
a diagonal form
\beq
(\Phi\circ\Phi^{\prime})^{n}(a^{\soprod n})
~=~\sum_{k=0}^{n} b^{n}_{k}~\Phi^{n-k+1}
\left(\Phi^{\prime k}(a^{\soprod k})\soprod a^{\soprod (n-k)}\right)
~,~~~~~~~~~~\epsilon(a)~=~\epsilon~.
\label{productaaa}
\eeq
That \eq{productaaa} follows from the product definition \e{operadproduct}
is trivial. The other way follows by collecting the string 
\mb{a_{1}\soprod\ldots\soprod a_{n}} of arguments into a linear combination
\beq
  a~=~\sum_{k=0}^{n}\lambda_{k} a_{k}
~,~~~~~~~~~~~~~~~~~~~~~~~\epsilon(a)~=~\epsilon~,
\label{alincomb}
\eeq
where the supernumbers \mb{\lambda_{k}}, \mb{k\in\{0,1,\ldots,n\}}, have
Grassmann parity \mb{\epsilon(\lambda_{k})=\epsilon(a_{k})+\epsilon}. The
product definition \e{operadproduct} then follows by inserting the linear
combination \e{alincomb} into \eq{productaaa}, considering terms proportional
to the \mb{\lambda_{1}\lambda_{2}\ldots\lambda_{n}} monomial, and using the
appropriate sign permutation rules. The general lesson to be learned is that
the diagonal carries all information. This polarization trick is not new, as
one can imagine; see for instance \Ref{aksz97} and \Ref{voronov04}.
The crucial point is that the nilpotency relations \e{nilprel} may be
considered on the diagonal only,
\beq
(\Phi\circ\Phi)^{n}(a^{\soprod n})~=~0
~,~~~~~~~~~~\epsilon(a)~=~\epsilon
~,~~~~~~~~~~n\in\{0,1,2,\ldots\}~.\label{nilprelaaa}
\eeq
Eq.\ \e{nilprelaaa} will be our starting point for subsequent investigations.

\noi
We mention in passing that a construction involving a pair of 
\mb{\bullet}-brackets \mb{\Phi^{a}}, \mb{a\in\{1,2\}}, sometimes referred to
as an ``\mb{Sp(2)}-formulation'' \cite{bda96}, can always be deduced from 
polarization of \mb{\Phi=\sum_{a=1}^{2}\lambda_{a}\Phi^{a}}, 
\mb{\epsilon(\lambda_{a})=0}.

\section{The Koszul Bracket Hierarchy}
\label{seckoszul}

\noi
The heart of the following construction goes back to Koszul
\cite{koszul85,akman97,ad96} and was later proven to be a \hla in \Ref{bda96}.

\subsection{Basic Settings}
\label{secbasicassosettings}

\noi
Consider a graded algebra \mb{({\cal A},\cdot)} of suspension parity
\mb{\epsilon\in\{0,1\}}, satisfying bi-linearity and associativity,
\bea
\epsilon(a\cdot b)&=&\epsilon_{a}+\epsilon_{b}+\epsilon~,\label{dotparity} \\
(\lambda a)\cdot(b\mu)&=&\lambda (a\cdot b) \mu~, \\
(a\lambda)\cdot b&=&(-1)^{\epsilon\epsilon_{\lambda}}a \cdot (\lambda b)~, \\
(a\cdot b)\cdot c&=&a\cdot (b\cdot c)~,
\eea  
where \mb{\lambda,\mu} are supernumbers and \mb{a,b,c\in{\cal A}} are algebra
elements. Let there also be given a fixed algebra element \mb{e} of Grassmann
parity \mb{\epsilon(e)=\epsilon} and a Grassmann-odd, linear operator 
\mb{\Delta:{\cal A}\to{\cal A}}, also known as a Grassmann-odd endomorphism
\mb{\Delta\in{\rm End}({\cal A})}. Note that the algebra product 
``\mb{\cdot}'' carries Grassmann parity, \cf \eq{dotparity}. This implies
for instance that a power \mb{a^{\cdot n}:=a\cdot\ldots\cdot a} of an element
\mb{a\in{\cal A}} has Grassmann parity 
\mb{\epsilon(a^{\cdot n})=n\epsilon(a)+(n\!-\!1)\epsilon}, 
\mb{n\in\{1,2,3,\ldots\}}. Let \mb{L_{a},R_{a}:{\cal A}\to{\cal A}} denote
the left and the right multiplication map \mb{L_{a}(b):=a\cdot b} and
\mb{R_{a}(b):=b\cdot a} with an algebra element \mb{a\in{\cal A}},
respectively. The Grassmann parity of the multiplication maps
\mb{L_{a},R_{a}\in{\rm End}({\cal A})} is in both cases
\mb{\epsilon(L_{a})=\epsilon(a)+\epsilon=\epsilon(R_{a})}.

\subsection{Review of the Commutative Case}
\label{seccommutative}

\noi
In this Subsection~\ref{seccommutative} we assume that the algebra
\mb{{\cal A}} is commutative. 

\begin{definition} If the algebra \mb{{\cal A}} is commutative, the {\bf
Koszul \mb{n}-bracket} \mb{\Phi_{\Delta}^{n}} is defined \cite{bda96} as
multiple, nested commutators acting on the algebra element \mb{e}, 
\beq
\Phi_{\Delta}^{n}(a_{1},\ldots, a_{n})
~:=~\underbrace{[[ \ldots [\Delta, L_{a_{1}}],\ldots ], L_{a_{n}}]}_{
n~{\rm commutators}}e
~,~~~~~~~~~~~~~\Phi_{\Delta}^{0}:=\Delta(e)~.
\label{koszulconstructabelian}
\eeq
\end{definition}

\noi
Here \mb{[S,T]:=S T-(-1)^{\epsilon_{S}\epsilon_{T}}T S} denotes the commutator
of two endomorphisms \mb{S,T\in{\rm End}({\cal A})} under composition.
One easily verifies that this definition is symmetric in the arguments
\mb{(a_{1},\ldots, a_{n})} by using the Jacobi identity for the 
commutator-bracket in \mb{{\rm End}({\cal A})}.

\begin{proposition} In the commutative case the Koszul \mb{\bullet}-bracket
\mb{\Phi_{\Delta}} satisfies a recursion relation with only three terms
\cite{akman97}
\bea
 \Phi_{\Delta}^{n+1}(a_{1},\ldots,a_{n},a_{n+1})&=&
\Phi_{\Delta}^{n}(a_{1},\ldots,a_{n}\cdot a_{n+1})
-\Phi_{\Delta}^{n}(a_{1},\ldots,a_{n})\cdot a_{n+1} \cr
&&-(-1)^{(\epsilon_{n}+\epsilon)(\epsilon_{n+1}+\epsilon)}
\Phi_{\Delta}^{n}(a_{1},\ldots,a_{n+1})\cdot a_{n}
\label{koszulabelianrecursion}
\eea
for \mb{n\in\{1,2,3,\ldots\}}.
\label{propositionabelianrecursivekoszul}
\end{proposition}

\noi
Note that the one-bracket \mb{\Phi_{\Delta}^{1}} can {\em not} be expressed
recursively in terms of the zero-bracket \mb{\Phi_{\Delta}^{0}} alone.

\noi
{\it Proof of Proposition~\ref{propositionabelianrecursivekoszul}:}~~
Observe that
\bea
\Phi_{[\Delta,L_{a}L_{b}]}^{n}(a_{1},\ldots, a_{n})
-\Phi_{[[\Delta,L_{a}],L_{b}]}^{n}(a_{1},\ldots, a_{n})
&=&(-1)^{\epsilon_{a}+\epsilon}
\Phi_{L_{a}[\Delta,L_{b}]}^{n}(a_{1},\ldots, a_{n}) \cr
&&+(-1)^{(\epsilon_{a}+\epsilon)(\epsilon_{b}+\epsilon)}(a \leftrightarrow b)
\label{comrecursionhelp}
\eea
for \mb{n\in\{0,1,2,\ldots\}}. The wanted \eq{koszulabelianrecursion} emerges
after relabelling of \eq{comrecursionhelp} and use of the definition 
\e{koszulconstructabelian}. We mention for later that \eq{comrecursionhelp}
also makes sense in a non-commutative setting.
\proofbox

\noi
The main example of the Koszul construction is with a bosonic suspension parity
\mb{\epsilon\!=\!0}, \cf \eq{permutationsign}, with \mb{e} being an algebra
unit, and where \mb{\Delta\in{\rm End}({\cal A})} is a nilpotent,
Grassmann-odd, linear operator, \mb{\Delta^{2}=0}, \mb{\epsilon(\Delta)=1}.
This is called a generalized Batalin-Vilkovisky algebra by Akman 
\cite{akman97,bbd96,bbd06}. If furthermore the higher brackets vanish, \ie
\mb{\Phi_{\Delta}^{n}=0}, \mb{n\geq 3}, then the \mb{\Delta} operator is by
definition an operator of at most second order, and \mb{({\cal A},\Delta)}
becomes a Batalin-Vilkovisky algebra \cite{getzler94,schwarz94}. We give an
explicit example in \eq{courantdelta2}. The zero-bracket
\mb{\Phi_{\Delta}^{0}=\Delta(e)} typically vanishes in practice. {}For a
Batalin-Vilkovisky algebra with a non-vanishing zero-bracket
\mb{\Phi_{\Delta}^{0}}, see \Ref{b06}.

\subsection{The Intermediate Case: \mb{{\rm Im}(\Delta)~\subseteq~Z({\cal A})}}
\label{secimdeltainza}

\noi
We would like to address the following two questions:
\begin{enumerate}
\item
Which choices of the bracket product coefficients \mb{c^{n}_{k}} turn the
Koszul construction \e{koszulconstructabelian} into a (generalized) 
homotopy Lie Algebra?
\item
Does there exist a non-commutative version of the Koszul construction?
\end{enumerate}

\noi
As we shall see in Subsection~\ref{secgeneralkoszul} the answer to the second
question is yes. {}For practical purposes, it is of interest to seek out
intermediate cases that are no longer purely commutative, but where the
non-commutative obstacles are manageable.
In this and the following Subsections~\ref{secimdeltainza}-\ref{secoffshell}
we make the simplifying assumption that the image of the \mb{\Delta} operator
lies in the center of the algebra, \ie
\beq
 {\rm Im}(\Delta)~\subseteq~ Z({\cal A})~,
\label{ansatzimdeltainza}
\eeq
where the center \mb{Z({\cal A})} is, as usual, 
\beq
  Z({\cal A})~:=~\left\{ a\in{\cal A}\left| \forall b\in{\cal A}:~
b\cdot a~=~(-1)^{(\epsilon_{a}+\epsilon)(\epsilon_{b}+\epsilon)}
a\cdot b\right. \right\}~.
\eeq
The full non-commutative case is postponed until 
Subsection~\ref{secgeneralkoszul}. 
The case \e{ansatzimdeltainza} clearly includes the commutative case, and
it turns out that the treatment of the first question from this intermediate
perspective is completely parallel to the purely commutative case.

\begin{definition} If the assumption \e{ansatzimdeltainza} is fulfilled, the
{\bf Koszul \mb{n}-bracket} is defined as symmetrized, nested commutators
acting on the algebra element \mb{e},
\beq
\Phi_{\Delta}^{n}(a_{1},\ldots, a_{n})
~:=~\frac{1}{n!}\sum_{\pi\in S_{n}}(-1)^{\epsilon_{\pi,a}}
\underbrace{[[ \ldots [\Delta, L_{a_{\pi(1)}} ],\ldots  ], L_{a_{\pi(n)}}]}_{
n~{\rm commutators}}e~,~~~~~~~~~~~~~\Phi_{\Delta}^{0}:=\Delta(e)~.
\label{koszulconstruct}
\eeq
\end{definition}

\noi
In general, all the information about the higher brackets is carried by the
diagonal, 
\beq
\Phi_{\Delta}^{n}(a^{\soprod n})~=~
\underbrace{[[\ldots [\Delta, L_{a}],\ldots ], L_{a} ]}_{n~{\rm commutators}}e
~=~\sum_{k=0}^{n}\twobyone{n}{k}(-L_{a})^{k}\Delta L_{a}^{n-k}(e)
~,~~~~~~~~~~\epsilon(a)~=~\epsilon~.
\label{koszulconstructaaa}
\eeq

\begin{proposition} If the assumption \e{ansatzimdeltainza} is fulfilled, the
Koszul brackets \mb{\Phi_{\Delta}} satisfy the recursion relation
\bea
\Phi_{\Delta}^{n}(a_{1},\ldots,a_{n})&=&\frac{1}{n(n\!-\!1)}
\sum_{1\leq i<j \leq n}\left[\prod_{k=i+1}^{n}
(-1)^{(\epsilon_{i}+\epsilon)(\epsilon_{k}+\epsilon)}\right]
\left[\prod_{\ell=j+1}^{n}
(-1)^{(\epsilon_{j}+\epsilon)(\epsilon_{\ell}+\epsilon)}\right]\cr
&&\Phi_{\Delta}^{n-1}\left(a_{1},\ldots,\widehat{a}_{i},\ldots,
\widehat{a}_{j},\ldots,a_{n},~ a_{j}\cdot a_{i}
+(-1)^{(\epsilon_{i}+\epsilon)(\epsilon_{j}+\epsilon)}
a_{i}\cdot a_{j}\right)\cr
&&-\frac{2}{n}\sum_{1\leq i\leq n} \left[\prod_{k=i+1}^{n}
(-1)^{(\epsilon_{i}+\epsilon)(\epsilon_{k}+\epsilon)}\right]
\Phi_{\Delta}^{n-1}\left(a_{1},\ldots,
\widehat{a}_{i},\ldots,a_{n}\right)\cdot a_{i}
\label{koszulrecursive}
\eea
for \mb{n\in\{2,3,4,\ldots\}}.
\label{propositionrecursivekoszul}
\end{proposition}

\noi
{}For instance the two-bracket \mb{\Phi_{\Delta}^{2}} can be defined via
the one-bracket \mb{\Phi_{\Delta}^{1}} as
\beq
 \Phi_{\Delta}^{2}(a,b)
~=~\Hf \Phi_{\Delta}^{1}(a\cdot b)-\Phi_{\Delta}^{1}(a)\cdot b
+(-1)^{(\epsilon_{a}+\epsilon)(\epsilon_{b}+\epsilon)}(a\leftrightarrow b)~.
\eeq 
The recursion relations are more complicated than in the commutative case.
Whereas the commutative recursion relations \e{koszulabelianrecursion} involve
only three terms, the number of terms now grows quadratically with the
number \mb{n} of arguments. Loosely speaking, one may say that the recursion
relations dissolve as one moves towards full-fledged non-commutativity,
\cf Subsection~\ref{secgeneralkoszul}. This is fine since recursion
relations are anyway not an essential ingredient of a homotopy Lie
algebra, although at a practical level they can be quite useful.

\noi
{\it Proof of Proposition~\ref{propositionrecursivekoszul}:}~~
Note that \eq{comrecursionhelp} still holds in this case: 
\beq
\Phi_{[\Delta,L_{a}^{2}]}^{n}(a^{\soprod n})
-\Phi_{[[\Delta,L_{a}],L_{a}]}^{n}(a^{\soprod n})
~=~ \Phi_{2L_{a}[\Delta,L_{a}]}^{n}(a^{\soprod n})
~,~~~~~~~~~~\epsilon(a)~=~\epsilon~,
\label{recursionhelp}
\eeq
for \mb{n\in\{0,1,2,\ldots\}}. This leads to 
\bea
\Phi_{\Delta}^{n+1}(a^{\soprod (n+1)})
&=&\Phi_{\Delta}^{n}((a \cdot a) \soprod a^{\soprod (n-1)})
-2a \cdot \Phi_{\Delta}^{n}(a^{\soprod n}) \cr
&=&\Phi_{\Delta}^{n}(a^{\soprod(n-1)}\soprod(a\cdot a))
-2 \Phi_{\Delta}^{n}(a^{\soprod n}) \cdot a~,~~~~~~~~~~
\epsilon(a)~=~\epsilon~,\label{koszulrecursiveaaa}
\eea
for \mb{n\in\{1,2,3,\ldots\}}. The recursion relation \e{koszulrecursive} 
now follows from polarization of \eq{koszulrecursiveaaa}, 
\cf Subsection~\ref{secpolarization}.
\proofbox

\subsection{Nilpotency Relations}
\label{seckoszulnilprel}

\noi
We now return to the first question in Subsection~\ref{secimdeltainza}. More
precisely, we ask which coefficients \mb{c^{n}_{k}} could guarantee the 
nilpotency relations \e{nilprelaaa}, if one is only allowed to additionally
assume that the \mb{\Delta} operator is nilpotent in the sense that
\beq
  \Delta R_{e}\Delta~=~0~? \label{deltanilp}
\eeq
Note that the criterion \e{deltanilp} reduces to the usual nilpotency
condition \mb{\Delta^{2}=0} if \mb{e} is a right unit for the algebra
\mb{{\cal A}}. The generic answer to the above question is given by the 
following Theorem~\ref{theoremkoszul}.

\begin{theorem}
Let there be given a set of \mb{c^{n}_{k}} product coefficients, 
\mb{n\geq k\geq 0}. The nilpotency relations  \e{nilprelaaa} are satisfied for
all Koszul brackets \mb{\Phi_{\Delta}} that have a nilpotent \mb{\Delta} 
operator (in the sense of \eq{deltanilp}), if and only if 
\beq
\forall n\in\{0,1,\ldots\}:~~c^{n}_{k}~=~c^{n} \label{koszulsol}
\eeq
is independent of \mb{k\in\{0,\ldots,n\}}.
\label{theoremkoszul}
\end{theorem}

\noi
Bearing in mind the trivial rescaling \e{trivialscaling}, this solution 
\e{koszulsol} is essentially \mb{c^{n}_{k}=1} in perfect alignment with the
requirement \e{stasheffchoice} in the original definition of Lada
and Stasheff \cite{ladastasheff93}. So there is no call for a new bracket
product ``\mb{\circ}'' to study the Koszul bracket hierarchy. This no-go 
statement obviously remains valid when considering the general non-commutative
case, \cf Subsection~\ref{secgeneralkoszul}, since the general 
\mb{\Phi_{\Delta}} bracket \e{genkoszulconstruct} should in particular 
reproduce all the severely limiting situations where the condition
\e{ansatzimdeltainza} holds.

\noi
{\it Proof of Theorem~\ref{theoremkoszul} when assuming  
\eq{ansatzimdeltainza}:}~~We start with the ``only if'' part. To see 
\eq{koszulsol}, first note that for two mutually commuting elements 
\mb{a,b \in{\cal A}} with \mb{\epsilon(a)=\epsilon},
\bea
\Phi_{\Delta}^{n+1}(b\soprod a^{\soprod n})
&=&\Phi_{[\Delta,L_{b}]}^{n}(a^{\soprod n}) \cr
&=&\sum_{i=0}^{n}\twobyone{n}{i}(-L_{a})^{i}\Delta L_{a}^{n-i}(b\cdot e)
-(-1)^{\epsilon_{b}+\epsilon}b \cdot \Phi_{\Delta}^{n}(a^{\soprod n})~.
\eea
Putting \mb{b=\Phi_{\Delta}^{k}(a^{\soprod k})} with Grassmann parity
\mb{\epsilon_{b}=1-\epsilon}, the element \mb{b} commutes with \mb{a}
because of \eq{ansatzimdeltainza}, and the \nth square bracket becomes
\bea
(\Phi_{\Delta}\circ\Phi_{\Delta})^{n}(a^{\soprod n})
&=&\sum_{k=0}^{n} b^{n}_{k}\sum_{i=0}^{n-k}
\twobyone{n\!-\!k}{i}(-L_{a})^{i}\Delta L_{a}^{n-k-i}
\left(\Phi_{\Delta}^{k}(a^{\soprod k})\cdot e\right)\cr
&&+\sum_{k=0}^{n} b^{n}_{k}~\Phi_{\Delta}^{k}(a^{\soprod k})\cdot
\Phi_{\Delta}^{n-k}(a^{\soprod (n-k)})~.
\label{koszulnilprel1}
\eea
The two sums on the \rhs of \eq{koszulnilprel1} are of different algebraic
natures, because the two \mb{\Delta}'s are nested in the first sum, while in
the second sum they are not. In general, to ensure the nilpotency relations
\e{nilprelaaa}, one should therefore impose that the two sums vanish 
separately. (To make this argument sound one uses that nilpotency relations 
\e{nilprelaaa} hold for all possible choices of \mb{{\cal A}}, \mb{\Delta}
and \mb{e} satisfying \eq{deltanilp}.) The vanishing of the second sum just
imposes a symmetry
\beq
   b^{n}_{k}~=~b^{n}_{n-k} \label{bsymmetric}
\eeq
among the \mb{b^{n}_{k}} coefficients, because the family of brackets
\mb{\Phi_{\Delta}^{k}(a,\ldots,a)}, \mb{k\in\{0,1,\ldots,n\}}, mutually
commute in a graded sense, which in plain English means: anti-commute.
We shall see shortly that the symmetry \e{bsymmetric} is superseded by
stronger requirements coming from the first sum.
After some elementary manipulations the first sum reads
\beq
\sum_{i=0}^{n}(-L_{a})^{i}\sum_{j=0}^{n-i}\Delta L_{a}^{n-i-j}
\twobyone{n}{i,j,n\!-\!i\!-\!j}
[\sum_{k=0}^{n-i-j}c^{n}_{k+j}\twobyone{n\!-\!i\!-\!j}{k}(-1)^{k}]
R_{e}\Delta L_{a}^{j}(e)~.
\label{firstsum}
\eeq
Terms where non-zero powers of \mb{L_{a}} are sandwiched between the two
\mb{\Delta}'s are bad, as the nilpotency condition \e{deltanilp} does not
apply to them. Accordingly, the expression inside the square brackets in
\eq{firstsum} must vanish for such terms. In detail, there should exist
complex numbers \mb{c^{n}_{(j)}}, \mb{0 \leq j \leq n}, such that
\beq
\forall i,j,n:~~~~0\leq i,j\leq n~~~~ \Rightarrow~~~~
\sum_{k=0}^{n-i-j}c^{n}_{k+j}\twobyone{n\!-\!i\!-\!j}{k}(-1)^{k}~=
~c^{n}_{(j)}\delta_{n,i+j}~.
\label{bigorthogonalcondition}
\eeq
(The complex numbers \mb{c^{n}_{k}} and \mb{c^{n}_{(j)}} should not be 
confused.) Putting \mb{j=0} and \mb{m=n-i\in \{0,\ldots, n\}}, the 
\eq{bigorthogonalcondition} reduces to
\beq
\forall m,n:~~~~0\leq m\leq n~~~~ \Rightarrow~~~~
 \sum_{k=0}^{m}c^{n}_{k}\twobyone{m}{k}(-1)^{k}~=~c^{n}_{(0)}\delta_{m,0}~.
\label{orthogonalcondition}
\eeq
This in turn implies that \mb{c^{n}_{k}} can only depend on \mb{n},
\beq
\forall k\in\{0,\ldots,n\}:~~~~c^{n}_{k}~=~c^{n}_{(0)}~,\label{koszulsol0}
\eeq
which establishes the claim \e{koszulsol0}. {}For completeness let us mention
that if one inserts the solution 
\e{koszulsol0} back into \eq{bigorthogonalcondition} one gets similarly,
\beq
\forall j\in\{0,\ldots,n\}:~~~~c^{n}_{(j)}~=~c^{n}_{(0)}~.\label{koszulsol1}
\eeq
The ``if'' part of the proof follows easily by going through above reasoning 
in reversed order, but it is also a consequence of
Theorem~\ref{theoremkoszulsquare} below.
\proofbox

\subsection{Off-Shell \wrt the Nilpotency Condition}
\label{secoffshell}

\noi
One may summarize the discussions of the last 
Subsection~\ref{seckoszulnilprel} in the following 
Theorem~\ref{theoremkoszulsquare}.

\begin{theorem} A Koszul \mb{\bullet}-bracket \mb{\Phi_{\Delta}} satisfies
a square identity \cite{bda96}
\beq
\Phi_{\Delta}\circ\Phi_{\Delta}~=~\Phi_{\Delta R_{e}\Delta}~,
\label{koszulsquare}
\eeq
where ``\mb{\circ}'' here refers to the ordinary bracket product
\e{operadproduct} with \mb{c^{n}_{k}=1}. 
\label{theoremkoszulsquare}
\end{theorem}

\noi
We stress that this identity holds without assuming the nilpotency condition
\e{deltanilp}. It is instructive to see a direct proof of this square identity
\e{koszulsquare} that uses a generating function and polarization to minimize
the combinatorics. In the case of the Koszul \mb{\bullet}-bracket
\mb{\Phi_{\Delta}} the generating function is just the ordinary exponential
function ``\mb{\exp}''. This is implemented as a formal series of 
``exponentiated brackets'',
\beq
\sum_{n=0}^{\infty}\frac{1}{n!}\Phi_{\Delta}^{n}(a^{\soprod n})
~=~\left(e^{-[L_{a},\cdot]}\Delta\right)e~=~e^{-L_{a}}\Delta e^{L_{a}}(e)
~,~~~~~~~~~~\epsilon(a)~=~\epsilon~.
\label{exponentiatedbracket}
\eeq
Conversely, one may always extract back the \nth bracket 
\mb{\Phi_{\Delta}^{n}} by identifying terms in \eq{exponentiatedbracket} that
has homogeneous scaling degree \mb{n} under scaling \mb{a\to\lambda a} 
of the argument \mb{a}.

\noi
{\it Proof of Theorem~\ref{theoremkoszulsquare} when assuming
\eq{ansatzimdeltainza}}:~~ {}First note that for two 
mutually commuting elements \mb{a,b \in{\cal A}} with 
\mb{\epsilon(a)=\epsilon},
\bea
\sum_{n=0}^{\infty}\frac{1}{n!}\Phi_{\Delta}^{n+1}(b\soprod a^{\soprod n})&=&
\sum_{n=0}^{\infty}\frac{1}{n!}\Phi_{[\Delta,L_{b}]}^{n}(a^{\soprod n}) \cr
&=&e^{-L_{a}}\Delta e^{L_{a}}(b\cdot e) 
-(-1)^{\epsilon_{b}+\epsilon}b \cdot e^{-L_{a}}\Delta e^{L_{a}}(e)~.
\eea
Putting \mb{b=e^{-L_{a}}\Delta e^{L_{a}}(e)} with Grassmann parity
\mb{\epsilon_{b}=1-\epsilon}, the element \mb{b} commutes with \mb{a} because
of \eq{ansatzimdeltainza}, the element \mb{b} is nilpotent
\mb{b\cdot b=0}, and the exponentiated \lhs of \eq{koszulsquare} becomes
\bea
\sum_{n=0}^{\infty}\frac{1}{n!}
(\Phi_{\Delta}\circ\Phi_{\Delta})^{n}(a^{\soprod n})
&=&\sum_{n=0}^{\infty}\sum_{k=0}^{n}\frac{1}{k!(n\!-\!k)!}
\Phi_{\Delta}^{n-k+1}\left(
\Phi_{\Delta}^{k}(a^{\soprod k})\soprod a^{\soprod (n-k)}\right)\cr
&=&e^{-L_{a}}\Delta e^{L_{a}}\left(e^{-L_{a}}\Delta e^{L_{a}}(e)\cdot e\right)
+e^{-L_{a}}\Delta e^{L_{a}}(e) \cdot e^{-L_{a}}\Delta e^{L_{a}}(e) \cr
&=&e^{-L_{a}}\Delta R_{e}\Delta e^{L_{a}}(e)
~=~\sum_{n=0}^{\infty}\frac{1}{n!}
\Phi_{\Delta R_{e}\Delta}^{n}(a^{\soprod n})~,
\eea
which is just the exponentiated \rhs of \eq{koszulsquare}.
\proofbox

\subsection{The General Non-Commutative Case}
\label{secgeneralkoszul}

\noi 
We now consider the general case without the assumption \e{ansatzimdeltainza}.

\begin{definition} In the general non-commutative case 
the {\bf Koszul \mb{n}-brackets} is defined as
\bea
\Phi_{\Delta}^{n}(a_{1},\ldots, a_{n})
&:=&\!\!\!\!\!\!\!\!\sum_{\footnotesize\begin{array}{c} i,j,k\geq 0 \cr
 i\!+\!j\!+\!k=n \end{array}}\frac{B_{i,k}}{i!j!k!}
\sum_{\pi\in S_{n}}
(-1)^{\epsilon_{\pi,a}+\epsilon_{\pi(1)}+\ldots+\epsilon_{\pi(i)}+i\epsilon}
a_{\pi(1)}\cdot\ldots\cdot a_{\pi(i)}\cr
&&~~~~~~~~\cdot\Delta\left(a_{\pi(i+1)}\cdot\ldots\cdot a_{\pi(i+j)}\cdot
e\right)\cdot a_{\pi(i+j+1)}\cdot\ldots\cdot a_{\pi(n)}~,
\label{genkoszulconstruct}
\eea
where the \mb{B_{k,\ell}} coefficients are given through the generating
function
\bea
B(x,y)&=&\sum_{i,j=0}^{\infty}B_{k,\ell}\frac{x^{k}}{k!}\frac{y^{\ell}}{\ell!}
~=~\frac{x-y}{e^{x}-e^{y}}~=~B(y,x) \cr
&=&1-\left(\frac{x}{2}+\frac{y}{2}\right)
+\Hf\left(\frac{x^{2}}{6}+\frac{2xy}{3}+\frac{y^{2}}{6}\right)
-\frac{1}{3!}\left(\frac{x^{2}y}{2}+\frac{xy^{2}}{2}\right) \cr
&&+\frac{1}{4!}\left(-\frac{x^{4}}{30}+\frac{2x^{3}y}{15}
+\frac{4x^{2}y^{2}}{5}+\frac{2xy^{3}}{15}-\frac{y^{4}}{30}\right)+\ldots~.
\label{xybernoullidef} 
\eea
\end{definition}

\noi
The \mb{B_{k,\ell}} coefficients are related to the Bernoulli numbers
\mb{B_{k}} via
\beq
B(x,y)~=~e^{-x}B(y\!-\!x)~=~e^{-y}B(x\!-\!y)~,
\eeq
\cf \eq{bernoullidef}, or in detail,
\beq
B_{k,\ell}~=~(-1)^{k}\sum_{i=0}^{k}\twobyone{k}{i}B_{i+\ell}~=~B_{\ell,k}
~,~~~~~~~~~~~~k,\ell\in\{0,1,2,\ldots\}~.
\eeq
The first few brackets read
\bea
\Phi_{\Delta}^{0}&=&\Delta(e)~, \\
\Phi_{\Delta}^{1}(a)&=&\Delta(a\cdot e)
-\Hf \Delta(e)\cdot a-\Hf(-1)^{\epsilon_{a}+\epsilon}a\cdot \Delta(e)~, \\
\Phi_{\Delta}^{2}(a,b)&=&\Hf\Delta(a\cdot b\cdot e)-\Hf\Delta(a\cdot e)\cdot b
-\Hf (-1)^{\epsilon_{a}+\epsilon}a\cdot\Delta(b\cdot e)
+\frac{1}{12}\Delta(e)\cdot a\cdot b  \cr
&&+\frac{1}{3}(-1)^{\epsilon_{a}+\epsilon} a\cdot\Delta(e)\cdot b 
+\frac{1}{12}(-1)^{\epsilon_{a}+\epsilon_{b}} a\cdot b\cdot\Delta(e)
+(-1)^{(\epsilon_{a}+\epsilon)(\epsilon_{b}+\epsilon)}(a \leftrightarrow b)~.
\eea
In general, all the information about the higher brackets is carried by the
diagonal, 
\beq
\Phi_{\Delta}^{n}(a^{\soprod n})
~=~\!\!\!\!\!\!\!\!\sum_{\footnotesize\begin{array}{c}i,j,k\geq 0 \cr
i\!+\!j\!+\!k=n \end{array}}\twobyone{n}{i,j,k}B_{i,k}~
a^{\cdot i}\cdot \Delta(a^{\cdot j}\cdot e)\cdot a^{\cdot k} 
~,~~~~~~~~~~\epsilon(a)~=~\epsilon~.
\label{genkoszulconstructaaa}
\eeq
The definition \e{genkoszulconstructaaa} is consistent with the previous
definition \e{koszulconstructaaa} for the intermediate case
\e{ansatzimdeltainza}. This is because \mb{B(x,x)=e^{-x}}, or equivalently,
\beq 
\sum_{k=0}^{n}\twobyone{n}{k}B_{k,n-k}~=~(-1)^{n}~.
\eeq
The formal series of exponentiated brackets may be compactly written
\beq
\sum_{n=0}^{\infty}\frac{1}{n!}\Phi_{\Delta}^{n}(a^{\soprod n})
~=~B(L_{a},R_{a})\Delta e^{L_{a}}(e)
~=~B(R_{a}\!-\!L_{a})e^{-L_{a}}\Delta e^{L_{a}}(e)
~,~~~~~~~~~~\epsilon(a)~=~\epsilon~.
\eeq
The latter expression shows that all the ``\mb{\cdot}'' products in the
bracket definition \e{genkoszulconstruct} can be organized as commutators
from either \mb{{\cal A}} or \mb{{\rm End}({\cal A})}, except for the dot
``\mb{\cdot}'' in front of the fixed element \mb{e}. 

\noi
In the general non-commutative case the \mb{\Phi_{\Delta}^{n+1}} bracket can 
{\em not} be expressed recursively in terms of the \mb{\Phi_{\Delta}^{n}} 
bracket alone, although there are exceptions. Most notably, the three-bracket
\mb{\Phi_{\Delta}^{3}} can be expressed purely in terms of the two-bracket
\mb{\Phi_{\Delta}^{2}},
\bea
\Phi_{\Delta}^{3}(a_{1},a_{2},a_{3})&=&
 \frac{1}{6}\sum_{\pi\in S_{3}}(-1)^{\epsilon_{\pi,a}}\left[
\Phi_{\Delta}^{2}\left(a_{\pi(1)}\cdot 
a_{\pi(2)},a_{\pi(3)}\right) -\Hf\Phi_{\Delta}^{2}\left(a_{\pi(1)}, 
a_{\pi(2)}\right)\cdot a_{\pi(3)}\right. \cr
&&\left. -\Hf(-1)^{\epsilon_{\pi(1)}+\epsilon}
a_{\pi(1)}\cdot \Phi_{\Delta}^{2}
\left(a_{\pi(2)},a_{\pi(3)}\right)\right]~.
\eea
(Of course, one may always replace appearances of \mb{\Delta} in definition
\e{genkoszulconstruct} with zero and one-brackets, \ie
\mb{\Delta(e)=\Phi_{\Delta}^{0}}, and 
\mb{\Delta(a\cdot e)=\Phi_{\Delta}^{1}(a)+\Hf\Phi_{\Delta}^{0}\cdot a
+\Hf(-1)^{\epsilon_{a}+\epsilon}a\cdot \Phi_{\Delta}^{0}}, and in this way 
express the \mb{n}-bracket \mb{\Phi_{\Delta}^{n}} in terms of lower brackets,
in this case \mb{\Phi_{\Delta}^{0}} and \mb{\Phi_{\Delta}^{1}}.)

\noi
Our main assertion is that the square identity \e{koszulsquare} in 
Theorem~\ref{theoremkoszulsquare} holds for the fully non-commutative
\mb{\Phi_{\Delta}} bracket definition \e{genkoszulconstruct}, \ie without
assuming \eq{ansatzimdeltainza}. The Theorem~\ref{theoremkoszul} is also valid
in the general situation. 

\noi
{\it Proof of Theorem~\ref{theoremkoszulsquare} in the general case}:~~
{}First note that for two elements \mb{a,b \in{\cal A}} with 
\mb{\epsilon(a)=\epsilon},
\bea
\sum_{n=0}^{\infty}\frac{1}{n!}\Phi_{\Delta}^{n+1}(b\soprod a^{\soprod n})
&=&B(L_{a},R_{a})\Delta(E(L_{a},R_{a}) b\cdot e)
+\sum_{i,j,k=0}^{\infty}B^{(R)}_{i,j,k}~
a^{\cdot i}\cdot\Delta e^{L_{a}}(e)\cdot a^{\cdot j}\cdot b\cdot a^{\cdot k}\cr
&&+(-1)^{\epsilon_{b}+\epsilon}\sum_{i,j,k=0}^{\infty}B^{(L)}_{i,j,k}~
a^{\cdot i}\cdot b\cdot a^{\cdot j}\cdot\Delta e^{L_{a}}(e)\cdot a^{\cdot k}~,
\label{koszulhelp}
\eea
where
\bea
E(x,y)&:=&\frac{e^{x}-e^{y}}{x-y}~=~
\sum_{k,\ell=0}^{\infty}\frac{x^{k}y^{\ell}}{(k\!+\!\ell\!+\!1)!}~, \\
B^{(R)}_{i,j,k}&:=&\frac{{B}_{i,j+k+1}}{i!(j\!+\!k\!+\!1)!}~, \\
B^{(R)}(x,y,z)&:=&\sum_{i,j,k=0}^{\infty}
B^{(R)}_{i,j,k}x^{i}y^{j}z^{k}~=~\frac{B(x,y)-B(x,z)}{y-z}~,\\
B^{(L)}_{i,j,k}&:=&\frac{{B}_{i+j+1,k}}{(i\!+\!j\!+\!1)!k!}~, \\
B^{(L)}(x,y,z)&:=&\sum_{i,j,k=0}^{\infty}
B^{(L)}_{i,j,k}x^{i}y^{j}z^{k}~=~\frac{B(x,z)-B(y,z)}{x-y}~. 
\eea
Putting \mb{b=B(L_{a},R_{a})\Delta e^{L_{a}}(e)} with Grassmann parity
\mb{\epsilon_{b}=1-\epsilon}, the exponentiated \lhs of \eq{koszulsquare} 
becomes 
\bea
\sum_{n=0}^{\infty}\frac{1}{n!}
(\Phi_{\Delta}\circ\Phi_{\Delta})^{n}(a^{\soprod n})
&=&\sum_{n=0}^{\infty}\sum_{k=0}^{n}\frac{1}{k!(n\!-\!k)!}
\Phi_{\Delta}^{n-k+1}\left(\Phi_{\Delta}^{k}(a^{\soprod k})
\soprod a^{\soprod (n-k)}\right)\cr
&=&B(L_{a},R_{a})\Delta\left(E(L_{a},R_{a})B(L_{a},R_{a})\Delta e^{L_{a}}(e)
\cdot e\right) \cr
&&+\sum_{i,j,k=0}^{\infty}B^{(R)}_{i,j,k}~
a^{\cdot i}\cdot\Delta e^{L_{a}}(e)\cdot a^{\cdot j}\cdot 
 B(L_{a},R_{a})\Delta e^{L_{a}}(e)\cdot a^{\cdot k} \cr
&&-\sum_{i,j,k=0}^{\infty}B^{(L)}_{i,j,k}~
a^{\cdot i}\cdot B(L_{a},R_{a})\Delta e^{L_{a}}(e)\cdot a^{\cdot j}\cdot 
\Delta e^{L_{a}}(e)\cdot a^{\cdot k}~. 
\label{genkoszullhs}
\eea
This should be compared with  the exponentiated \rhs of \eq{koszulsquare},
\beq
\sum_{n=0}^{\infty}\frac{1}{n!}\Phi_{\Delta R_{e}\Delta}^{n}(a^{\soprod n})
~=~B(L_{a},R_{a})\Delta R_{e}\Delta e^{L_{a}}(e)~.\label{genkoszulrhs}
\eeq
The two sides \es{genkoszullhs}{genkoszulrhs} are equal provided that the 
following two conditions are met
\bea
E(x,y)B(x,y)&=&1~,\label{1stsum} \\
B^{(R)}(x,y,z)B(y,z)&=&B(x,y)B^{(L)}(x,y,z)\label{2nd3rdsum}~.
\eea
The first equation \e{1stsum} has a unique solution for \mb{B(x,y)} given by
\mb{1/E(x,y)}, leaving no alternative to \eq{xybernoullidef}. It is remarkable
that this unique solution \e{xybernoullidef} satisfies the non-trivial second
criterion \e{2nd3rdsum} as well, as one may easily check by inspection,
thereby ensuring the existence of the non-commutative Koszul construction.
\proofbox

\section{The Derived Bracket Hierarchy} 
\label{secderoperad}

\noi 
In this Section we consider an important class of \mb{\bullet}-brackets that
naturally requires a non-trivial bracket product \e{operadproduct} in order to
satisfy the nilpotency relations \e{nilprel}, namely the so-called derived 
\mb{\bullet}-brackets. As we shall soon see in \eq{derbracket} below, the 
derived brackets are composed of nested Lie brackets in a simple
manner. This should be contrasted with the non-trivial definition
\e{genkoszulconstruct} of the non-commutative Koszul hierarchy
that -- among other things -- involved the Bernoulli numbers. Nevertheless,
in a strange twist, while the \mb{c^{n}_{k}} coefficients in the nilpotency
relations \e{nilprel} are all simply \mb{1} for the non-commutative Koszul
hierarchy, the \mb{c^{n}_{k}} coefficients will be considerably more
complicated for the derived hierarchy and involve -- of all things -- 
the Bernoulli numbers!

\subsection{Definitions}

\noi
We abandon the associative ``\mb{\cdot}'' structure considered in
Subsection~\ref{secbasicassosettings}, and consider instead a Lie algebra
\mb{({\cal A},[~,~])} of parity \mb{\epsilon\in\{0,1\}}, satisfying
bi-linearity, skewsymmetry and the Jacobi identity,
\bea
\epsilon([a,b])&=&\epsilon_{a}+\epsilon_{b}+\epsilon~,\\
{} [ \lambda a,b \mu ]&=&\lambda [ a,b] \mu~, \\
{} [a\lambda, b]&=&(-1)^{\epsilon\epsilon_{\lambda}}[a,\lambda b]~, \\
{} [b,a]&=&-(-1)^{(\epsilon_{a}+\epsilon)(\epsilon_{b}+\epsilon)}[a,b]~,\\
0&=&\sum_{a,b,c~{\rm cycl.}}
(-1)^{(\epsilon_{a}+\epsilon)(\epsilon_{c}+\epsilon)}[[a,b],c]~,\label{jacid}
\eea  
where \mb{\lambda,\mu} are supernumbers.
Let there be given a fixed Lie algebra element \mb{Q \in {\cal A}}.

\begin{definition}
The {\bf derived \mb{n}-bracket} \mb{\Phi_{Q}^{n}}, 
\mb{n\in\{0,1,2,\ldots\}}, is defined as \cite{bm98}
\beq
\Phi_{Q}^{n}(a_{1},\ldots, a_{n})
~:=~\frac{1}{n!}\sum_{\pi\in S_{n}}(-1)^{\epsilon_{\pi,a}}
\underbrace{[[ \ldots [ Q, a_{\pi(1)} ], \ldots ], a_{\pi(n)}]}_{
n~{\rm Lie~brackets}}
~,~~~~~~~~~~~~~\Phi_{Q}^{0}~:=~Q~.
\label{derbracket}
\eeq
\end{definition}

\noi
Note that the zero-bracket \mb{\Phi_{Q}^{0}=Q} is just the fixed Lie algebra
element \mb{Q} itself. 
Since we are ultimately interested in \mb{\bullet}-brackets \mb{\Phi_{Q}} 
that carry an odd intrinsic Grassmann parity \mb{\epsilon(\Phi_{Q})\!=\!1}, 
\cf \eq{nilprel}, we shall demand from now on that  the Grassmann parity 
\mb{\epsilon_{Q}} of the fixed Lie algebra element \mb{Q} is opposite of the 
suspension parity \mb{\epsilon}, 
\beq
   \epsilon_{Q}~=~1-\epsilon~.
\eeq 
All the information is again carried by the diagonal, 
\beq
\Phi_{Q}^{n}(a^{\soprod n})~=~[\Phi_{Q}^{n-1}(a^{\soprod (n-1)}),a]
~=~\underbrace{[[\ldots [Q,a],\ldots ], a]}_{n~{\rm Lie~brackets}} 
~=~(-\ad a)^{n}Q~,~~~~~~~~~~\epsilon(a)~=~\epsilon~,
\label{derbracketaaa}
\eeq
where we have defined the adjoint action 
\mb{\ad:~{\cal A}\to {\rm End}({\cal A})} by \mb{(\ad a) (b):=[a,b]}.

\begin{proposition} A derived \mb{\bullet}-bracket \mb{\Phi_{Q}}
satisfies the recursion relation
\beq
\Phi_{Q}^{n}(a_{1},\ldots,a_{n})~=~
\frac{1}{n}\sum_{i=1}^{n} \left[\prod_{k=i+1}^{n}
(-1)^{(\epsilon_{i}+\epsilon)(\epsilon_{k}+\epsilon)}\right]
 \left[\Phi_{Q}^{n-1}\left(a_{1},\ldots,
\hat{a}_{i},\ldots,a_{n}\right), a_{i}\right]
\label{derivedrecursive}
\eeq
for  \mb{n\in\{1,2,3,\ldots\}}.
\label{propositionderivedrecursive}
\end{proposition}

\noi
{\it Proof of Proposition~\ref{propositionderivedrecursive}:}~~
The recursion relation \e{derivedrecursive} follows from polarization
of \eq{derbracketaaa}, \cf Subsection~\ref{secpolarization}.
\proofbox

\noi
As we saw in Subsection~\ref{seclikealg} there is a Lie-like bracket of
opposite parity \mb{\epsilon_{Q}=1-\epsilon} given by
\beq
[a,b]_\QQ~:=~(-1)^{\epsilon_{a}+\epsilon_{Q}}\Phi_{Q}^{2}(a,b)
~=~  \Hf [[a,Q],b] + \Hf [a,[Q,b]]
~=~-(-1)^{(\epsilon_{a}+\epsilon_{Q})(\epsilon_{b}+\epsilon_{Q})}[b,a]_\QQ~.
\label{dertwobracket}
\eeq
Thus the derived \mb{\bullet}-bracket \mb{\Phi_{Q}} gives rise to an 
interesting duality \mb{[~,~]\to [~,~]_\QQ} between Lie-like brackets of even
and odd parity \cite{bm99dual}. The suspension parity \mb{\epsilon}
was introduced in the first place in \eq{permutationsign} to bring the even
and odd brackets on equal footing, and we see that the formalism 
embraces this symmetry. The bracket \e{dertwobracket} is known as a
(skewsymmetric, inner) derived bracket \cite{yks96,yks04,yks05}. The outer,
derived \mb{\bullet}-bracket hierarchies are modeled after the properties
of the inner hierarchies, and will be discussed elsewhere.

\subsection{Nilpotency versus Square Relations}

\noi
We would like to analyze which coefficients \mb{c^{n}_{k}} could guarantee
the nilpotency relations \e{nilprelaaa}, if we are only allowed to
additionally assume that \mb{Q} is nilpotent in the Lie bracket sense,
\beq
  [Q,Q]~=~0~. \label{qnilp}
\eeq
The \nth square bracket 
\mb{(\Phi_{Q} \circ \Phi_{Q})^{n}(a^{\soprod n})} is nothing but a linear
combination of terms built out of \mb{n\!+\!1} nested Lie brackets
\mb{[~,~]}, whose \mb{n\!+\!2} arguments consist of two \mb{Q}'s and 
\mb{n} \mb{a}'s. The only such term that the nilpotency condition \e{qnilp}
annihilates, is the term
\bea
\Phi_{[Q,Q]}^{n}(a^{\soprod n})&=&
\underbrace{[[\ldots [[Q,Q],a],\ldots ], a]}_{n+1~{\rm Lie~brackets}}
~=~(-\ad a)^{n}[Q,Q] \cr
&=&\sum_{k=0}^{n}\twobyone{n}{k}
\left[\Phi_{Q}^{n-k}(a^{\soprod (n-k)}),\Phi_{Q}^{k}(a^{\soprod k})\right]
~,~~~~~~~~~~\epsilon(a)~=~\epsilon~,
\label{allowedterm}
\eea
for \mb{n\in\{0,1,2,\ldots\}}.
Therefore, instead of imposing the nilpotency condition \e{qnilp}, it is
equivalent to let the \nth square bracket 
\mb{(\Phi_{Q} \circ \Phi_{Q})^{n}(a^{\soprod n})} be proportional to the
term \e{allowedterm}, \ie
\beq
(\Phi_{Q} \circ \Phi_{Q})^{n}(a^{\soprod n})
~=~\alpha_{n}~\Phi_{[Q,Q]}^{n}(a^{\soprod n})~,\label{derivedsquarealpha}
\eeq
where \mb{\alpha_{n}} is a proportionality factor that depends on
\mb{n\in \{0,1,2,\ldots\}}. This off-shell strategy \wrt the nilpotency
condition \e{qnilp} has also been promoted in \Ref{bm99nonilp} in a similar
context.
Since one may trivially scale the nilpotency relations \e{nilprelaaa} with a
non-zero complex number, \cf Subsection~\ref{secrenormalization}, it is 
enough to study the square relation \e{derivedsquarealpha} with a 
proportionality factor equal to either \mb{\alpha_{n}\!=\!1} or 
\mb{\alpha_{n}\!=\!0}. The case \mb{\alpha_{n}\!=\!1} is a set of coupled, 
non-homogeneous linear (also known as affine) equations in the \mb{c^{n}_{k}}
product coefficients. The analogous homogeneous problem corresponds to
letting the proportionality factor be \mb{\alpha_{n}\!=\!0}, while continuing
{\em not} to require nilpotency \e{qnilp} of \mb{Q}.

\subsection{Solution}
\label{secsolution}

\noi
In this Subsection we present the complete solution to the square relation 
\e{derivedsquarealpha}. To this end, let \mb{B_{k}} be the Bernoulli numbers,
\mb{k\in\{0,1,2,\ldots\}}, generated by
\beq
B(x)~=~\frac{x}{e^{x}-1}~=~\sum_{k=0}^{\infty}\frac{B_{k}}{k!}x^{k}
~=~1-\frac{x}{2}+\frac{1}{6}\frac{x^{2}}{2!}-\frac{1}{30}\frac{x^{4}}{4!}
+{\cal O}(x^{6})~,
\label{bernoullidef}
\eeq
and let us for later convenience define the negative Bernoulli numbers as
zero,
\beq
0~=~B_{-1}~=~B_{-2}~=~B_{-3}~=~\ldots~. \label{negbernoullidef}
\eeq

\begin{theorem}
Let there be given a set of \mb{c^{n}_{k}} coefficients with
\mb{n\!\geq\! k\!\geq\! 0}. The square relations 
\beq
\Phi_{Q} \circ \Phi_{Q}~=~\Phi_{[Q,Q]}~\label{derivedsquareaaa}
\eeq
 are satisfied for all derived \mb{\bullet}-brackets \mb{\Phi_{Q}}, if and
only if 
\beq
 c^{n}_{k}~=~B_{k}+\delta_{k,1}+c^{n(H)}_{k}
~,~~~~~~~~~~~~n\geq k\geq 0~,
\label{derivedcsol}
\eeq
where the homogeneous part \mb{c^{n(H)}_{k}} solves the corresponding 
homogeneous equation 
\beq
\Phi_{Q} \circ \Phi_{Q}~=~0~.\label{derivednilprel}
\eeq
{}For a given \mb{n\in \{0,1,2,\ldots\}}, the solution space for the 
homogeneous problem \e{derivednilprel} is \mb{[\frac{n+1}{2}]} dimensional.
A basis \mb{c^{n(H)}_{k(m)}} of solutions, labelled by an integer 
\mb{m\in \{0,1,\ldots,[\frac{n-1}{2}]\}}, is
\beq
c^{n(H)}_{k(m)}~=~\twobyone{k}{m}B_{k-m}-\twobyone{k}{n\!-\!m}B_{k-n+m}
~=~-(m \leftrightarrow n\!-\!m)~.
\label{derivedhomcsol}
\eeq
\label{theoremderived}
\end{theorem}

\noi
In practice, it is easier to let the label \mb{m\in \{0,1,\ldots,n\}} run
all the way to \mb{n}, and work with an over-complete set of solutions.
Also introduce a generating polynomial
\beq
 c^{n(H)}_{k}(t):=\sum_{m=0}^{n}c^{n(H)}_{k(m)}t^{m}~.
\eeq
One may reformulate the solution \eqs{derivedcsol}{derivedhomcsol} \wtho 
a generating function
\beq
c(x,y)~:=~\sum_{k,\ell=0}^{\infty}c^{k+\ell}_{k}
\frac{x^{k}}{k!}\frac{y^{\ell}}{\ell!}~\equiv~
\sum_{k,\ell=0}^{\infty}\frac{b^{k+\ell}_{k}}{(k\!+\!\ell)!}x^{k}y^{\ell}~.
\label{defcxy}
\eeq
The particular solution \mb{c^{n}_{k}=B_{k}+\delta_{k,1}} is then
\bea
 c(x,y)&=&B(x)e^{x+y}~=~B(-x)e^{y}~=~(B(x)\!+\!x)e^{y} \cr
&=&1+\left(\frac{x}{2}+y\right)
+\Hf\left( \frac{x^{2}}{6}+xy+y^{2}\right)
+\frac{1}{3!}\left( \frac{x^{2}y}{2}
+\frac{3xy^{2}}{2}+y^{3}\right) \cr
&&+\frac{1}{4!}\left(- \frac{x^{4}}{30}+x^{2}y^{2}
+2xy^{3}+y^{4}\right)+\ldots~,\label{genparsolution}
\eea
and the homogeneous solution is
\beq 
c^{(H)}(x,y,t)~=~B(x)e^{xt+y}-B(xt)e^{x+yt}~,\label{genhomsolution}
\eeq
where \mb{B} is the Bernoulli generating function, \cf \eq{bernoullidef}.
Eqs.\ \es{genparsolution}{genhomsolution} are our main result of 
Section~\ref{secderoperad}. We emphasize that the square relation 
\e{derivedsquareaaa} and its homogeneous counterpart \e{derivednilprel} are 
satisfied with these \mb{c^{n}_{k}} product solutions {\em without} assuming
the nilpotency condition \e{qnilp}. Also note that none of these solutions are
consistent with the ordinary product \e{stasheffchoice}. The solution shows 
that attempts to fit the derived bracket hierarchy \e{derbracket} into 
the original \hla definition \e{stasheffchoice} are bound to be unnatural 
and will only work in special situations.

\noi
{\it Proof of Theorem~\ref{theoremderived}:}~~
To derive the solution \eqs{derivedcsol}{derivedhomcsol}, first note that
for two elements \mb{a,b \in{\cal A}} with \mb{\epsilon(a)=\epsilon},
\bea
\Phi_{Q}^{n+1}(b\soprod a^{\soprod n})&=&\frac{1}{n\!+\!1}\sum_{\ell=0}^{n}
\underbrace{[[\ldots [[\Phi_{Q}^{\ell}(a^{\soprod \ell}),b],a],\ldots ], a]}_{
n-\ell+1~{\rm Lie~brackets}} \cr
&=&\frac{1}{n\!+\!1}\sum_{\ell=0}^{n}(-\ad a)^{n-\ell}
[\Phi_{Q}^{\ell}(a^{\soprod \ell}),b]
\label{derivedhelp}
\eea
for \mb{n\in \{0,1,2,\ldots\}}. 
Letting \mb{b=\Phi_{Q^{\prime}}^{k}(a^{\soprod k})} this becomes
\beq
\Phi_{Q}^{n+1}\left(\Phi_{Q^{\prime}}^{k}(a^{\soprod k})
\soprod a^{\soprod n}\right)
~=~\sum_{i=0}^{n}\frac{1}{i\!+\!1}\twobyone{n}{i} 
\left[\Phi_{Q}^{n-i}(a^{\soprod (n-i)}),
\Phi_{Q^{\prime}}^{k+i}(a^{\soprod (k+i)})\right]~.
\label{derhelp}
\eeq
After some elementary manipulations the \nth bracket product reads
\bea
(\Phi_{Q} \circ \Phi_{Q^{\prime}})^{n}(a^{\soprod n})
&=&\sum_{k=0}^{n} b^{n}_{k}~\Phi_{Q}^{n-k+1}
\left(\Phi_{Q^{\prime}}^{k}(a^{\soprod k})\soprod a^{\soprod (n-k)}\right)\cr
&=&\sum_{k=0}^{n}\sum_{i=0}^{k}
\frac{b^{n}_{k-i}}{i\!+\!1}\twobyone{n\!-\!k\!+\!i}{i} 
\left[\Phi_{Q}^{n-k}(a^{\soprod (n-k)}),
\Phi_{Q^{\prime}}^{k}(a^{\soprod k})\right] \cr
&=&\sum_{k=0}^{n} \twobyone{n}{k}\sum_{i=0}^{k}
\frac{c^{n}_{k-i}}{i\!+\!1}\twobyone{k}{i} 
\left[\Phi_{Q}^{n-k}(a^{\soprod (n-k)}),
\Phi_{Q^{\prime}}^{k}(a^{\soprod k})\right]~.
\label{derivedsquare1}
\eea
Combining eqs.\ \e{allowedterm}, \es{derivedsquareaaa}{derivedsquare1}
with \mb{Q^{\prime}\!=\!Q}, one derives
\beq
\forall k,n:~~~0\leq k\leq n~~~~~~\Rightarrow~~~~~~
\sum_{i=0}^{k}\frac{c^{n}_{k-i}}{i\!+\!1}\twobyone{k}{i}
-1~=~-(k \leftrightarrow n\!-\!k)~.\label{nonhomlinceq}
\eeq
The \rhs of \eq{nonhomlinceq} comes from the symmetry
\beq
\left[\Phi_{Q}^{k}(a^{\soprod k}),\Phi_{Q}^{\ell}(a^{\soprod \ell})\right]
~=~(k \leftrightarrow \ell)~,~~~~~~~~~~~~~k,\ell\in \{0,1,2,\ldots\}~.
\eeq
This symmetry is the origin of the non-trivial homogeneous solutions.
Let us first assume that the \lhs of \e{nonhomlinceq} vanishes. The unique
solution for this case reads
\beq
   c^{n}_{k}~=~B_{k}+\delta_{k,1}~,\label{particularcsol}
\eeq
which establishes \eq{derivedcsol}.
We now focus on the homogeneous part of \eq{nonhomlinceq}. {}First note 
that for a fixed \mb{m\in\{0,\ldots,n\}}, the equation
\beq
\sum_{i=0}^{k}\frac{c^{n}_{k-i}}{i\!+\!1}\twobyone{k}{i}~=~\delta_{k,m}~,
\label{kroneckerdeltasourse}
\eeq
with a Kronecker delta function source on the right-hand side, has the unique
solution
\beq
   c^{n}_{k}~=~\twobyone{k}{m}B_{k-m}~.
\eeq
Hence a \mb{(k \leftrightarrow n\!-\!k)} skewsymmetric version of the
Kronecker delta function source, \ie
\beq
\sum_{i=0}^{k}\frac{c^{n}_{k-i}}{i\!+\!1}\twobyone{k}{i}
~=~\delta_{k,m}-\delta_{k,n-m}
\eeq
will have the unique solution \eq{derivedhomcsol}.
\proofbox

\subsection{Ward-like and Jacobi-like Identities}

\noi
We now discuss particular useful solutions, \ie non-trivial identities with as
few terms as possible. The \nth square bracket 
\mb{(\Phi_{Q} \circ \Phi_{Q})^{n}(a,\ldots, a)} typically
has \mb{n+1} terms on the diagonal. Here we shall use the freedom in the 
homogeneous part to kill most of these terms.

\noi
Consider first the Ward solution
\beq
c^{n}_{k}~=~B_{k}+\delta_{k,1}-c^{n(H)}_{k(0)}
~=~\delta_{k,1}+\delta_{k,n}~,~~~~~~~~~~~~n\geq k\geq 0~,
\label{wardsol1}
\eeq
or equivalently using \eq{defcxy},
\beq
 c(x,y)~=~xe^{y}+e^{x}~.\label{wardsol2}
\eeq
It corresponds to a hierarchy of Ward identities
\beq
n~\Phi_{Q}^{n}\left(\Phi_{Q}^{1}(a)\soprod a^{\soprod (n-1)} \right)
+\Phi_{Q}^{1}\left(\Phi_{Q}^{n}(a^{\soprod n})\right)~=~
\Phi_{[Q,Q]}^{n}(a^{\soprod n})
~,~~~~~~~\epsilon(a)~=~\epsilon~,\label{wardidaaa}
\eeq
where \mb{n\in \{0,1,2,\ldots\}}. (The name ``Ward identity'' refers to a
similar identity encountered in String Theory.) After polarization the Ward 
identities are
\bea
\sum_{k=1}^{n}(-1)^{\epsilon_{1}+\ldots+\epsilon_{k-1}+(k-1)\epsilon}
~\Phi_{Q}^{n}\left(a_{1},\ldots,
a_{k-1},\Phi_{Q}^{1}(a_{k}),a_{k+1},\ldots, a_{n}\right) &&\cr
+\Phi_{Q}^{1}\left(\Phi_{Q}^{n}(a_{1},\ldots, a_{n})\right)&=&
\Phi_{[Q,Q]}^{n}(a_{1},\ldots, a_{n})~.\label{wardid}
\eea
The first few Ward identities read,
\bea
\Phi_{Q}^{1}(\Phi_{Q}^{0})&=&\Phi_{[Q,Q]}^{0}~,\label{wardid0} \\
2\Phi_{Q}^{1}\left(\Phi_{Q}^{1}(a)\right)
&=&\Phi_{[Q,Q]}^{1}(a)~,\label{dernilp1}\label{wardid1} \\
\Phi_{Q}^{2}\left(\Phi_{Q}^{1}(a),b\right)
+(-1)^{\epsilon_{a}+\epsilon}
\Phi_{Q}^{2}\left(a,\Phi_{Q}^{1}(b)\right)
+\Phi_{Q}^{1}\left(\Phi_{Q}^{2}(a,b)\right)
&=&\Phi_{[Q,Q]}^{2}(a,b)~.\label{derleibnitz1}
\eea
Consider next the solution
\bea
c^{n}_{k}&=&B_{k}+\delta_{k,1}-\frac{2}{n}c^{n(H)}_{k(1)}
+\sum_{m=0}^{n}\frac{m}{n}c^{n(H)}_{k(m)} \cr
&=&\frac{4}{n}\delta_{k,2}+\frac{2}{n}\delta_{k,n-1}-\delta_{k,n}~,
~~~~~~~~~~~~0\neq n\geq k\geq 0~,\label{derjacsol1}
\eea
or equivalently using \eq{defcxy},
\bea
 c(x,y)-1&=&2x^{2}E^{\prime}(y)+(2y-x) E(x)\cr
&=&2\left(\frac{x}{y}\right)^{2}\left(1+(y-1)e^{y}\right)
+\left(2\frac{y}{x}-1\right)\left(e^{x}-1\right)~.\label{derjacsol2}
\eea
Here we have defined
\beq
E(x)~:=~\frac{e^{x}-1}{x}~=~\sum_{k=0}^{\infty}\frac{x^{k}}{(k\!+\!1)!}
~=~\frac{1}{B(x)}~.
\eeq
The solution \e{derjacsol1} corresponds to generalized Jacobi identities
\beq
2n~ \Phi_{Q}^{n}\left(\Phi_{Q}^{2}(a\soprod a)
\soprod a^{\soprod (n-1)}\right)
+2\Phi_{Q}^{2}\left(\Phi_{Q}^{n}(a^{\soprod n})\soprod a\right)
- \Phi_{Q}^{1}\left(\Phi_{Q}^{n+1}(a^{\soprod (n+1)})\right)
~=~\Phi_{[Q,Q]}^{n+1}(a^{\soprod (n+1)})~,\label{derjacid}
\eeq
where \mb{n\in \{0,1,2,\ldots\}}. The first few read,
\bea
2\Phi_{Q}^{2}\left(\Phi_{Q}^{0},a\right)
-\Phi_{Q}^{1}\left(\Phi_{Q}^{1}(a)\right)
&=&\Phi_{[Q,Q]}^{1}(a)~, \\
\Phi_{Q}^{2}\left(\Phi_{Q}^{1}(a),b\right)
+(-1)^{\epsilon_{a}+\epsilon}
\Phi_{Q}^{2}\left(a,\Phi_{Q}^{1}(b)\right)
+\Phi_{Q}^{1}\left(\Phi_{Q}^{2}(a,b)\right)
&=&\Phi_{[Q,Q]}^{2}(a,b)~,\label{derleibnitz2} \\
2~{\rm Jac}(a,b,c)
-\Phi_{Q}^{1}\left(\Phi_{Q}^{3}(a,b,c)\right)
&=&\Phi_{[Q,Q]}^{3}(a,b,c)~,\label{derjacid3}
\eea
where  \mb{{\rm Jac}(a,b,c)} is the Jacobiator, \cf \eq{jacobiator1}. 
It is worth mentioning that the zero-bracket \mb{\Phi_{Q}^{0}}, which
normally complicates a homotopy Lie algebra, \cf eqs.\ \e{nilprel1}, 
\e{nilprel2} and \e{nilprel3}, here decouples, \cf the corresponding eqs.\ 
\e{dernilp1}, \e{derleibnitz1} and \e{derjacid3}. 
Therefore, in the nilpotent case \mb{[Q,Q]=0}, the Grassmann-odd one-bracket 
\mb{\Phi_{Q}^{1}} is nilpotent \e{dernilp1}, and it obeys a Leibniz rule
\e{wardid} \wrt the \mb{n}-bracket \mb{\Phi_{Q}^{n}}. And perhaps most
importantly, the two-bracket satisfies a generalized Jacobi identity 
\e{derjacid3} that only differs from the original Jacobi identity by a 
\mb{\Phi_{Q}^{1}}-exact term.

\noi
By transcribing the work of Courant \cite{courant90} to this situation, one
may define the notion of a Dirac subalgebra.

\begin{definition} A {\bf Dirac subalgebra} is a subspace 
\mb{{\cal L}\subseteq{\cal A}} that is: 
\begin{enumerate}
\item
Maximal Abelian \wrt the original Lie bracket \mb{[~,~]}, 
\item
Closed under the two-bracket \mb{\Phi_{Q}^{2}}, \ie
\mb{\Phi_{Q}^{2}({\cal L},{\cal L})\subseteq {\cal L}}, or equivalently,
\mb{[{\cal L},{\cal L}]_{Q}\subseteq {\cal L}}, \cf \eq{dertwobracket}.
\end{enumerate}
\label{defdiracsubalg}
\end{definition}

\noi
It follows immediately from the bracket definition \e{derbracket} that all the
higher brackets \mb{\Phi_{Q}^{n}}, \mb{n\in\{3,4,5,\ldots\}} vanish on a Dirac
subalgebra \mb{{\cal L}}. In particular, the Jacobi identity for the 
two-bracket \mb{\Phi_{Q}^{2}}  is satisfied in a Dirac subalgebra 
\mb{{\cal L}}, \cf \eq{derjacid3}. We also point out a connection to 
Courant algebroids, where the generalized Jacobi identity \e{derjacid3} 
translates into the first (out of five) defining properties for the Courant
algebroid, \cf \Ref{xu97}.

\section{The Courant Bracket}
\label{seccourant}

\noi
The Courant bracket \cite{courant90} has received much interest in recent years
primarily due to Hitchin generalized complex geometry \cite{hitchin03}. A broad
introduction to the subject can be found in the PhD theses of Roytenberg
\cite{roytenberg99} and Gualtieri \cite{gualtieri04}.
In this Section we give three different constructions of the (skewsymmetric)
Courant bracket as a derived bracket, and we elaborate on its connection to 
\hlas \cite{roytenberg98,roytenberg99}. One well-known construction
\cite{yks96,yks04,yks05,roytenberg99} relies on an operator representation, see
Subsection~\ref{secschroedinger}, and two partially new constructions rely 
on an even and an odd Poisson bracket
\cite{lyakhovich04,roytenberg99,roytenberg02}, respectively, see
Subsections~\ref{seccansympl}-\ref{seccourantderivedbracket}.

\noi
The Courant bracket is defined on vectors and exterior forms as
\bea
[X,Y]_\HH&=&[X,Y]+(-1)^{\epsilon_{Y}}i_\XX i_\YY H
~=~-(-1)^{\epsilon_{X}\epsilon_{Y}}[Y,X]_\HH
~,~~~~~~~~~~X,Y\in\Gamma(TM)~,\label{couxy} \\
{}[X,\eta]_\HH &=&\Hf ({\cal L}_\XX+ i_\XX d)\eta~=~
(i_\XX d+\frac{(-1)^{\epsilon_{X}}}{2} d i_\XX)\eta
~=~({\cal L}_\XX-\frac{(-1)^{\epsilon_{X}}}{2} d i_\XX)\eta \cr
&=&-(-1)^{\epsilon_{X}\epsilon_{\eta}}[\eta,X]_\HH~,
\label{couxeta}  \\
{}[\xi,\eta]_\HH&=&0~,~~~~~~~~~~
\xi,\eta\in\Gamma(\bigwedge{}^{\bullet}(T^{*}M))~,
\label{couxieta}
\eea
with a closed twisting form \mb{H\in\Gamma(\bigwedge(T^{*}M))}. (We shall
ignore the fact that the twisting form \mb{H} is zero in the original Courant
bracket \cite{courant90}.) The Courant bracket \e{couxy}-\e{couxieta} does not
satisfy the Leibniz rule nor the Jacobi identity. It is therefore natural to
ask what is the significance of these formulas, in particular \eq{couxeta}?
In hindsight the answer is, that there exists a \hla structure behind the
Courant bracket, that makes the Jacobi identity valid modulo \mb{Q}-exact
terms, \cf \eq{derjacid3}. And underneath the \hla structure, there is a
Grassmann-odd nilpotent Hamiltonian vector field that generalizes the de Rham
exterior derivative. As we shall see in
Subsection~\ref{sechighercourantbrackets} below, the higher Courant brackets
are naturally defined on multi-vectors and exterior forms via the derived
bracket hierarchy \e{derbracket}. When restricted to only vector fields,
functions and one-forms, the \mb{L_{\infty}} hierarchy truncates and becomes
an \mb{L_{3}} algebra \cite{ladamarkl95}. Roytenberg and Weinstein
\cite{roytenberg98,roytenberg99} define a related set of higher brackets 
through a homological resolution \cite{stassheff98}. 

\noi
{}Finally, let us mention for completeness that people often consider a
non-skewsymmetric version of the Courant bracket (which is also called a
Dorfman bracket and is related to Loday/Leibniz algebras), partly to avoid 
handling the Jacobi identity directly, and partly to simplify the axioms for a
Courant algebroid, \cf \Ref{roytenberg99}. Nevertheless, at the end of the
day, it is often the skewsymmetrized bracket that is relevant for applications.
More importantly, our underlying derived bracket definition \e{derbracket} is
manifestly skewsymmetric, and hence we shall here only treat skewsymmetric
brackets. We shall take advantage of polarization to shortcut the lengthy
calculations that is normally associated with the skewsymmetric bracket, \cf
Subsection~\ref{secpolarization}.

\subsection{Review of the Operator Representation}
\label{secschroedinger}

\noi
Perhaps the simplest realization of the Courant bracket as a (skewsymmetric, 
inner) derived bracket is the following operator construction due to
Kosmann-Schwarzbach \cite{yks96,yks04,yks05,roytenberg99}. Consider a
\mb{d}-dimensional bosonic manifold \mb{M}, and let \mb{x^{i}},
\mb{i\in\{1,2,\ldots,d\}}, denote local bosonic coordinates in some coordinate
patch \mb{U\subseteq M}.  To avoid cumbersome notation, we shall often take the
liberty to write local objects, which formally only live on \mb{U}, as if they
are living on the whole manifold \mb{M}. {}Furthermore, we shall regard
exterior forms 
\beq
\eta=\eta(x^{i},\cc^{i})~\in~\Gamma(\bigwedge{}^{\bullet}(T^{*}M))
~\cong~ C^{\infty}(E)~,~~~~~~~~~~~~~~~E\equiv\Pi TM~,
\label{etafcts}
\eeq 
as functions in the variable \mb{(x^{i},\cc^{i})} on the parity-inverted
tangent bundle \mb{E\equiv\Pi TM} by identifying the basis of one-forms 
\mb{dx^{i}\equiv \cc^{i}} with the fermionic variables \mb{\cc^{i}}. 
Notice that an \mb{n}-form \mb{\eta=\frac{1}{n!}\eta_{i_{1}\ldots i_{n}}
\cc^{i_{1}}\ldots\cc^{i_{n}}} has Grassmann parity \mb{\epsilon_{\eta}=n}
modulo \mb{2}, if the coordinate functions
\mb{\eta_{i_{1}\ldots i_{n}}=\eta_{i_{1}\ldots i_{n}}(x)} are bosonic. 
Now let the Lie algebra \mb{{\cal A}} of Section~\ref{secderoperad} be the Lie
algebra \mb{{\cal A} = {\rm End}(C^{\infty}(E))} of operators acting on
the above functions \e{etafcts}, and let the Lie bracket for \mb{{\cal A}} be
given by the commutator bracket \mb{[\cdot,\cdot]}. We shall often 
identify an exterior form \mb{\eta} with the left multiplication operator
\mb{L_{\eta}\!\in\!{\cal A}}, \ie the operator that multiplies from the left
with \mb{\eta}. {}For the fixed Lie algebra element of
Section~\ref{secderoperad}, one now chooses a twisted version 
\beq
 D~:=~d+H_{\odd}~\in~{\cal A}
\label{dcharge}
\eeq
of the de Rham exterior derivative \mb{d} on the manifold \mb{M},
\beq
d~=~\cc^{i}\frac{\partial}{\partial x^{i}}~,~~~~~~~~~\epsilon(d)~=~1~.
\label{rhamdiff}
\eeq
Here \mb{H_{\odd}\in\Gamma(\bigwedge^{\odd}(T^{*}M))} is an odd closed exterior
form, \ie linear combinations of forms of odd form-degree, that are closed.
The oddness condition ensures that \mb{D} carries definite Grassmann-parity.
The \mb{D} operator is nilpotent,
\beq
 [D,D]~=~2dH_{\odd}~=~0~,
\eeq
because the exterior form \mb{H_{\odd}} is closed. Parity-inverted vector
fields \mb{\tilde{X}} (where the tilde ``\mb{\sim}'' denotes parity-inversion,
\cf \eq{tildeiso} below) are now represented as first-order differential
operators,
\beq
i_\XX~=~X^{i}\frac{\partial^{l}}{\partial \cc^{i}}~,~~~~~~~~
\epsilon(i_\XX)~=~1\!-\!\epsilon_{X}~.
\label{ixcontraction}
\eeq
One may now construct the derived \mb{\bullet}-bracket hierarchy of \mb{D}.
The zero-bracket \mb{\Phi_{D}^{0}=D} is the generator \mb{D} itself. The
one-bracket \mb{\Phi_{D}^{1}=[D,\cdot]} is a nilpotent, Grassmann-odd operator,
\cf \eq{dernilp1}. It is just the de Rham exterior derivative on exterior
forms,
\beq
\Phi^{1}_{D}(\eta)~=~ [D,\eta] ~=~d\eta~,
\eeq
independently of \mb{H_{\odd}}. On vectors, the one-bracket \mb{\Phi^{1}_{D}}
becomes a sum of a Lie derivative and a contracted term,
\beq
(-1)^{\epsilon_{X}} \Phi^{1}_{D}(i_\XX)~=~[i_\XX,D]
~=~{\cal L}_\XX+i_\XX H_{\odd}~.
\eeq
Here and below, we repeatedly make use of the Cartan relations
\beq 
2d^2~=~[d,d]~=~0~,~~~~~~[i_\XX,i_\YY]~=~0~,~~~~~~
{\cal L}_\XX~=~[i_\XX,d]~,~~~~~~i_{[X,Y]}^{}~=~[{\cal L}_\XX,i_\YY]~.
\eeq
The two-bracket \mb{\Phi^{2}_{D}} gives rise to an odd version
\mb{[\cdot,\cdot]_\DD} of the Courant bracket \cite{yks96,yks04,yks05},
\cf \eq{dertwobracket}.  

\begin{definition} The odd {\bf Courant bracket} \mb{[\cdot,\cdot]_\DD} is
defined as the derived bracket,
\bea
[a,b]_\DD&:=&(-1)^{\epsilon_{a}+1}\Phi^{2}_{D}(a,b)
~=~\Hf [[a,D],b]+\Hf [a,[D,b]] \cr
&=&-(-1)^{(\epsilon_{a}+1)(\epsilon_{b}+1)}[b,a]_\DD
~,~~~~~~~~~~~~~a,b~\in~{\cal A}={\rm End}(C^{\infty}(E))~.
\eea 
\end{definition}

\noi
In this formulation, the Courant bracket \e{couxy}-\e{couxieta} reads 
 \bea
[i_\XX,i_\YY]_\DD&=&i_{[X,Y]}^{}+(-1)^{\epsilon_{Y}}
i_\XX i_\YY H_{\odd}~,~~~~~~~~~~X,Y\in\Gamma(TM)~,
\label{anticourantdhxy} \\
{}[i_\XX,\eta]_\DD &=&\Hf ({\cal L}_\XX+ i_\XX d)\eta~,
\label{anticourantdhxeta}  \\
{}[\xi,\eta]_\DD&=&0~,~~~~~~~~~~
\xi,\eta\in\Gamma(\bigwedge{}^{\bullet}(T^{*}M))~.
\label{anticourantdhxieta}
\eea

\subsection{Symplectic Structure}
\label{seccansympl}

\noi
It is also possible to build the Courant bracket \e{couxy}-\e{couxieta} as a
derived bracket \wtho a Poisson bracket \mb{\{\cdot,\cdot\}}. Consider a
\mb{d}-dimensional bosonic base manifold \mb{M}, let \mb{x^{i}}, 
\mb{i\in\{1,2,\ldots,d\}}, denote local bosonic coordinates
in some coordinate patch, and let \mb{p_{i}} be the local bosonic basis vectors
\mb{p_{i}\equiv\partial_{i}\in \Gamma(TM)} for symmetric multi-vectors,
\mb{i\in\{1,2,\ldots,d\}}. We now define a Poisson bracket
on the cotangent bundle \mb{T^{*}M} as
\beq
\begin{array}{rcccl}
\{p_{i},x^{j}\}&=&\delta_{i}^{j}&=&-\{x^{j},p_{i}\}~, \cr
\{x^{i},x^{j}\}&=&0~,
\end{array}
\label{poisson1}
\eeq
where we are going to fix the fundamental Poisson bracket \mb{\{p_{i},p_{j}\}} 
a little bit later. Within the \mb{(x^{i},p_{j})} sector, it is consistent to
put \mb{\{p_{i},p_{j}\}} to zero, but an extension below to other sectors of
the Whitney sum \mb{\cE\equiv T^{*}M\oplus E \oplus E^{*}\supseteq T^{*}M} 
will complicate matters, \cf \eq{poisson4} and Table~\ref{poissontable}. (The 
sign of the Poisson bracket \e{poisson1}, which is opposite of the standard
physics conventions, has been chosen to minimize appearances of minus signs.)

\begin{table}[b]
\caption{The Poisson bracket \mb{\{\cdot,\cdot\}}.} 
\label{poissontable}
\begin{center}
\begin{tabular}{|c||c|c|c|c|}  \hline
&$x^{j}$&$p_{j}$&$\cc^{j}$&$\bb_{j}$\\ \hline\hline
$x^{i}$&$0$&$-\delta^{i}_{j}$&$0$&$0$\\ \hline
$p_{i}$&$\delta^{j}_{i}$&$-\tilde{R}_{ij}$&$-\Gamma_{ik}^{j}\cc^{k}$&
$\Gamma_{ij}^{k}\bb_{k}$\\ \hline
$\cc^{i}$&$0$&$\Gamma_{jk}^{i}\cc^{k}$&$0$&$\delta^{i}_{j}$\\ \hline
$\bb_{i}$&$0$&$-\Gamma_{ji}^{k}\bb_{k}$&$\delta^{j}_{i}$&$0$\\ \hline
\end{tabular}
\end{center}
\end{table}

\begin{table}[t]
\caption{The Poisson bracket \mb{\{\cdot,\cdot\}} in Darboux coordinates.} 
\label{canpoissontable}
\begin{center}
\begin{tabular}{|c||c|c|c|c|}  \hline
&$x^{j}$&$\PP_{j}$&$\cc^{j}$&$\bb_{j}$\\ \hline\hline
$x^{i}$&$0$&$-\delta^{i}_{j}$&$0$&$0$\\ \hline
$\PP_{i}$&$\delta^{j}_{i}$&$0$&$0$&$0$\\ \hline
$\cc^{i}$&$0$&$0$&$0$&$\delta^{i}_{j}$\\ \hline
$\bb_{i}$&$0$&$0$&$\delta^{j}_{i}$&$0$\\ \hline
\end{tabular}
\end{center}
\end{table}

\begin{table}[b] 
\caption{The symplectic two-form \mb{\omega} in \eq{symplecticform} consists of
the inverse matrix of Table~\ref{poissontable}.} 
\label{invpoissontable}
\begin{center}
\begin{tabular}{|c||c|c|c|c|}  \hline
&$\ddd x^{j}$&$\ddd p_{j}$&$\ddd\cc^{j}$&$\ddd\bb_{j}$
\\ \hline\hline
$\ddd x^{i}$&$-\partial_{[i}^{}\Gamma_{j]\ell}^{k}\cc^{\ell}\bb_{k}$&
$\delta^{j}_{i}$&$-\Gamma_{ij}^{k}\bb_{k}$&$\Gamma_{ik}^{j}\cc^{k}$
\\ \hline
$\ddd p_{i}$&$-\delta^{i}_{j}$&$0$&$0$&$0$\\ \hline
$\ddd\cc^{i}$&$-\Gamma_{ji}^{k}\bb_{k}$&$0$&$0$&$\delta^{j}_{i}$\\ \hline
$\ddd\bb_{i}$&$\Gamma_{jk}^{i}\cc^{k}$&$0$&$\delta^{i}_{j}$&$0$\\ \hline
\end{tabular}
\end{center}
\end{table}

\noi
We next introduce the notation
\mb{\bb_{i}\equiv\tilde{\partial}_{i}\in\Gamma(E)} for the local fermionic
basis of skewsymmetric multi-vectors, where as before the tilde ``\mb{\sim}''
represents a parity-inversion, \cf \eq{tildeiso}. The skewsymmetric
multi-vectors
\beq
\pi=\pi(x^{i},\bb_{i})~\in~\Gamma(\bigwedge{}^{\bullet}(TM))
~\cong~ C^{\infty}(E^{*})~,~~~~~~~~~~~~~~~E^{*}\cong\Pi T^{*}M~,
\eeq
can be identified with functions on the parity-inverted cotangent bundle
\mb{\Pi T^{*}M\cong E^{*}}. Similarly, as explained in
Subsection~\ref{secschroedinger}, we write 
\mb{\cc^{i}\equiv dx^{i}\in\Gamma(E^{*})} for the local fermionic one-forms
that constitute a basis for the exterior forms. Recall also that exterior forms
\e{etafcts} can be regarded as functions on the parity-inverted tangent bundle
\mb{E\equiv\Pi TM}. Note that the natural symmetric pairing 
\mb{\langle\tilde{\partial}_{i}, dx^{j}\rangle_{+}=\delta_{i}^{j}
=\langle dx^{j},\tilde{\partial}_{i}\rangle_{+}} between exterior forms and 
multi-vectors can equivalently be viewed as a canonical Poisson bracket of
fermions,
\beq
\begin{array}{rcccl}
\{\bb_{i},\cc^{j}\}&=&\delta_{i}^{j}&=&\{\cc^{j},\bb_{i}\}~,\cr
\{\cc^{i},\cc^{j}\}&=&0&=&\{\bb_{i},\bb_{j}\}~.
\end{array}
\label{poisson2}
\eeq
The idea is now to regard the two Poisson brackets \es{poisson1}{poisson2} as
part of the same symplectic structure on a \mb{4d} dimensional manifold, 
which we take to be the total space
\mb{\cE\equiv T^{*}M\oplus E\oplus E^{*}} of the \mb{3d} dimensional
vector bundle \mb{\cE\to M}
with local coordinates \mb{(x^{i};p_{i},\cc^{i},\bb_{i})}, and
let the combined symplectic structure play the r\^ole of the Lie algebra
structure \mb{[\cdot,\cdot]} of Section~\ref{secderoperad}. Obviously this idea
implies that one should fix the Poisson bracket in the cross-sectors between
the even and the odd coordinates. In the end, it turns out that the
cross-sector assignments do not matter, as they do not enter the Courant
bracket in pertinent sectors. However to be specific, we shall model the
cross-sectors over an arbitrary connection
\mb{\nabla:\Gamma(TM) \times \Gamma(TM) \to \Gamma(TM)}. In detail, the Poisson
brackets between bosonic and fermionic variables are \cite{lyakhovich04} 
\beq
\begin{array}{rcccl}
\{x^{i},\bb_{j}\}&=&0&=&\{\bb_{j},x^{i}\}~,\cr
 \{x^{i},\cc^{j}\}&=&0&=&\{\cc^{j},x^{i}\}~,\cr
\{p_{i},\bb_{j}\}&=&\Gamma_{ij}^{k}\bb_{k}&=&-\{\bb_{j},p_{i}\}~,\cr
-\{p_{i},\cc^{j}\}&=&\Gamma_{ik}^{j}\cc^{k}&=&\{\cc^{j},p_{i}\}~,
\end{array}
\label{poisson3}
\eeq
\cf Table~\ref{poissontable}. To ensure the Jacobi identity for the
\mb{\{\cdot,\cdot\}} bracket, one finally defines the \mb{\{p_{i},p_{j}\}}
sector to be
\beq
-\{p_{i},p_{j}\}~=~R^{k}{}_{\ell ij}\cc^{\ell}\bb_{k}
~=:~\tilde{R}_{ij}~,
\label{poisson4}
\eeq
where
\beq
R^{k}{}_{\ell ij}~=~\frac{\partial \Gamma^{k}_{j\ell}}{\partial x^{i}}
+\Gamma^{k}_{im}\Gamma^{m}_{j\ell}-(i\leftrightarrow j)
\label{riemanntensor}
\eeq
is the Riemann curvature tensor. {}For instance, the second Bianchi identity
\beq
0~=~\sum_{i,j,k~{\rm cycl.}}(\frac{\partial R^{m}{}_{n ij}}{\partial x^{k}}
+\Gamma^{m}_{k\ell}R^{\ell}{}_{n ij}-\Gamma^{\ell}_{kn}R^{m}{}_{\ell ij})
\label{2bianchiid}
\eeq
guarantees that the Jacobi identity holds in the \mb{(p_{i},p_{j},p_{k})}
sector.

\noi
A simpler picture emerges if one introduces the momentum variables
\cite{voronov99,voronov02}
\beq
 \PP_{i}~:=~p_{i}+\Gamma^{k}_{ij}\cc^{j}\bb_{k}~.
\label{ppvariable}
\eeq
Remarkably, the quadruple \mb{(x^{i},\PP_{i},\cc^{i},\bb_{i})} are local
Darboux coordinates for the Poisson bracket \mb{\{\cdot,\cdot\}}, \cf
Table~\ref{canpoissontable}. The corresponding symplectic two-form is just
the canonical two-form
\beq
\omega~=~\ddd x^{i}\wedge\ddd\PP_{i}+\ddd\cc^{i}\wedge\ddd\bb_{i}
~,~~~~~~~~~~~~\ddd \omega~=~0~, 
\label{symplecticform}
\eeq
\cf Table~\ref{invpoissontable}. Here \mb{\ddd} denotes the de Rham exterior
derivative on the Whitney sum \mb{\cE},
\beq
 \ddd~=~\ddd x^{i}\papal{x^{i}}+\ddd p_{i}\papal{p_{i}}
+\ddd \cc^{i}\papal{\cc^{i}}+\ddd \bb_{i}\papal{\bb_{i}}
~,~~~~~~~~~~\epsilon(\ddd)~=~0~,
\eeq
which should not be confused with the de Rham exterior derivative \mb{d} on 
the base manifold \mb{M}, \cf \eq{rhamdiff}. One may choose the symplectic
potential \mb{\vartheta} to be
\beq
 -\vartheta~=~\PP_{i}\ddd x^{i}-\bb_{i}\ddd\cc^{i}
~=~\ddd x^{i}~\PP_{i}+\ddd\cc^{i}~\bb_{i}
~,~~~~~~~~~~\ddd\vartheta~=~\omega~.
\label{symplecticpot}
\eeq
Let us address the issue of coordinate transformations in the base manifold
\mb{M}. Because of the presence of the \mb{\Gamma^{k}_{ij}} symbol in the
definition \e{ppvariable}, the momentum variables \mb{\PP_{i}} do not have the
simple co-vector transformation law that for instance the variables \mb{p_{i}}
and \mb{\bb_{i}} enjoy. Nevertheless, the local expression \e{symplecticpot}
for the one-form \mb{\vartheta} is invariant. Hence \mb{\vartheta} is a
globally defined symplectic potential for the \mb{(2d|2d)} symplectic manifold
\mb{\cE} with an exact symplectic two-form \mb{\omega=\ddd\vartheta}.
Moreover, the Whitney sum \mb{\cE\cong T^{*}E} may be identified with the
cotangent bundle \mb{T^{*}E}, where \mb{E\equiv\Pi TM} is the parity-inverted
tangent bundle with local coordinates \mb{(x^{i},\cc^{i})}. More precisely,
the momentum variables \mb{(\PP_{i},\bb_{i})} can be identified with the fiber
coordinates of \mb{T^{*}E}, and \mb{(\PP_{i},\bb_{i})} are co-vectors in that
sense \cite{voronov99,voronov02}. On the other hand, if one instead had started
with the cotangent bundle \mb{T^{*}E} rather than the Whitney sum \mb{\cE}, one
would have gotten the symplectic structure \e{symplecticform} for free, without
the use of a connection \mb{\nabla}. Depending on the application, it is useful
to do just that. The catch, is, that the \mb{\PP_{i}} variables follow a more
complicated set of transformation rules than the \mb{p_{i}} variables.

\noi 
There is a natural tilde isomorphism \mb{\sim:\Gamma(TM)\to \Gamma(\Pi TM)},
which maps vectors to parity-inverted vectors,
\beq
\begin{array}{rclcrcl}
 \Gamma(TM)&\ni& X=X^{i}p_{i}&~~\stackrel{\sim}{\longmapsto}~~&
X^{i}\bb_{i}=:\tilde{X}&\in&\Gamma(\Pi TM)~, \cr
&&\epsilon_{X} &&\epsilon_{\tilde{X}}&=&1-\epsilon_{X}~.
\end{array}
\label{tildeiso}
\eeq
In particular, vectors \mb{X} and parity-inverted vectors \mb{\tilde{X}} are
bosons and fermions, respectively, if the coordinate functions
\mb{X^{i}\!=\!X^{i}(x)} are bosonic, as is normally the case. More generally,
we define the tilde operation \mb{\sim:C^{\infty}(\cE)\to C^{\infty}(\cE)} to
be the right derivation
\beq
 \tilde{a}~:=(a\papar{p_{i}})\bb_{i}+(a\papar{\bb_{i}})p_{i}
~,~~~~~~~~~a=a(x^{i},p_{i},\cc^{i},\bb_{i})~\in~C^{\infty}(\cE)
~,~~~~~~~~~\epsilon(\sim)=1~.
\label{tildederiv}
\eeq
The Poisson bracket on vectors and exterior forms mimics the Lie bracket, the 
covariant derivative and the interior product (=contraction),
\bea
\{X,Y\}&=&\left(X^{i}\frac{\partial Y^{j}}{\partial x^{i}}p_{j}
-(-1)^{\epsilon_{X}\epsilon_{Y}}(X\leftrightarrow Y)\right)
- X^{i}Y^{j}\tilde{R}_{ij} \cr
&=:&[X,Y]- R(X,Y)^{\sim}~, \label{poissonxy}\\
\{X,\tilde{Y}\}&=&X^{i}(\frac{\partial Y^{j}}{\partial x^{i}}\bb_{j}
+ \Gamma_{ij}^{k}Y^{j}\bb_{k})~=:~(\nabla_\XX Y)^{k}\bb_{k}
~=:~(\nabla_\XX Y)^{\sim}~,\label{poissonxty}\\
\{\tilde{X},\tilde{Y}\}&=&0~,~~~~~~~~~~X,Y\in\Gamma(TM)~,\label{poissontxty}\\
\{X,\eta\}&=&X^{i}(\papal{x^{i}}
-\Gamma_{ik}^{j}\cc^{k}\papal{\cc^{j}})\eta
~=:~\nabla_\XX \eta~,\label{poissonxeta}\\
\{\tilde{X},\eta\} &=&X^{i}\frac{\partial^{l} \eta}{\partial \cc^{i}}
~=:~i_\XX\eta~,\label{poissontxeta}  \\
\{\xi,\eta\}&=&0~,~~~~~~~~~~
\xi,\eta\in\Gamma(\bigwedge{}^{\bullet}(T^{*}M))~.\label{poissonxieta}
\eea
The sign conventions are,
\beq
\epsilon(i_\XX)~=~1-\epsilon_{X}~=~\epsilon_{\tilde{X}}~,~~~~~~~
i_{\lambda X}^{}\eta~=~\lambda~ i_\XX\eta~,~~~~~~~
(\lambda X)^{\sim}~=~\lambda \tilde{X}~,~~~~~~~X\lambda 
~=~(-1)^{\epsilon_{\lambda}\epsilon_{X}} \lambda X~,
\eeq
where \mb{\lambda} is a supernumber.

\subsection{Anti-Symplectic Structure}
\label{seccanantisympl}

\begin{table}[b] 
\caption{The anti-bracket \mb{(\cdot,\cdot)}.} 
\label{antitable}
\begin{center}
\begin{tabular}{|c||c|c|c|c|}  \hline
&$x^{j}$&$\bb_{j}$&$\cc^{j}$&$p_{j}$\\ \hline\hline
$x^{i}$&$0$&$-\delta^{i}_{j}$&$0$&$0$\\ \hline
$\bb_{i}$&$\delta^{j}_{i}$&$-R_{ij}$&$-\Gamma_{ik}^{j}\cc^{k}$&
$\Gamma_{ij}^{k}p_{k}$\\ \hline
$\cc^{i}$&$0$&$\Gamma_{jk}^{i}\cc^{k}$&$0$&$-\delta^{i}_{j}$\\ \hline
$p_{i}$&$0$&$-\Gamma_{ji}^{k}p_{k}$&$\delta^{j}_{i}$&$0$\\ \hline
\end{tabular}
\end{center}
\end{table}

\noi
There is a dual formulation in terms of an odd Poisson bracket, also known as
an anti-bracket and traditionally denoted as \mb{(\cdot,\cdot)} in the physics
literature. The symbol \mb{\{\cdot,\cdot\}} will be reserved for the even
Poisson bracket introduced in the last Subsection~\ref{seccansympl}. The
anti-bracket \mb{(\cdot,\cdot)} on \mb{E^{*}\subseteq\cE} is given as
\cite{lyakhovich04} 
\beq
\begin{array}{rcccl}
  (\bb_{i},x^{j})&=&\delta_{i}^{j}&=&-(x^{j},\bb_{i})~, \cr
(x^{i},x^{j})&=&0~, \cr
-(\bb_{i},\bb_{j})&=&R^{k}{}_{\ell ij}\cc^{\ell}p_{k}&=:&R_{ij}~.
\end{array}
\label{schoutennijenhuis1}
\eeq
It is well-known that in the flat case \mb{R^{k}{}_{\ell ij}\!=\!0}, this 
anti-bracket \e{schoutennijenhuis1} is just the Schouten-Nijenhuis bracket
\beq
  (\pi,\rho)_{SN}~:=~\pi(\papar{\bb_{i}}\papal{x^{i}}
-\papar{x^{i}}\papal{\bb_{i}})\rho~,
\label{schoutennijenhuis0}
\eeq
when restricting to skewsymmetric multi-vectors
\mb{\pi,\rho\in\Gamma(\bigwedge{}^{\bullet}(TM))\cong C^{\infty}(E^{*})}. 
Similar to \eq{poisson2}, the natural skewsymmetric pairing 
\mb{\langle\partial_{i}, dx^{j}\rangle_{-}=\delta_{i}^{j}
=-\langle dx^{j},\partial_{i}\rangle_{-}}
can be modelled over an anti-bracket,
\beq
\begin{array}{rcccl}
  (p_{i},\cc^{j})&=&\delta_{i}^{j}&=&-(\cc^{j},p_{i})~, \cr
(\cc^{i},\cc^{j})&=&0&=&(p_{i},p_{j})~.
\end{array}
\label{schoutennijenhuis2}
\eeq
The antibracket \mb{(\cdot,\cdot)} is extended to the Whitney sum \mb{\cE} by
fixing the cross-sectors as 
\beq
\begin{array}{rcccl}
(x^{i},p_{j})&=&0&=&(p_{j},x^{i})~,\cr
 (x^{i},\cc^{j})&=&0&=&(\cc^{j},x^{i})~,\cr
(\bb_{i},p_{j})&=&\Gamma_{ij}^{k}p_{k}&=&-(p_{j},\bb_{i})~,\cr
-(\bb_{i},\cc^{j})&=&\Gamma_{ik}^{j}\cc^{k}
&=&(\cc^{j},\bb_{i})~,
\end{array}
\label{schoutennijenhuis3}
\eeq
\cf Table~\ref{antitable}.

\begin{table}[t] 
\caption{The anti-bracket \mb{(\cdot,\cdot)} in Darboux coordinates.} 
\label{canantitable}
\begin{center}
\begin{tabular}{|c||c|c|c|c|}  \hline
&$x^{j}$&$\BB_{j}$&$\cc^{j}$&$p_{j}$\\ \hline\hline
$x^{i}$&$0$&$-\delta^{i}_{j}$&$0$&$0$\\ \hline
$\BB_{i}$&$\delta^{j}_{i}$&$0$&$0$&$0$\\ \hline
$\cc^{i}$&$0$&$0$&$0$&$-\delta^{i}_{j}$\\ \hline
$p_{i}$&$0$&$0$&$\delta^{j}_{i}$&$0$\\ \hline
\end{tabular}
\end{center}
\end{table}

\begin{table}[b]
\caption{The anti-symplectic two-form \mb{\tilde{\omega}} in
\eq{antisymplecticpot} consists of the inverse matrix of
Table~\ref{antitable}.}
\label{invantitable}
\begin{center}
\begin{tabular}{|c||c|c|c|c|}  \hline
&$\ddd x^{j}$&$\ddd\bb_{j}$&$\ddd\cc^{j}$&$\ddd p_{j}$
\\ \hline\hline
$\ddd x^{i}$&$-\partial_{[i}^{}\Gamma_{j]\ell}^{k}\cc^{\ell}p_{k}$&
$\delta^{j}_{i}$&$\Gamma_{ij}^{k}p_{k}$&$\Gamma_{ik}^{j}\cc^{k}$\\ \hline
$\ddd\bb_{i}$&$-\delta^{i}_{j}$&$0$&$0$&$0$\\ \hline
$\ddd\cc^{i}$&$-\Gamma_{ji}^{k}p_{k}$&$0$&$0$&$\delta^{j}_{i}$\\ \hline
$\ddd p_{i}$&$-\Gamma_{jk}^{i}\cc^{k}$&$0$&$-\delta^{i}_{j}$&$0$\\ \hline
\end{tabular}
\end{center}
\end{table}

\noi
When one defines the anti-field variables
\beq
 \BB_{i}~:=~\bb_{i}+\Gamma^{k}_{ij}\cc^{j}p_{k}~,
\label{bbvariable}
\eeq
the quadruple \mb{(x^{i},\BB_{i},\cc^{i},p_{i})} become local Darboux
coordinates for the anti-bracket \mb{(\cdot,\cdot)}, \cf
Table~\ref{canantitable}. The corresponding anti-symplectic two-form reads
\beq
\tilde{\omega}~=~\ddd x^{i}\wedge\ddd \BB_{i}+\ddd\cc^{i}\wedge\ddd p_{i}
~,~~~~~~~~~~~~\ddd \tilde{\omega}~=~0~, 
\label{antisymplecticform}
\eeq
\cf Table~\ref{invantitable}. One may choose the following globally defined
anti-symplectic potential
\beq
 -\tilde{\vartheta}~=~\BB_{i}\ddd x^{i}+p_{i}\ddd\cc^{i}
~=~\ddd x^{i}~\BB_{i}+\ddd\cc^{i}~p_{i}
~,~~~~~~~~~~\ddd\tilde{\vartheta}~=~\tilde{\omega}~.
\label{antisymplecticpot}
\eeq
Note that the anti-symplectic potential \mb{\tilde{\vartheta}} is -- as the
name suggests -- equal to the symplectic potential \e{symplecticpot} acted
upon with the tilde operator ``\mb{\sim}'', \cf \eq{tildederiv}. The
\mb{(2d|2d)} anti-symplectic manifold \mb{\cE\cong\Pi T^{*}E} may be
identified with the parity-inverted cotangent bundle \mb{\Pi T^{*}E}. More
precisely, the anti-field variables \mb{(\BB_{i},p_{i})} can be identified
with the fiber coordinates of \mb{\Pi T^{*}E}.

\noi
We here display a commutative diagram of various bundle isomorphisms, so-called
Legendre transformations and canonical projection maps possible
\cite{roytenberg99,roytenberg02,voronov99,voronov02} 
\beq
\begin{array}{ccccccccccc}
&&&&&&&&{\rm Anti}&& \cr 
&&{\rm Legendre}&&({\rm if}~\nabla)&&
({\rm if}~\nabla)&&{\rm Legendre}&& \cr 
&T^{*}(E^{*})&\cong&T^{*}E&\cong&\cE&
\cong&\Pi T^{*}E&\cong&\Pi T^{*}(T^{*}M)&\cr\cr
&\parallel&\searrow&\downarrow&\swarrow&\downarrow&
\searrow&\downarrow&\swarrow&\parallel& \cr\cr
&\Pi T(E^{*})&\rightarrow&E\oplus E^{*}&\rightarrow&M&
\leftarrow&T^{*}M\oplus E&\leftarrow&\Pi T(T^{*}M)&
\end{array}
\label{identificationdiagram}
\eeq
\cf Table~\ref{identificationtable}. We shall actually only use the bundle
isomorphism \mb{T^{*}E\cong(\cE;\nabla)\cong\Pi T^{*}E}, corresponding to the
upper middle part of the diagram \e{identificationdiagram}. These bundle
identifications will from now on often be used without explicitly mentioning
it, as it will be clear from the context whether they have been applied or not.
The rest of the diagram \e{identificationdiagram} is shown for the sake of
completeness. The Grassmann-odd tilde transformation \e{tildederiv} (which
exchanges the fiber coordinates \mb{p_{i}\leftrightarrow\bb_{i}} and
\mb{\PP_{i}\leftrightarrow\BB_{i}}), is responsible for the apparent reflection
symmetry along a vertical symmetry axis in the diagram
\e{identificationdiagram}. 

\begin{table}[t]
\caption{A list of manifolds in diagram \e{identificationdiagram}.}
\label{identificationtable}
\begin{center}
\begin{tabular}{|c|c|c|}  \hline
Manifold & Local coordinates & Structure \\ \hline\hline
\rule[-1.5ex]{0ex}{4ex} Base manifold $M$&$(x^{i})$&\\ \hline \hline 
\rule[-1.5ex]{0ex}{4ex} Cotangent bundle $T^{*}M$&
$(x^{i},p_{i})$&Sympl.~pot.\ $-p_{i}\ddd x^{i}$ \\ \hline 
\rule[-1.5ex]{0ex}{4ex} Parity-inv.~tang.~bdl.\ $E\equiv \Pi TM$&
$(x^{i},\cc^{i})$&\\ \hline 
\rule[-1.5ex]{0ex}{4ex} Parity-inv.~cot.~bdl.\
$\Pi T^{*}M\cong E^{*} $&$(x^{i},\bb_{i})$
&Anti-sympl.~pot.\ $-\bb_{i}\ddd x^{i}$\\ \hline \hline
Parity-inverted tangent bundle&& \\
$\Pi T(E^{*})$&
$(x^{i},\bb_{i};\cc^{i},-P_{i})$\rule{10ex}{0ex}&\\ \hline
Cotangent bundle&&Symplectic potential\\
$T^{*}(E^{*})$&
$(x^{i},\bb_{i};P_{i},\cc^{i})$\rule{10ex}{0ex}&
$-\PP_{i}\ddd x^{i}+\cc^{i}\ddd\bb_{i}$\\ \hline
Cotangent bundle&&Symplectic potential\\
$T^{*}E$&
$(x^{i},\cc^{i};P_{i},\bb_{i})$\rule{10ex}{0ex}&
$ \vartheta=-\PP_{i}\ddd x^{i}+\bb_{i}\ddd\cc^{i}$\\ \hline
Whitney sum $\cE\equiv T^{*}M\oplus E\oplus E^{*}$&
\rule{5ex}{0ex}$\updownarrow \PP_{i}=p_{i}+\Gamma^{k}_{ij}\cc^{j}\bb_{k}$&\\
equipped with a connection&$(x^{i};p_{i},\cc^{i},\bb_{i})$\rule{10ex}{0ex}&\\
$\nabla:\Gamma(TM) \times \Gamma(TM) \to \Gamma(TM)$
&\rule{5ex}{0ex}$\updownarrow 
\BB_{i}=\bb_{i}+\Gamma^{k}_{ij}\cc^{j}p_{k}$&\\ \hline 
Parity-inverted cotangent bundle&
$(x^{i},\cc^{i};\BB_{i},p_{i})$\rule{10ex}{0ex}&Anti-symplectic potential\\
$\Pi T^{*}E$&&
$\tilde{\vartheta}=-\BB_{i}\ddd x^{i}-p_{i}\ddd\cc^{i}$\\ \hline
Parity-inverted cotangent bundle&
$(x^{i},p_{i};-\BB_{i},\cc^{i})$\rule{10ex}{0ex}&Anti-symplectic potential\\
$\Pi T^{*}(T^{*}M)$&&
$-\BB_{i}\ddd x^{i}+\cc^{i}\ddd p_{i}$\\ \hline
Parity-inverted tangent bundle&
$(x^{i},p_{i};\cc^{i},\BB_{i})$\rule{10ex}{0ex}&\\
$\Pi T(T^{*}M)$&&\\ \hline
\end{tabular}
\end{center}
\end{table}

\noi
We mention in passing that the anti-bracket may be encoded in a commutative
Koszul hierarchy
\beq
(-1)^{\epsilon_{a}}(a,b)~=~\Phi^{2}_{\Delta}(a,b)~=~[[\Delta,L_{a}],L_{b}]1~,
~~~~~~~~~~~~a,b~\in~C^{\infty}(\Pi T^{*}E)~,
\eeq
of an odd, nilpotent, second-order \mb{\Delta} operator
\bea
\Delta&=& -\papal{x^{i}}\papal{\BB_{i}}+ \papal{\cc^{i}}\papal{p_{i}}
\label{courantdelta1} \\
&=&(-\papal{x^{i}}-\Gamma_{ij}^{j}-\Gamma_{ij}^{k}p_{k}\papal{p_{j}}
+\Gamma_{ik}^{j}\cc^{k}\papal{\cc^{j}}
+\Hf R_{ij}\papal{\bb_{j}})\papal{\bb_{i}}
+ \papal{\cc^{i}}\papal{p_{i}}~,
\label{courantdelta2}
\eea
\cf Subsection~\ref{seccommutative}. Both \mb{\Delta} formulas
\es{courantdelta1}{courantdelta2} are invariant under coordinate
transformations in the base manifold \mb{M} without the use of a volume form
\cite{b06}. This is due to a balance of bosonic and fermionic degrees of
freedom. The latter \mb{\Delta} formula \e{courantdelta2}, which uses the 
bundle isomorphism \mb{\Pi T^{*}E\cong\cE}, \cf diagram
\e{identificationdiagram}, is presumably new. Note that the higher Koszul
brackets and the Koszul zero-bracket vanish, \ie \mb{\Phi_{\Delta}^{n}\!=\!0}
for \mb{n\!\geq\!3} and for \mb{n\!=\!0}, so this is an example of a
commutative Batalin-Vilkovisky algebra. 

\noi  
The anti-bracket on vectors and exterior forms mimics the covariant
derivative, the Lie bracket and the interior product, as was the case for the
even Poisson bracket \mb{\{\cdot,\cdot\}}. However in the anti-bracket case
the r\^oles of \mb{X} and \mb{\tilde{X}} are exchanged, 
\bea
(X,Y)&=&0~,~~~~~~~~~~X,Y\in\Gamma(TM)~,\label{antixy}\\
(\tilde{X},Y)&=&X^{i}(\frac{\partial Y^{j}}{\partial x^{i}}p_{j}
+ \Gamma_{ij}^{k}Y^{j}p_{k})~=:~(\nabla_\XX Y)^{k}p_{k}
~=:~\nabla_{X}Y~,\label{antitxy}\\
(\tilde{X},\tilde{Y})
&=&\left(X^{i}\frac{\partial Y^{j}}{\partial x^{i}}\bb_{j}
-(-1)^{\epsilon_{X}\epsilon_{Y}}(X\leftrightarrow Y)\right)
- X^{i}Y^{j}R_{ij} \cr
&=&[X,Y]^{\sim} - R(X,Y)~,\label{antitxty}\\
(X,\eta)&=&X^{i}\frac{\partial^{l} \eta}{\partial \cc^{i}}
~=:~i_\XX\eta~,\label{antixeta}  \\
(\tilde{X},\eta)&=&X^{i}(\papal{x^{i}}
-\Gamma_{ik}^{j}\cc^{k}\papal{\cc^{j}})\eta
~=:~\nabla_\XX\eta~,\label{antitxeta}\\
(\xi,\eta)&=&0~,~~~~~~~~~~
\xi,\eta\in\Gamma(\bigwedge{}^{\bullet}(T^{*}M))~.\label{antixieta}
\eea

\subsection{Derived Brackets}
\label{seccourantderivedbracket}

\noi
We now use the even and odd symplectic structure \mb{\{\cdot,\cdot\}} and
\mb{(\cdot,\cdot)} from Subsections~\ref{seccansympl}-\ref{seccanantisympl} to
define a \hla structure on \mb{C^{\infty}(T^{*}E)} and
\mb{C^{\infty}(\Pi T^{*}E)}, respectively. Using the bundle isomorphisms of
diagram \e{identificationdiagram}, we then obtain two \hla structures on the
same algebra \mb{C^{\infty}(\cE)}. The even and odd symplectic structure
\mb{\{\cdot,\cdot\}} and \mb{(\cdot,\cdot)} will here both play the r\^{o}le of
the Lie algebra structure \mb{[\cdot,\cdot]} of Section~\ref{secderoperad}. It
comes in handy that we have left the parity \mb{\epsilon\in\{0,1\}} of the
\mb{[\cdot,\cdot]} bracket open, so that we readily can model both even and odd
brackets. Similarly, we will have two generators for the derived
\mb{\bullet}-brackets, a Grassmann-odd generator \mb{Q} and a Grassmann-even
generator \mb{S}, which (despite the notation) both will play the r\^{o}le of
the fixed Lie algebra element \mb{Q} of Section~\ref{secderoperad}. The new
\mb{Q} and \mb{S} are defined as \cite{lyakhovich04}
\beq
\begin{array}{rclclcrcl}
Q&:=&\cc^{i} \PP_{i}+H_{\odd}
&=&\cc^{i}p_{i}+\Hf\cc^{i}\cc^{j}T^{k}_{ij}\bb_{k}+H_{\odd}~,
&~~~~~~~&\epsilon_{Q}&=&1~,\cr\cr
S&:=&\cc^{i} \BB_{i}+H_{\even}
&=&\cc^{i}\bb_{i}+\Hf\cc^{i}\cc^{j}T^{k}_{ij}p_{k}+H_{\even}~,
&&\epsilon_{S}&=&0~,
\end{array}
\label{charges}
\eeq
where the exterior forms \mb{H_{\odd}\in\Gamma(\bigwedge^{\odd}(T^{*}M))}
(resp.\ \mb{H_{\even}\in\Gamma(\bigwedge^{\even}(T^{*}M))}) are closed
Grassmann-odd (resp.\ Grassmann-even) forms. The local expressions for
\mb{Q\in C^{\infty}(T^{*}E)} and \mb{S\in C^{\infty}(\Pi T^{*}E)} are invariant
under coordinate transformations in the base manifold \mb{M}, and hence define
global scalars. Here \mb{T^{k}_{ij}} in \eq{charges} is the torsion tensor,
\beq
T^{k}_{ij}~=~\Gamma^{k}_{ij}-\Gamma^{k}_{ji}~.
\label{torsiontensor}
\eeq
The generators \mb{Q} and \mb{S} are nilpotent in the Poisson and anti-bracket
sense,
\bea
\{Q,Q\}&=&2\cc^{i}\frac{\partial H_{\odd}}{\partial x^{i}}
~=~2dH_{\odd}~=~0~,  \label{qcme} \\
(S,S)&=&2\cc^{i}\frac{\partial H_{\even}}{\partial x^{i}}
~=~2dH_{\even}~=~0~, \label{bvcme}
\eea
respectively, because \mb{H_{\odd}} and \mb{H_{\even}} are closed. (The
nilpotency \es{qcme}{bvcme} are also due to the first Bianchi identity,
\beq
\sum_{i,j,k~{\rm cycl.}}R^{\ell}{}_{k ij}~=~
\sum_{i,j,k~{\rm cycl.}}(\frac{\partial T^{\ell}_{ij}}{\partial x^{k}}
+ \Gamma_{km}^{\ell} T^{m}_{ij})~,
\label{1bianchiid}
\eeq
if one starts from the latter expressions in \eq{charges} that depends on
\mb{p_{i}} and \mb{\bb_{i}} explicitly.) We stress that the nilpotent
generators \mb{Q} and \mb{S}, and similarly, the even and odd Poisson
structures \mb{\{\cdot,\cdot\}} and \mb{(\cdot,\cdot)}, are on completely equal
footing, regardless of what the notation might suggest. (The notation is
inspired by the physics literature on constrained dynamics; for instance the
\eq{bvcme} resembles the Classical Master Equation of Batalin and Vilkovisky.)
The generators \mb{Q} and \mb{S} turn \mb{C^{\infty}(T^{*}E)} and 
\mb{C^{\infty}(\Pi T^{*}E)} into \hlas with derived \mb{\bullet}-brackets
\mb{\Phi_{Q}} and \mb{\Phi_{S}}, respectively. The zero-bracket
\mb{\Phi_{Q}^{0}=Q} is the generator \mb{Q} itself. The one-bracket
\mb{\Phi_{Q}^{1}=\{Q,\cdot\}} is a nilpotent, Grassmann-odd, Hamiltonian vector
field on \mb{T^{*}E}, \cf \eq{dernilp1}. Similarly for the \mb{\Phi_S} bracket
hierarchy. {}For exterior forms \mb{\eta\in C^{\infty}(E)} (which can be viewed
as functions on any of the three bundles \mb{\cE}, \mb{T^{*}E} and
\mb{\Pi T^{*}E} mentioned in Table~\ref{identificationtable}), the one-brackets
are just the de Rham exterior derivative \mb{d},
\beq 
\left. \begin{array}{rcl}
\Phi_{Q}^{1}(\eta)&=&\{Q,\eta\} \cr \cr
\Phi_{S}^{1}(\eta)&=&(S,\eta)\end{array}\right\}
~=~\cc^{i}\frac{\partial \eta}{\partial x^{i}}~=~d\eta~,
\label{derhamabama}
\eeq
independently of \mb{H_{\odd}}, \mb{H_{\even}} and \mb{\nabla}.
On vectors \mb{X\in\Gamma(TM)}, the one-bracket \mb{\Phi_{Q}^{1}} reads
\bea
\Phi_{Q}^{1}(X)~=~\{Q,X\}
&=&dX + \cc^{i}(\Gamma^{t})^{k}_{ij}X^{j}\PP_{k}   
-c^{i}c^{j} \partial_{i}(\Gamma^{t})^{k}_{j\ell}~\bb_{k}X^{\ell}
-(-1)^{\epsilon_{X}}\nabla_\XX H_{\odd} \cr
&=&\nabla^{t} X-(R^{t})^{k}{}_{\ell}~\bb_{k}X^{\ell}
-(-1)^{\epsilon_{X}}\nabla_\XX H_{\odd}~, \\
\Phi_{S}^{1}(\tilde{X})~=~(S,\tilde{X})
&=&dX^{\sim} + \cc^{i}(\Gamma^{t})^{k}_{ij}X^{j}B_{k}   
-(-1)^{\epsilon_{X}}c^{i}c^{j}\partial_{i}(\Gamma^{t})^{k}_{j\ell}~
p_{k}X^{\ell}-(-1)^{\epsilon_{X}}\nabla_\XX H_{\even} \cr
&=&\nabla^{t}X^{\sim}-(-1)^{\epsilon_{X}}(R^{t})^{k}{}_{\ell}~p_{k}X^{\ell}
-(-1)^{\epsilon_{X}}\nabla_\XX H_{\even}~, \\
\Phi_{Q}^{1}(\tilde{X})~=~\{Q,\tilde{X}\}
&=&(-1)^{\epsilon_{X}}X^{i}\PP_{i}+dX^{\sim}
+(-1)^{\epsilon_{X}}i_{X}H_{\odd}~\in~C^{\infty}(T^{*}E) \cr
&=&(-1)^{\epsilon_{X}}X+\nabla^{t}X^{\sim}
+(-1)^{\epsilon_{X}}i_{X}H_{\odd}~, \\
\Phi_{S}^{1}(X)~=~(S,X)
&=&\BB_{i}X^{i}+dX+(-1)^{\epsilon_{X}}i_\XX H_{\even}
~\in~C^{\infty}(\Pi T^{*}E) \cr
&=&(-1)^{\epsilon_{X}}\tilde{X}+\nabla^{t}X
+(-1)^{\epsilon_{X}}i_\XX H_{\even}~,
\eea
where \mb{\nabla\!=\!\cc^{i}\nabla_{i}} is the connection one-form, 
\mb{R^{k}{}_{\ell}\!=\!\Hf \cc^{i}\cc^{j}R^{k}{}_{\ell ij}} is the curvature
two-form, and \mb{\nabla^{t}} denotes the transposed connection, which is
defined as \mb{(\Gamma^{t})^{k}_{ij}\!:=\!\Gamma^{k}_{ji}}. The geometric
importance of these one-brackets is underscored by the fact that their adjoint
action (in the Poisson or anti-bracket sense) reproduce the Lie derivative on
exterior forms
\beq 
\left. \begin{array}{rcccl}
\{\Phi_{Q}^{1}(\tilde{X}),\eta\}
&=&\{Q,\{\tilde{X},\eta\}\}
&+&(-1)^{\epsilon_{X}}\{\tilde{X},\{Q,\eta\}\}\cr \cr
(\Phi_{S}^{1}(X),\eta)&=&(S,(X,\eta))&+&(-1)^{\epsilon_{X}}(X,(S,\eta))
\end{array} \right\}
~=~[d,i_\XX]\eta~=~(-1)^{\epsilon_{X}}{\cal L}_\XX\eta~,
\eeq
independently of \mb{H_{\odd}}, \mb{H_{\even}} and \mb{\nabla}. The
two-brackets \mb{\Phi_{Q}^{2}} and \mb{\Phi_{S}^{2}} give rise to an odd and
an even Courant bracket, \mb{(\cdot,\cdot)_\QQ} and \mb{\{\cdot,\cdot\}_\SS},
respectively, \cf \eq{dertwobracket}. 

\begin{definition} The odd and even {\bf Courant brackets}
\mb{(\cdot,\cdot)_\QQ} and \mb{\{\cdot,\cdot\}_\SS} are defined as the derived
brackets,
\bea
(a,b)_\QQ&:=&(-1)^{\epsilon_{a}+1}\Phi_{Q}^{2}(a,b)
~=~\Hf\{\{a,Q\},b\}+\Hf\{a,\{Q,b\}\} \cr
&=&-(-1)^{(\epsilon_{a}+1)(\epsilon_{b}+1)}(b,a)_\QQ
~,~~~~~~~~~~~~~~a,b~\in~C^{\infty}(T^{*}E)~,
\label{courantantibracket} \\
\{a,b\}_\SS&:=&(-1)^{\epsilon_{a}}\Phi_{S}^{2}(a,b)
~=~\Hf((a,S),b)+\Hf(a,(S,b)) \cr
&=&-(-1)^{\epsilon_{a}\epsilon_{b}}\{b,a\}_\SS
~,~~~~~~~~~~~~~~a,b~\in~C^{\infty}(\Pi T^{*}E)~,
\label{courantbracket}
\eea
respectively.
\label{defcourantbracketqs}
\end{definition}

\noi
Notice the complete democracy among the even and odd brackets in the
Definition~\ref{defcourantbracketqs}. By restricting to vector fields and
exterior forms one finds the celebrated formulas \e{couxy}-\e{couxieta} for the
Courant bracket, 
\bea
\{X,Y\}_\SS&=&[X,Y]+(-1)^{\epsilon_{Y}}i_\XX i_\YY H_{\even}
~,~~~~~~~~~~X,Y\in\Gamma(TM)~,\label{poissoncourantxy} \\
(\tilde{X},\tilde{Y})_{Q}&=&[X,Y]^{\sim}+(-1)^{\epsilon_{Y}}
i_\XX i_\YY H_{\odd}~,\label{anticouranttxty} \\
\{X,\eta\}_\SS~=~
(\tilde{X},\eta)_\QQ &=&\Hf ({\cal L}_\XX + i_\XX d)\eta
\label{courantxeta}~,  \\
\{\xi,\eta\}_\SS&=&0~, \label{poissoncourantxieta} \\
(\xi,\eta)_\QQ&=&0~,~~~~~~~~~~
\xi,\eta\in\Gamma(\bigwedge{}^{\bullet}(T^{*}M))~,
\label{anticourantxieta}
\eea
which do not use the bundle isomorphisms of diagram
\e{identificationdiagram}, and hence are independent of \mb{\nabla}; plus one
finds the following formulas
\bea
(X,Y)_\QQ&=&\Hf\nabla_\XX [\nabla^{t}Y]-\Hf\nabla^{t}_{i}X(\nabla^{t}Y)^{i}
-\frac{(-1)^{\epsilon_{Y}}}{2}  \nabla_\XX \nabla_\YY H_{\odd} \cr
&&+{\cal O}(c^{2}b)
-(-1)^{(\epsilon_{X}+1)(\epsilon_{Y}+1)}(X\leftrightarrow Y)~,
\label{anticourantxy} \\
\{\tilde{X},\tilde{Y}\}_\SS&=&\Hf\nabla_\XX [\nabla^{t}\tilde{Y}]
+\frac{(-1)^{\epsilon_{Y}}}{2}\left(\nabla^{t}_{i}\tilde{X}(\nabla^{t}Y)^{i}
-\nabla_\XX\nabla_\YY H_{\even}\right) \cr
&&+{\cal O}(c^{2}p)
-(-1)^{(\epsilon_{X}+1)(\epsilon_{Y}+1)}(X\leftrightarrow Y)~,
\label{poissoncouranttxty} \\
(X,\eta)_\QQ~=~
 \{\tilde{X},\eta\}_\SS &=&\Hf ([\nabla_\XX ,d]+ \nabla_\XX d)\eta
~=~\nabla_\XX d\eta-\frac{(-1)^{\epsilon_{X}}}{2} d\nabla_\XX \eta~,
\label{courantxeta2} 
\eea
which do use the bundle isomorphisms of diagram \e{identificationdiagram},
and whose right-hand sides are second-order differential operators and not 
particularly illuminating. The lesson to be learned, is, that one should stick
to vector fields without parity-inversion 
\mb{X\in \Gamma(TM)\subseteq C^{\infty}(\Pi T^{*}E)} for the \mb{\Phi_{S}}
brackets on the algebra \mb{C^{\infty}(\Pi T^{*}E)}, and to vector fields with
parity-inversion \mb{\tilde{X}\in\Gamma(E)\subseteq C^{\infty}(T^{*}E)} for the
\mb{\Phi_{Q}} brackets on the algebra \mb{C^{\infty}(T^{*}E)}.

\subsection{Discussion}
\label{secoutlook}

\noi
Obviously, we have only scratched the surface. One can for instance also 
calculate what the Courant brackets should be on skewsymmetric multi-vectors.
{}For instance, the one-bracket \mb{\Phi_{Q}^{1}} on a multi-vector 
\mb{\pi\in\Gamma(\bigwedge{}^{\bullet}(TM))
\cong C^{\infty}(E^{*})\subseteq C^{\infty}(T^{*}E)} reads
\bea
\Phi_{Q}^{1}(\pi)&=&d\pi+(\PP_{i}+H_{\odd}\papar{\cc^{i}})(\papal{\bb_{i}}\pi) 
~\in~C^{\infty}(T^{*}E) \cr
&=&\nabla^{t}\pi+(p_{i}+H_{\odd}\papar{\cc^{i}})(\papal{\bb_{i}}\pi)
~,~~~~~~~~~~\pi\in\Gamma(\bigwedge{}^{\bullet}(TM))~.
\eea
The Courant two-bracket \mb{(\cdot,\cdot)_\QQ} becomes a twisted version of the
Schouten-Nijenhuis bracket \e{schoutennijenhuis0},
\beq
(\pi,\rho)_\QQ~=~(\pi,\rho)_{SN}^{}+(\pi\papar{\bb_{i}})
(\papal{\cc^{i}}H_{\odd}\papar{\cc^{j}})
(\papal{\bb_{j}}\rho)~\in~C^{\infty}(T^{*}E)~,~~~~~~~~~~
\pi,\rho\in\Gamma(\bigwedge{}^{\bullet}(TM))~,
\eeq
which does not depend on the \mb{\PP_{i}} variables. However, the Courant 
bracket \mb{(\pi,\eta)_{Q}} between a higher-order skewsymmetric multi-vector, 
\mb{\pi}, and a higher-order form, \mb{\eta}, does not close on such objects, 
but will in general also depend on the \mb{\PP_{i}} variables. A complete
treatment of the Courant bracket would therefore include an investigation of
form-valued multi-vectors, where the word ``multi-vector'' here should be
understood in a generalized sense that depends on both the bosonic and
fermionic generators, \mb{p_{i}\equiv\partial_{i}\in\Gamma(TM)} and
\mb{\bb_{i}\equiv\tilde{\partial}_{i}\in\Gamma(\Pi TM)}, or the analogues
obtained via the bundle identifications of diagram \e{identificationdiagram}. 
It is then natural, in turn, to allow the twisting \mb{H_{\odd}} (resp.\ 
\mb{H_{\even}}) to be an odd (resp.\ even) form-valued multi-vector as well.

\noi
The even or odd Poisson construction may be generalized further to an arbitrary
Poisson or anti-Poisson manifold with a nilpotent function \mb{Q} or \mb{S}
satisfying \mb{\{Q,Q\}=0} or \mb{(S,S)=0}, respectively. 

\noi
{}Finally, let us mention that the odd bracket \mb{(\cdot,\cdot)_\QQ} may be
viewed as a classical counterpart of the odd operator bracket
\mb{[\cdot,\cdot]_\DD} in the spirit of deformation quantization. In detail, 
one defines a hat quantization map ``\mb{\wedge}'' that takes functions
\mb{a} (also known as symbols) to normal-ordered differential operators
\mb{\hat{a}},
\bea
C^{\infty}(T^{*}E)&\ni&a=a(x^{i},\cc^{i},\PP_{i},\bb_{i})~~~~~~~~~
\stackrel{\wedge}{\longmapsto} \cr\cr
\hat{a}=~:a(x^{i},\cc^{i},\papal{x^{i}},\papal{\cc^{i}}):&=&
\left. a~\exp(\papar{\PP_{i}}\papal{x^{i}}
+\papar{\bb_{i}}\papal{\cc^{i}})\right|_{\PP_{i},b_{i}=0}
~\in~{\rm End}(C^{\infty}(E))~.
\label{hatdef}
\eea
The colons in \eq{hatdef} indicate normal-ordering, which means that the
derivatives are ordered to the rightmost position. Examples of the
``\mb{\wedge}'' quantization map are
\beq
\hat{\eta}~=~\eta~,~~~~~~\hat{\tilde{X}}~=~i_{X}~,~~~~~~\hat{Q}~=~D~,~~~~~~
\hat{p}_{i}~=~\papal{x^{i}}-\Gamma^{k}_{ij}\cc^{j}\papal{\cc^{i}}
~=~\nabla_{i}~,
\eeq
\cf eqs.\ \e{etafcts}, \e{ixcontraction}, \es{charges}{dcharge}.
The algebra isomorphism \mb{\hat{a}\circ\hat{b}=\widehat{a\star b}} between
corresponding associative algebras is given by the \mb{\star} product,
\beq
  a\star b~=~(a~\exp(\papar{\PP_{i}}\papal{x^{i}}
+\papar{\bb_{i}}\papal{\cc^{i}})~b)~\in~C^{\infty}(T^{*}E)
~,~~~~~~~~~~~~a,b~\in~C^{\infty}(T^{*}E)~.
\label{starproduct}
\eeq
The outer parenthesis on the \rhs of \eq{starproduct} indicates that the
expression should be interpreted as a function, \ie a zeroth-order differential
operator. The ``\mb{\wedge}'' quantization map \e{hatdef} is clearly related
to the symplectic structure \e{symplecticform}. In fact, the \mb{\star} 
commutator
\mb{[a,b]_{\star}\equiv a\star b-(-1)^{\epsilon_{a}\epsilon_{b}}b\star a}
is a quantum deformation of the Poisson bracket \mb{\{a,b\}}, where the Planck
constant here is set equal to the value \mb{-2\pi i}, that is, \mb{i\hbar=1}.
In the same way we get the new result, that the Courant operator bracket
\beq
 [\hat{a},\hat{b}]_\DD=\Hf\left([[a,Q]_{\star},b]_{\star}
+[a,[Q,b]_{\star}]_{\star}\right)^{\wedge}
~,~~~~~~~~~~~~a,b~\in~C^{\infty}(T^{*}E)~,
\eeq
is a quantum deformation of the classical Courant bracket
\mb{(a,b)_\QQ=\Hf\{\{a,Q\},b\}+\Hf\{a,\{Q,b\}\}}.

\noi
Recall that the bracket \mb{(\cdot,\cdot)_\QQ} and the operator counterpart
\mb{[\cdot,\cdot]_\DD} are both odd brackets. They both realize the Courant
bracket \e{couxy}-\e{couxieta} \wtho parity-inverted vectors, either directly
via \mb{\tilde{X}\in C^{\infty}(T^{*}E)}, or via the contraction
\mb{i_\XX\in {\rm End}( C^{\infty}(E))}. They are also both capable of
reproducing the same twisting with an odd closed form \mb{H_{\odd}}. On the
other hand, the possibility of twisting the Courant bracket with an even closed
form \mb{H_{\even}} seems to have gone unnoticed so far in the literature. The
even bracket \mb{\{\cdot,\cdot\}_\SS} realizes just this possibility.

\subsection{Higher Brackets}
\label{sechighercourantbrackets}

\noi
It is a major point that the derived bracket construction \e{derbracket} of the
Courant bracket automatically provides us with an infinite tower of higher
Courant brackets and a host of nilpotency relations \e{derivedsquarealpha}
corresponding to all the allowed values of the \mb{c^{n}_{k}} product
coefficients found in Section~\ref{secderoperad}, like for instance the Ward
identity \e{wardid} and the generalized Jacobi identity \e{derjacid}. In this
Subsection we calculate the higher Courant brackets in pertinent sectors. The
connection \mb{\nabla} and the bundle isomorphisms of diagram
\e{identificationdiagram} are not used in the rest of the paper.

\begin{proposition} The higher Courant brackets among vectors read
\bea
\Phi_{S}^{n}(X_{(1)},\ldots,X_{(n)})&=&
i_{X_{(1)}}^{}\ldots i_{X_{(n)}}^{}H_{\even}
~,~~~~~~~~X_{(1)},\ldots,X_{(n)}\in\Gamma(TM)~, \cr \cr
\left. \begin{array}{r}
\Phi_{Q}^{n}(\tilde{X}_{(1)},\ldots,\tilde{X}_{(n)})\cr
\Phi_{D}^{n}(i_{X_{(1)}}^{},\ldots,i_{X_{(n)}}^{})
\end{array}\right\}
&=&i_{X_{(1)}}^{}\ldots i_{X_{(n)}}^{}H_{\odd}
~,~~~~~~~~\epsilon(X_{(1)})=0,~~\ldots,~~\epsilon(X_{(n)})=0~,
\eea
for \mb{n\in\{3,4,5,\ldots\}}. The Courant three-bracket between two vectors 
\mb{X,Y\in\Gamma(TM)} with Grassmann-parity \mb{\epsilon_{X}=0=\epsilon_{Y}} 
and one exterior form \mb{\eta\in\Gamma(\bigwedge^{\bullet}(T^{*}M))}, is
\bea
\Phi_{S}^{3}(X,Y,\eta)&=&
\Phi_{Q}^{3}(\tilde{X},\tilde{Y},\eta)~=~\Phi_{D}^{3}(i_\XX,i_\YY,\eta) \cr
&=&\frac{1}{3}\left(i_{[X,Y]}^{}+\Hf (i_\XX{\cal L}_\YY
- i_\YY{\cal L}_\XX)+i_\XX i_\YY d \right)\eta \cr
&=&\left(\Hf (i_\XX {\cal L}_\YY- i_\YY {\cal L}_\XX)
+\frac{1}{3}di_\XX i_\YY \right)\eta\cr
&=&\Hf\left(i_\XX i_\YY d-\frac{1}{3}di_\XX i_\YY+i_{[X,Y]}^{}\right)\eta~.
\eea
The higher Courant brackets of vectors and one exterior form are given
recursively as
\bea
\Phi_{S}^{n+1}(X_{(1)},\ldots,X_{(n)},\eta)&=&
\Phi_{Q}^{n+1}(\tilde{X}_{(1)},\ldots,\tilde{X}_{(n)},\eta)
~=~\Phi_{D}^{n+1}(i_{X_{(1)}}^{},\ldots,i_{X_{(n)}}^{},\eta)\cr
&=&\frac{1}{n\!+\!1}\sum_{1\leq i\leq n}(-1)^{i-1}i_{X_{(i)}}^{}
\Phi_{Q}^{n}(\tilde{X}_{(1)},\ldots,\widehat{\tilde{X}}_{(i)},
\ldots,\tilde{X}_{(n)},\eta)
\label{highercourantrecur}
\eea
for \mb{n\in\{3,4,5,\ldots\}} and \mb{\epsilon(X_{(1)})=0}, \mb{\ldots},
\mb{\epsilon(X_{(n)})=0} bosonic, or directly as
\bea
\Phi_{Q}^{n+1}(\tilde{X}_{(1)},\ldots,\tilde{X}_{(n)},\eta) 
&=&\frac{6}{(n\!+\!1)n(n\!-\!1)}\sum_{1\leq i<j \leq n}(-1)^{(n-1-i)+(n-j)}\cr
&&\times i_{X_{(1)}}^{}\ldots \widehat{i}_{X_{(i)}}^{}\ldots 
\widehat{i}_{X_{(j)}}^{}\ldots i_{X_{(n)}}^{}
\Phi_{Q}^{3}\left(\tilde{X}_{(i)},\tilde{X}_{(j)},\eta\right)
\label{highercourant}
\eea
for \mb{n\in\{2,3,4,\ldots\}}. Brackets between vectors and exterior forms with
more than one exterior form as argument vanish.
\label{propositionhighercourant}
\end{proposition}

\noi
We note that the \mb{(n\!+\!1)}-brackets in 
\eqs{highercourantrecur}{highercourant} vanish if the form degree of \mb{\eta}
is \mb{\leq n\!-\!2}.

\noi
{\it Proof of Proposition~\ref{propositionhighercourant}:}~~The calculations
are most efficiently done along the diagonal \mb{X^{\soprod n}} with the 
vector \mb{X} taken to be a fermion, \mb{\epsilon_{X}=1}. Hence the
parity-inverted vector \mb{\tilde{X}} and \mb{i_\XX} are bosons. One finds
\bea
\Phi_{S}^{2}(X\soprod X) &=& -[X,X]+i_{X}^{2}H_{\even}~, \\
\Phi_{Q}^{2}(\tilde{X}\soprod\tilde{X})
&=& -[X,X]^{\sim}+i_{X}^{2}H_{\odd}~, \\
\Phi_{D}^{2}(i_\XX\soprod i_\XX)
&=& -i_{[X,X]}^{}+i_{X}^{2}H_{\odd}~, \\
\Phi_{S}^{n}(X^{\soprod n}) 
&=&(-i_{X})^{n}H_{\even}~,\\
\left. \begin{array}{r}
\Phi_{Q}^{n}(\tilde{X}^{\soprod n}) \cr
\Phi_{D}^{n}(i_\XX{}^{\soprod n}) \end{array}\right\}
&=&(-i_\XX)^{n}H_{\odd}~,~~~~~~~~~~~\epsilon_{X}~=~1~,
\eea
for \mb{n\in\{3,4,5,\ldots\}}.
These facts imply that the expansion \eq{derivedhelp} truncates 
after only three terms,
\bea
 \Phi_{Q}^{n}(\tilde{X}^{\soprod(n-1)} \soprod \eta)
&=&\frac{1}{n}\sum_{k=0}^{n-1}(-\ad\tilde{X})^{n-1-k}
\left\{ \Phi_{Q}^{k}(\tilde{X}^{\soprod k}), \eta \right\} \cr
&=&\frac{1}{n}\sum_{k=0}^{2}(-i_\XX)^{n-1-k}
\left\{ \Phi_{Q}^{k}(\tilde{X}^{\soprod k}), \eta \right\} \cr
&=&\frac{3}{n}(-i_\XX)^{n-3}
\Phi_{Q}^{3}(\tilde{X}\soprod\tilde{X}\soprod\eta)~,
\eea
with \mb{\epsilon_{X}=1} and \mb{n\in\{3,4,5,\ldots\}}. Similarly for the
\mb{S} and \mb{D} hierarchies. In the \mb{n\!=\!3} case this reads
\bea 
\Phi_{S}^{3}(X\soprod X\soprod\eta)&=&
\Phi_{Q}^{3}(\tilde{X}\soprod\tilde{X}\soprod\eta)
~=~\Phi_{D}^{3}(i_\XX\soprod i_\XX\soprod\eta)
~=~\frac{1}{3}\left(i_{X}^{2}d+i_\XX{\cal L}_\XX-i_{[X,X]}^{}\right)\eta \cr
&=&\left(i_\XX{\cal L}_\XX+\frac{1}{3}di_{X}^{2} \right)\eta
~=~\Hf\left(i_{X}^{2}d-\frac{1}{3}di_{X}^{2}-i_{[X,X]}^{}\right)\eta
~,~~~~~~~~~\epsilon_{X}~=~1~.
\eea
Now apply polarization \mb{X=\sum_{i=1}^{n}\lambda^{(i)} X_{(i)}} with 
\mb{\epsilon(\lambda^{(i)})=1} and \mb{\epsilon(X_{(i)})=0}, 
\cf Subsection~\ref{secpolarization}. The vanishing of higher brackets with
two or more exterior forms follows from \eqs{poissonxieta}{antixieta}.
\proofbox

\subsection{\mb{B}-transforms}

\noi
In this Subsection the \mb{B}-transforms \cite{gualtieri04,hitchin03} is 
generalized to the higher brackets. The \mb{B}-transforms in
this context are canonical or anti-canonical transformations generated by even
or odd exterior forms \mb{B_{\even}} or \mb{B_{\odd}}, respectively. It
follows immediately from the derived bracket definition \e{derbracket} that
\bea
e^{-[B_{\even},\cdot]}\Phi_{D}^{n}\left(a_{1},\ldots,a_{n}\right)
&=&\Phi_{D^{\prime}}^{n}\left(e^{-[B_{\even},\cdot]}a_{1},\ldots,
e^{-[B_{\even},\cdot]}a_{n}\right)~, \cr
e^{-\{B_{\even},\cdot\}}\Phi_{Q}^{n}\left(a_{1},\ldots,a_{n}\right)
&=&\Phi_{Q^{\prime}}^{n}\left(e^{-\{B_{\even},\cdot\}}a_{1},\ldots,
e^{-\{B_{\even},\cdot\}}a_{n}\right)~, \cr
e^{-(B_{\odd},\cdot)}\Phi_{S}^{n}\left(a_{1},\ldots,a_{n}\right)
&=&\Phi_{S^{\prime}}^{n}\left(e^{-(B_{\odd},\cdot)}a_{1},\ldots,
e^{-(B_{\odd},\cdot)}a_{n}\right)~,
\eea 
where we have defined
\beq
D^{\prime}~:=~e^{-[B_{\even},\cdot]}D~,~~~~~~
Q^{\prime}~:=~e^{-\{B_{\even},\cdot\}}Q~,~~~~~~
S^{\prime}~:=~e^{-(B_{\odd},\cdot)}S~.
\eeq
The \mb{B}-transforms are a symmetry of the derived brackets, \ie 
\beq
D^{\prime}~=~D~,~~~~~~~~~~~~Q^{\prime}~=~Q~,~~~~~~~~~~~~S^{\prime}~=~S~,  
\eeq
if the \mb{B}-forms 
are closed, 
\beq
[D,B_{\even}]~=~dB_{\even}~=~0~,~~~~~~ 
\{Q,B_{\even}\}~=~dB_{\even}~=~0~,~~~~~~ 
(S,B_{\odd})~=~dB_{\odd}~=~0~.  
\eeq
In this way it becomes obvious, that closed \mb{B}-transforms are algebra
automorphisms for the full Courant \mb{\bullet}-bracket hierarchy
\cite{gualtieri04,hitchin03}.

\section{Supplementary Formalism}
\label{secsuppl}

\noi
Section~\ref{secsuppl} is an open-ended investigation, where we make contact to
some notions, that could be useful for future studies. More specifically, we go
back to the original setup of Section~\ref{secdefshliealg} and touch on some of
the theoretical aspects of a \hla \mb{{\cal A}}, such as properties of the
bracket product ``\mb{\circ}'' and  co-algebraic structures on
\mb{\Sym^{\bullet}_{\epsilon}{\cal A}}.

\subsection{Pre-Lie Products}
\label{secprelie}

\noi
What properties should one demand of the ``\mb{\circ}'' product? Associativity
is too strong: This is not even fulfilled for the ordinary product with
coefficients \mb{c^{n}_{k}=1}. The next idea is to let the product
``\mb{\circ}'' be pre-Lie. To measure the non-associativity one usually defines
the associator
\beq 
{\rm ass}(\Phi,\Phi^{\prime},\Phi^{\prime\prime})
~:=~(\Phi \circ \Phi^{\prime})\circ\Phi^{\prime\prime}
-\Phi \circ(\Phi^{\prime}\circ\Phi^{\prime\prime})~.
\label{associator}
\eeq

\begin{definition}
The bracket product ``\mb{\circ}'' is {\bf pre-Lie} if for all 
\mb{\bullet}-brackets
\mb{\Phi,\Phi^{\prime},\Phi^{\prime\prime}:~
\Sym^{\bullet}_{\epsilon}{\cal A} \to {\cal A}}
the associator is symmetric in the last two entries,
\beq
{\rm ass}(\Phi,\Phi^{\prime},\Phi^{\prime\prime})
~=~(-1)^{\epsilon_{\Phi^{\prime}}\epsilon_{\Phi^{\prime\prime}}}
{\rm ass}(\Phi,\Phi^{\prime\prime},\Phi^{\prime})~.
\label{defprelie}
\eeq
\end{definition}

\noi
{}For a pre-Lie product ``\mb{\circ}'' the commutator
\beq
  [\Phi,\Phi^{\prime}]~:=~\Phi\circ\Phi^{\prime}
-(-1)^{\epsilon_{\Phi}\epsilon_{\Phi^{\prime}}}\Phi^{\prime}\circ\Phi
\label{commutatorofoperads}
\eeq
becomes a Lie bracket that satisfies the Jacobi identity, hence the name
``pre-Lie''. One may simplify the pre-Lie condition \e{defprelie} by
polarization into
\beq
{\rm ass}(\Phi,\Phi^{\prime},\Phi^{\prime})~=~0
~,~~~~~~~~~~~~~\epsilon_{\Phi^{\prime}}=1~,
\label{defpreliepol}
\eeq
\cf Subsection~\ref{secpolarization}.
We now give a necessary and sufficient condition in terms of the
\mb{c^{n}_{k}} coefficients for the  ``\mb{\circ}'' product to be pre-Lie.

\begin{proposition}
A  ``\mb{\circ}'' product is pre-Lie if and only if the \mb{c^{n}_{k}} 
product coefficients satisfy the following two conditions
\beq
\forall k,\ell,m\geq 0:~~\left\{
\begin{array}{rcl}
c^{k+\ell+m}_{k}c^{\ell+m+1}_{\ell+1}
&=& c^{k+\ell+m}_{k+\ell}c^{k+\ell}_{k}~, \cr \cr
c^{k+\ell+m}_{k}c^{\ell+m+1}_{\ell}
&=&(k\leftrightarrow \ell)~. \end{array} 
\right.
\label{cpreliecond}
\eeq
\label{propositionprelieconditions}
\end{proposition}

\noi
{\it Proof of Proposition~\ref{propositionprelieconditions}:}~~
To see \eq{cpreliecond}, first note that for two vectors \mb{a,b \in{\cal A}}
with \mb{\epsilon(a)=\epsilon},
\bea
(\Phi\circ\Phi^{\prime})^{n+1}(b\soprod a^{\soprod n})
&=&\sum_{k=0}^{n}\frac{k\!+\!1}{n\!+\!1}b^{n+1}_{k+1} 
\Phi^{n-k+1}\left(\Phi^{\prime k+1}(b\soprod a^{\soprod k})
\soprod a^{\soprod (n-k)}\right)  \cr
&&+\sum_{k=0}^{n}\frac{n\!-\!k\!+\!1}{n\!+\!1} b^{n+1}_{k}
\Phi^{n-k+2}\left(\Phi^{\prime k}(a^{\soprod k})
\soprod b\soprod a^{\soprod (n-k)} \right)
\label{preliehelp}
\eea
for \mb{n\in\{0,1,2,\ldots\}}. Therefore
\bea
\left((\Phi\circ\Phi^{\prime})\circ
\Phi^{\prime\prime}\right)^{n}(a^{\soprod n}) 
&=& 
\!\!\!\!\!\!\!\!
\sum_{\footnotesize \begin{array}{c}k,\ell,m\geq 0 \cr 
k\!+\!\ell\!+\!m=n\end{array}}
b^{n}_{k} \left[ b^{\ell+m+1}_{\ell+1}
\frac{\ell\!+\!1}{\ell\!+\!m\!+\!1}
\Phi^{m+1}\left(\Phi^{\prime \ell +1}\left(
\Phi^{\prime\prime k}(a^{\soprod k})\soprod
a^{\soprod \ell}\right)\soprod a^{\soprod m}\right) \right. \cr
&&\left. ~~~~~~~~~+b^{\ell+m+1}_{\ell}
\frac{m\!+\!1}{\ell\!+\!m\!+\!1}
\Phi^{m+2}\left(\Phi^{\prime\ell}(a^{\soprod \ell})
\soprod\Phi^{\prime\prime k}(a^{\soprod k})
\soprod a^{\soprod m}\right)\right]~.
\label{asso1}
\eea
On the other hand,
\beq
\left(\Phi\circ(\Phi^{\prime}\circ
\Phi^{\prime\prime})\right)^{n}(a^{\soprod n})
~=~\!\!\!\!\!\!\!\!
\sum_{\footnotesize\begin{array}{c}k,\ell,m\geq 0 \cr 
k\!+\!\ell\!+\!m=n\end{array}}
b^{n}_{k+\ell}b^{k+\ell}_{k}
\Phi^{m+1}\left(\Phi^{\prime \ell +1}\left(
\Phi^{\prime\prime k}(a^{\soprod k})\soprod
a^{\soprod \ell}\right)\soprod a^{\soprod m}\right)~.
\label{asso2}
\eeq
Next insert the two expressions \es{asso1}{asso2} into the pre-Lie condition
\e{defpreliepol} with \mb{\Phi^{\prime}\!=\!\Phi^{\prime\prime}} odd. 
By comparing coefficients one derives
\beq
\forall k,\ell,m\geq 0:~~\left\{
\begin{array}{rcl}
b^{k+\ell+m}_{k} b^{\ell+m+1}_{\ell+1}\frac{\ell\!+\!1}{\ell\!+\!m\!+\!1}
&=& b^{k+\ell+m}_{k+\ell}b^{k+\ell}_{k}~, \cr \cr
b^{k+\ell+m}_{k}b^{\ell+m+1}_{\ell}
\frac{m\!+\!1}{\ell\!+\!m\!+\!1}
&=&(k\leftrightarrow \ell)~. \end{array} 
\right.
\label{bpreliecond}
\eeq
The two conditions \e{cpreliecond} follow by translating \e{bpreliecond}
into the ``\mb{c}'' picture \wtho \eq{bcdef}.
\proofbox

\noi
We would like to find the possible \mb{c^{n}_{k}} coefficients that solves
the two necessary and sufficient pre-Lie conditions \e{cpreliecond}. 
The full problem turns out to be quite involved.
{}For simplicity, we shall work within the following generic case
\beq
   c^{0}_{0}~\neq~ 0~\wedge~ c^{1}_{0}~\neq~ 0~.
\label{technicalansatz}
\eeq

\begin{theorem}
The ``\mb{\circ}'' product is pre-Lie and satisfies the condition
\e{technicalansatz}, if and only if there exist non-zero complex numbers 
\mb{\lambda_{n}}, \mb{n\in\{0,1,2,\ldots\}}, such that
\beq
c^{n}_{k}~=~\frac{\lambda_{k}\lambda_{n-k+1}}{\lambda_{n}}~.
\label{preliesol}
\eeq
\label{theoremprelie}
\end{theorem}

\noi
In other words, a generic pre-Lie product is essentially just an ordinary
product \mb{c^{n}_{k}=1} with rescaled brackets 
\mb{\Phi^{n}\to \lambda_{n}\Phi^{n}}. Note that the solution \e{preliesol}
is non-degenerate, \cf \eq{nondegenerateproduct}, and implies that 
\mb{c^{n}_{1}=c^{n}_{n}} does not depend on \mb{n\in\{0,1,2,\ldots\}}. 
It turns out that in the special case \mb{c^{0}_{0}=0~\vee~ c^{1}_{0}= 0},
there exists infinitely many disconnected solutions to be classified elsewhere.

\noi
{\it Proof of Theorem~\ref{theoremprelie}:}~~It is simple to check that the
solution \e{preliesol} satisfies the two pre-Lie conditions \e{cpreliecond}
and the technical condition \e{technicalansatz}. Now let us prove the other
direction. Putting \mb{k=0} in the first of the two conditions \e{cpreliecond},
one gets after relabelling
\beq
\forall k,n:~~~0\leq k\leq n~~~~~~\Rightarrow~~~~~~
c^{n}_{0}c^{n+1}_{k+1}
~=~c^{k}_{0}c^{n}_{k}~.
\label{cpreliecond1}
\eeq
One may apply \eq{cpreliecond1} twice to produce
\beq
 \forall n\in\{0,1,2,\ldots\}:~~
c^{n}_{0}c^{n+1}_{0}c^{n+2}_{2}~=~c^{n}_{0}c^{0}_{0}c^{1}_{0}~.
\label{twostep}
\eeq
Equation \e{twostep} and the assumption \e{technicalansatz} lead to
\beq
\forall n\in\{0,1,2,\ldots\}:~~ c^{n}_{0}~\neq ~0~
\label{prelienondeg}
\eeq
by an inductive argument. Therefore one may define non-zero numbers
\beq
   \lambda_{n}~:=~\prod_{k=0}^{n-1}c^{k}_{0}~=~c^{n-1}_{0}\lambda_{n-1}
~,~~~~~~~~~~~~\lambda_{0}~:=~1~.
\eeq
Repeated use of \e{cpreliecond1} leads to the solution \e{preliesol}.
Interestingly in the generic case \e{technicalansatz}, one does not need the
second of the two pre-Lie conditions \e{cpreliecond} to derive the solution
\e{preliesol}.
\proofbox

\subsection{A Co-Product}
\label{seccoproduct}

\noi 
Similar to the ``\mb{\circ}'' bracket product construction \e{operadproduct}
one may define a co-product \mb{\triangle} on 
\mb{\Sym^{\bullet}_{\epsilon}{\cal A}}.

\begin{definition}
Let there be given a set of complex numbers \mb{\gamma^{n}_{k}} with 
\mb{n\!\geq\! k\!\geq\! 0}. The {\bf co-product} 
\mb{
\triangle:~\Sym^{\bullet}_{\epsilon}{\cal A} ~\to~ 
\Sym^{\bullet}_{\epsilon}{\cal A} \oprod \Sym^{\bullet}_{\epsilon}{\cal A}}
is then defined as
\beq
 \triangle(a_{1}\soprod\ldots\soprod a_{n})
~:=~\sum_{k=0}^{n}\frac{\gamma^{n}_{k}}{k!(n\!-\!k)!}
\sum_{\pi\in S_{n}}(-1)^{\epsilon_{\pi,a}}
(a_{\pi(1)}\soprod\ldots\soprod a_{\pi(k)})\oprod
(a_{\pi(k+1)}\soprod\ldots\soprod a_{\pi(n)}) 
\label{coproddef}
\eeq
for \mb{n\in\{0,1,2, \ldots\}}.
\end{definition}

\noi 
We denote the coefficients \mb{\gamma^{n}_{k}}, \mb{n\geq k\geq 0}, with a 
Greek \mb{\gamma} to stress that they in general differ from the  
``\mb{\circ}'' bracket product coefficients \mb{c^{n}_{k}}. Recall that 
``\mb{\oprod}'' and ``\mb{\soprod}'' denote the un-symmetrized and
symmetrized tensor product in the tensor algebras \mb{T^{\bullet}{\cal A}} and 
\mb{\Sym^{\bullet}_{\epsilon}{\cal A}}, respectively, \cf \eq{ideal}. The
Grassmann parity of the co-product \mb{\triangle}, and the bracket product 
``\mb{\circ}'' are assumed to be bosonic,
\beq
\epsilon(\triangle)~=~\epsilon(\circ)~=~0~,
\eeq 
while the parity of the symmetrized and the un-symmetrized tensor products 
``\mb{\soprod}'' and ``\mb{\oprod}'' follows the suspension parity,
\beq
\epsilon(\soprod)~=~\epsilon(\oprod)~=~\epsilon~.
\eeq
The co-product definition \e{coproddef} is by polarization equivalent to
\beq
 \triangle(a^{\soprod n})
~=~\sum_{k=0}^{n}\beta^{n}_{k}~ a^{\soprod k}\oprod a^{\soprod(n-k)}
~,~~~~~~~~~~\epsilon(a)~=~\epsilon~, \label{coproddefaaa}
\eeq
where we assume that an analogue of \eq{bcdef} holds for the Greek co-product
coefficients \mb{\beta^{n}_{k}} and \mb{\gamma^{n}_{k}},
\beq
  \beta^{n}_{k}~\equiv~\twobyone{n}{k}\gamma^{n}_{k}
~,~~~~~~~~~~~~~~0\leq k\leq n~.
\label{betagammadef}
\eeq 
The standard co-product \mb{\triangle} on 
\mb{\Sym^{\bullet}_{\epsilon}{\cal A}} corresponds to coefficients 
\mb{\gamma^{n}_{k}=1}, \cf \Ref{penkava95}.

\begin{definition}
A co-product \mb{\triangle} is {\bf co-associative} if
\beq
 ({\bf 1} \oprod \triangle) \triangle
~=~(\triangle \oprod{\bf 1}) \triangle~.
\label{defcoasso}
\eeq
\end{definition}

\noi
Co-associativity is equivalent to 
\beq
\forall k,\ell,m\geq 0:~~
\beta^{k+\ell+m}_{k}\beta^{\ell+m}_{\ell}
~=~\beta^{k+\ell+m}_{k+\ell}\beta^{k+\ell}_{k}~,
\label{coassocond}
\eeq
which generically looks like 
\mb{\beta^{k+\ell}_{k}=\frac{\lambda_{k}\lambda_{\ell}}
{\lambda_{k+\ell}}}. 
Similarly, a co-product \mb{\triangle} is co-commutative 
if \mb{\beta^{n}_{k}=\beta^{n}_{n-k}}.

\begin{definition}
A linear operator \mb{\delta:~\Sym^{\bullet}_{\epsilon}{\cal A}~\to~
\Sym^{\bullet}_{\epsilon}{\cal A}} is a {\bf co-derivation} \cite{penkava95} if
\beq
\triangle\delta~=~
\left(\delta\oprod{\bf 1}+{\bf 1}\oprod\delta\right)\triangle~.
\label{defcoderivation}
\eeq
\end{definition}

\subsection{A Lifting}
\label{secprops}

\noi
We now describe a lifting map ``\mb{\sim}'': \mb{\Phi\mapsto\tilde{\Phi}} for
\mb{\bullet}-brackets.

\beq
\begin{array}{rcccl}
&& \tilde{\Phi} && \cr 
\Sym^{\bullet}_{\epsilon}{\cal A} && \longrightarrow
&&\Sym^{\bullet}_{\epsilon}{\cal A} \cr\cr
&\searrow &&   \cr 
\Phi&&&&\cr
&&{\cal A}&&
\end{array}  
\eeq
One co-product plays a special r\^ole in this lifting. This is the co-product
that corresponds to the ``\mb{\circ}'' bracket product itself, \ie when the
co-product coefficients \mb{\gamma^{n}_{k}} are equal to the ``\mb{\circ}''
product coefficients \mb{c^{n}_{k}}. Let this particular co-product be denoted
by a triangle \mb{\stackrel{\circ}{\triangle}} with a ``\mb{\circ}'' on top. 

\begin{definition} 
Let there be given a bracket product ``\mb{\circ}'' and a \mb{\bullet}-bracket
\mb{\Phi:~\Sym^{\bullet}_{\epsilon}{\cal A} \to {\cal A}}. The {\bf lifted
bracket} \mb{\tilde{\Phi}:~\Sym^{\bullet}_{\epsilon}{\cal A} \to
\Sym^{\bullet}_{\epsilon}{\cal A}} is defined as
\beq
\tilde{\Phi}~:=~\Sym^{\bullet}_{\epsilon}(\Phi\oprod {\bf 1})
\stackrel{\circ}{\triangle}~,
\eeq
or written out,
\beq
 \tilde{\Phi}(a_{1}\soprod\ldots\soprod a_{n})
~:=~\sum_{k=0}^{n}\frac{c^{n}_{k}}{k!(n\!-\!k)!}
\sum_{\pi\in S_{n}}(-1)^{\epsilon_{\pi,a}}
\Phi^{k}(a_{\pi(1)}\soprod\ldots\soprod a_{\pi(k)})\soprod
a_{\pi(k+1)}\soprod\ldots\soprod a_{\pi(n)} \label{propdef}
\eeq
for \mb{n\in\{0,1,2, \ldots\}}. 
\end{definition}

\noi
The definition \e{propdef} is by polarization equivalent to
\beq
 \tilde{\Phi}(e^{\soprod a})
~=~\sum_{k,\ell \geq 0}\frac{c^{k+\ell}_{k}}{k!\ell!}
\Phi^{k}(a^{\soprod k})\soprod a^{\soprod \ell}
~,~~~~~~~~~~\epsilon(a)~=~\epsilon~, \label{propexpa}
\eeq
where we have defined a formal exponentiated algebra element as
\beq
 e^{\soprod a}~:=~\sum_{n\geq 0} \frac{1}{n!}a^{\soprod n}~,~~~~~~~~~~
\epsilon(a)~=~\epsilon~.
\eeq
In case of a standard bracket product  ``\mb{\circ}'' with coefficients 
\mb{c^{n}_{k}=1}, the lifting \e{propdef} becomes the standard lifting 
\cite{penkava95} of a \mb{\bullet}-bracket \mb{\Phi} to a co-derivations
\mb{\delta=\tilde{\Phi}}.

\begin{proposition}
Let there be given a bracket product ``\mb{\circ}'' and a co-product
\mb{\triangle}. All lifted brackets \mb{\tilde{\Phi}} are co-derivations, if
and only if
\beq
\forall k,\ell,m\geq 0:~~\left\{
\begin{array}{rcl}
c^{k+\ell+m}_{k}\gamma^{\ell+m+1}_{\ell+1}
&=& \gamma^{k+\ell+m}_{k+\ell}c^{k+\ell}_{k}~, \cr \cr
c^{k+\ell+m}_{k}\gamma^{\ell+m+1}_{\ell}
&=& \gamma^{k+\ell+m}_{\ell}c^{k+m}_{k}~. \end{array}
\right.
\label{ccoderivcond}
\eeq
\label{propositioncoderivation}
\end{proposition}

\noi
The two co-derivation conditions \e{ccoderivcond} become identical if the
co-product \mb{\triangle} is co-commutative. The
Proposition~\ref{propositioncoderivation} suggests that one should adjust 
the co-product \mb{\triangle} according to which 
``\mb{\circ}'' product one is studying. {}For instance, if the bracket product
coefficients \mb{c^{n}_{k}=\lambda_{k}\lambda_{n-k+1}/ \lambda_{n}} are of the
generic pre-Lie form \e{preliesol}, it is possible to satisfy the two 
co-derivation conditions \e{ccoderivcond} by choosing co-product coefficients
of the form \mb{\gamma^{n}_{k}=\lambda_{k}\lambda_{n-k}/ \lambda_{n}}. This
choice of co-product is at the same time both co-associative and 
co-commutative.

\noi
{\it Proof of Proposition~\ref{propositioncoderivation}:}~~{}First note that
for two vectors \mb{a,b \in{\cal A}} with \mb{\epsilon(a)=\epsilon},
\beq
\triangle(b\soprod  a^{\soprod n})~=~
\sum_{k=0}^{n}\frac{k\!+\!1}{n\!+\!1}\beta^{n+1}_{k+1}
(b\soprod a^{\soprod k})\oprod a^{\soprod(n-k)}
+\sum_{k=0}^{n}\frac{n\!-\!k\!+\!1}{n\!+\!1}\beta^{n+1}_{k}
a^{\soprod k}\oprod (b\soprod a^{\soprod(n-k)})
\label{coderivhelp}
\eeq
for \mb{n\in\{0,1,2,\ldots\}}. Therefore
\bea
\triangle\tilde{\Phi}(a^{\soprod n}) 
&=& \!\!\!\!\!\!\!\!
\sum_{\footnotesize \begin{array}{c}k,\ell,m\geq 0 \cr
k\!+\!\ell\!+\!m=n\end{array}}
b^{n}_{k} \left[ \beta^{\ell+m+1}_{\ell+1}
\frac{\ell\!+\!1}{\ell\!+\!m\!+\!1}
\left(\Phi^{k}(a^{\soprod k})\soprod 
a^{\soprod \ell}\right)\oprod a^{\soprod m}\right. \cr
&&\left. ~~~~~~~~~+\beta^{\ell+m+1}_{\ell}
\frac{m\!+\!1}{\ell\!+\!m\!+\!1}
a^{\soprod \ell}\oprod\left(\Phi^{k}(a^{\soprod k})\soprod
a^{\soprod m}\right)\right]~.
\label{codo1}
\eea
On the other hand,
\beq
(\tilde{\Phi}\oprod {\bf 1})(a^{\soprod n})
~=~\!\!\!\!\!\!\!\!
\sum_{\footnotesize\begin{array}{c}k,\ell,m\geq 0 \cr
k\!+\!\ell\!+\!m=n\end{array}}
\beta^{n}_{k+\ell}b^{k+\ell}_{k}
\left(\Phi^{k}(a^{\soprod k})\soprod 
a^{\soprod \ell}\right)\oprod a^{\soprod m}~,
\label{codo2}
\eeq
and
\beq
({\bf 1}\oprod\tilde{\Phi})(a^{\soprod n})
~=~\!\!\!\!\!\!\!\!
\sum_{\footnotesize\begin{array}{c}k,\ell,m\geq 0 \cr
k\!+\!\ell\!+\!m=n\end{array}}
\beta^{n}_{\ell}b^{k+m}_{k}
a^{\soprod \ell}\oprod\left(\Phi^{k}(a^{\soprod k})\soprod
a^{\soprod m}\right)~.
\label{codo3}
\eeq
Next insert the three expressions \e{codo1}, \es{codo2}{codo3} into the
co-derivative definition \e{defcoderivation} with \mb{\delta=\tilde{\Phi}}.
By comparing coefficients one derives
\beq
\forall k,\ell,m\geq 0:~~\left\{
\begin{array}{rcl}
b^{k+\ell+m}_{k}\beta^{\ell+m+1}_{\ell+1}
\frac{\ell\!+\!1}{\ell\!+\!m\!+\!1}
&=& \beta^{k+\ell+m}_{k+\ell}b^{k+\ell}_{k}~, \cr \cr
b^{k+\ell+m}_{k}\beta^{\ell+m+1}_{\ell}
\frac{m\!+\!1}{\ell\!+\!m\!+\!1}
&=& \beta^{k+\ell+m}_{\ell}b^{k+m}_{k}~. \end{array}
\right.
\label{bcoderivcond}
\eeq
The two conditions \e{ccoderivcond} follow by translating \e{bcoderivcond}
into the ``\mb{c}'' picture \wtho \eq{bcdef}.
\proofbox

\begin{proposition}
A  bracket product ``\mb{\circ}'' is pre-Lie, if and only if the lifting map
``\mb{\sim}'' is an algebra homomorphism, \ie for all brackets 
\mb{\Phi,\Phi^{\prime}:~\Sym^{\bullet}_{\epsilon}{\cal A} \to {\cal A}},
\beq
(\Phi\circ\Phi^{\prime})^{\sim}~=~\tilde{\Phi}\tilde{\Phi}^{\prime}~.
\label{algebrahomo}
\eeq
\label{propositiontildeprelie}
\end{proposition}

\noi
{\it Proof of Proposition~\ref{propositiontildeprelie}:}~~{}First note that
for two vectors \mb{a,b \in{\cal A}} with \mb{\epsilon(a)=\epsilon},
\bea
\tilde{\Phi}(b\soprod  e^{\soprod a})&=&\sum_{k,\ell \geq 0}
\frac{c^{k+\ell+1}_{k+1}}{k!\ell!}
\Phi^{k+1}(b\soprod  a^{\soprod k})\soprod a^{\soprod\ell}
+\sum_{k,\ell \geq 0}\frac{c^{k+\ell+1}_{k}}{k!\ell!}
\Phi^{k}(a^{\soprod k})\soprod b\soprod a^{\soprod\ell}~.
\label{propbexpa}
\eea
Therefore the composition of two lifted brackets \mb{\tilde{\Phi}} and
\mb{\tilde{\Phi}^{\prime}} is
\bea
\tilde{\Phi}\tilde{\Phi}^{\prime}(e^{\soprod a})
&=&\sum_{k,\ell,m \geq 0}
\frac{c^{\ell+m+1}_{\ell+1}c^{k+\ell+m}_{k}}{k!\ell!m!}
\Phi^{\ell+1}\left(\Phi^{\prime k}(a^{\soprod k})\soprod
a^{\soprod \ell }\right)\soprod a^{\soprod m} \cr
&&+\sum_{k,\ell,m \geq 0}\frac{c^{\ell+m+1}_{\ell}c^{k+\ell+m}_{k}}{k!\ell!m!}
\Phi^{\ell}(a^{\soprod \ell})\soprod
\Phi^{\prime k}(a^{\soprod k})\soprod a^{\soprod m}~.
\label{proppropexpa}
\eea
On the other hand,
\beq
(\Phi\circ\Phi^{\prime})^{\sim}(e^{\soprod a})~=~\sum_{k,\ell,m \geq 0}
\frac{c^{k+\ell+m}_{k+\ell}c^{k+\ell}_{k}}{k!\ell!m!}
\Phi^{\ell+1}\left(\Phi^{\prime k}(a^{\soprod k})
\soprod a^{\soprod \ell}\right)\soprod a^{\soprod m}~.
\label{phiphitilde}
\eeq
By comparing coefficients in \eqs{proppropexpa}{phiphitilde} one sees that the
condition \e{algebrahomo} is equivalent to the two pre-Lie conditions
\e{cpreliecond}.
\proofbox

\subsection{Normalization}

\noi
One may get back the \mb{\bullet}-bracket 
\mb{\Phi:~\Sym^{\bullet}_{\epsilon}{\cal A}\to{\cal A}} from its lifted 
bracket \mb{\tilde{\Phi}:~\Sym^{\bullet}_{\epsilon}{\cal A}
\to\Sym^{\bullet}_{\epsilon}{\cal A}} \wtho the projection maps
\mb{\pi_{n}:~\Sym^{\bullet}_{\epsilon}{\cal A}\to\Sym^{n}_{\epsilon}{\cal A}}.
In particular,
\beq
\pi_{1}\circ\tilde{\Phi}(a_{1}\soprod\ldots\soprod a_{n})
~=~c^{n}_{n}~\Phi(a_{1}\soprod\ldots\soprod a_{n})~.
\label{pione}
\eeq

\begin{definition} 
A ``\mb{\circ}'' bracket product is {\bf normalized} if
\beq
\forall n\in\{0,1,2,\ldots\}:~c^{n}_{n}~=~1~.
\label{normalizationcondition}
\eeq
\end{definition}

\noi
Obviously, a normalized ``\mb{\circ}'' product is non-degenerate, \cf 
\eq{nondegenerateproduct}. Moreover, in the normalized case one may sharpen 
\eq{pione} into 
\beq
\Phi~=~\pi_{1}\circ\tilde{\Phi}~,
\eeq
so that the following diagram is commutative.

\beq
\begin{array}{rcccl}
&& \tilde{\Phi} && \cr 
\Sym^{\bullet}_{\epsilon}{\cal A} && \longrightarrow
&&\Sym^{\bullet}_{\epsilon}{\cal A} \cr\cr
&\searrow && \swarrow  \cr 
\Phi&&&&\pi_{1}\cr
&&{\cal A}&&
\end{array}  
\eeq

\noi
In the normalized case the product of brackets \e{operadproduct} may be
related to composition of the lifted brackets as
\beq
 \Phi\circ\Phi^{\prime}~=~\Phi\tilde{\Phi}^{\prime}
~=~\pi_{1}\tilde{\Phi}\tilde{\Phi}^{\prime}~.
\eeq
This carries the advantage that composition, unlike the bracket product, is
born associative. It is natural to ask, what \mb{c^{n}_{k}} coefficients
would satisfy a nilpotency condition for a lifted bracket
\beq
 \tilde{\Phi} \tilde{\Phi}~=~0~,~~~~~~~~~~~\epsilon_{\Phi}~=~1~?
\label{strongnilpotency}
\eeq
Contrary to the nilpotency relations \e{nilprel} for the ``\mb{\circ}'' 
product, which are first-order equations in the \mb{c^{n}_{k}} coefficients,
the nilpotency conditions \e{strongnilpotency} are quadratic in the 
\mb{c^{n}_{k}} coefficients. We end this discussion with a corollary.

\begin{corollary}
{}For a normalized pre-Lie product ``\mb{\circ}'', the bracket \mb{\Phi} is 
nilpotent \wrt the bracket product ``\mb{\circ}'', if and only if the lifted
bracket \mb{\tilde{\Phi}} is nilpotent \wrt composition, \ie
\beq
\Phi\circ\Phi~=~0 ~~~~~~~\Leftrightarrow~~~~~~\tilde{\Phi}\tilde{\Phi}~=~0~.
\eeq
\label{corollarynihil}
\end{corollary}

\subsection{The Ward Solution Revisited}
\label{secwardrevisited}

\noi
As an example let us consider the Ward solution \e{wardsol1} of the derived
bracket hierarchy, but this time normalized according to the 
\eq{normalizationcondition},
\beq
c^{n}_{k}~=~\left\{\begin{array}{rcl}1 &&{\rm for}~k=1 \vee k=n~, \cr
   0 && {\rm otherwise}~,\end{array} \right. 
\label{wardsol3}
\eeq
or equivalently using \eq{defcxy},
\beq
 c(x,y)~=~x(e^{y}-1)+e^{x}~.\label{wardsol4}
\eeq
This solution is identical to the original solution 
\es{wardsol1}{wardsol2}, except for the fact that we have divided the first 
Ward identity \e{wardid1} with \mb{2}, which is always permissible, \cf 
Subsection~\ref{secrenormalization}.

\begin{proposition}
The Ward solution \e{wardsol3} is pre-Lie, normalized and satisfies the
nilpotency relations for the derived bracket hierarchy, \ie 
\mb{[Q,Q]=0~~ \Rightarrow~~ \Phi_{Q} \circ \Phi_{Q}=0}.
\label{wardproposition}
\end{proposition}

\noi
We conclude that there is a non-empty overlap between the derived solutions
found in Section~\ref{secderoperad}, the pre-Lie property \e{defprelie} and
the normalization condition \e{normalizationcondition}.

\noi
{\it Proof of Proposition~\ref{wardproposition}:}~~
The Ward solution \e{wardsol3} is obviously normalized, \cf 
\eq{normalizationcondition}. We saw in Section~\ref{secderoperad} that it
satisfies the nilpotency relations for the derived bracket hierarchy. The
pre-Lie property \e{cpreliecond} may either be checked directly, or perhaps
more enlightening, one may consider product coefficients
\mb{c^{n}_{k}=\lambda_{k}\lambda_{n-k+1}/ \lambda_{n}} of the generic 
pre-Lie form \e{preliesol} with 
\beq
\lambda_{k}~=~\left\{\begin{array}{rcl}1 &{\rm for}&k\in\{0,1\}~, \cr
   \epsilon^{k} &{\rm for}&k\in\{2,3,4,\ldots\}~,\end{array} \right. 
\label{lambdasol}
\eeq
where \mb{\epsilon} is a non-zero complex number.
One may easily see that the solution \e{preliesol} becomes the Ward solution
\e{wardsol3} in the limit \mb{\epsilon\to 0}. Hence the Ward solution
\e{wardsol3} is also pre-Lie by continuity.
\proofbox

\section{Conclusions}
\label{secconclusion}

\noi
The paper contains the following main new results:
\begin{itemize}
\item
We found a non-commutative generalization \e{genkoszulconstruct} of the 
higher Koszul brackets, such that they form a homotopy Lie algebra.
\item
We found the most general nilpotency relations for the derived bracket 
hierarchy, \cf Theorem~\ref{theoremderived}.
\item
We defined and calculated the higher Courant brackets, \cf 
Proposition~\ref{propositionhighercourant}.
\end{itemize}

\noi 
A common platform for all of these topics is provided by a (generalized) \hla
that allows for arbitrary non-degenerate prefactors \mb{c^{n}_{k}} in the
nilpotency relations. These prefactors can equivalently be viewed as a
non-standard bracket product ``\mb{\circ}'', \cf \eq{operadproduct}. The
generalization is desirable because the original definition \e{stasheffchoice}
of Lada and Stasheff \cite{ladastasheff93} excludes important systems, for
instance the derived bracket hierarchy, which in all other respects has the
hallmarks of an \mb{L_{\infty}} algebra, \cf Section~\ref{secderoperad}. In 
Section~\ref{secsuppl} we have analyzed the (generalized) \hlas further, and
we have displayed their co-algebra structures. The question remains whether
the (generalized) \hla definition \e{nondegenerateproduct} considered in this
paper is the final say. {}For instance, should one demand the
pre-Lie property \e{defprelie} of a \hla definition? As we saw in
Subsection~\ref{secwardrevisited} the pre-Lie property carries a small, but
non-empty, overlap with the solutions \es{genparsolution}{genhomsolution} 
to the derived bracket hierarchy. A complete answer will require further 
studies and definitely more examples.

\vspace{0.8cm}
\noindent
{\sc Acknowledgement:}~Special thanks go to J.D.~Stasheff and the referees for
remarks on the first version of the text. The work of K.B.\ is supported by
the Ministry of Education of the Czech Republic under the project
MSM 0021622409.

%\vspace{0.3cm}
%\newpage

\end{document}